\documentclass[12pt]{article}

\usepackage{natbib}
\usepackage[nottoc,notlof,notlot]{tocbibind}

\bibpunct{(}{)}{;}{a}{}{;}
\usepackage{breakcites}
\usepackage{graphicx,xcolor}
\graphicspath{ {images/} }
\usepackage[algoruled]{algorithm2e}
\usepackage[pagebackref=false,breaklinks=true]{hyperref} 
\hypersetup{
    colorlinks=true,
    citecolor=blue,
    filecolor=blue,
    linkcolor=blue,
    urlcolor=blue,
    bookmarksopen=true,
    pdfstartview=FitH
}
\usepackage{caption,threeparttable}

\usepackage{amsthm,amsmath}
\usepackage{bbm}
\usepackage{bm}
\usepackage{amsfonts}%
\usepackage{amssymb}%
\usepackage{mathrsfs}

\usepackage{geometry} 
\usepackage{graphicx} 
\usepackage{framed}
\usepackage{caption}
\usepackage{subcaption}

\usepackage{booktabs} 
\usepackage{array} 
\usepackage{paralist} 
\usepackage{verbatim} 

\usepackage{lipsum}
\usepackage{setspace}
\renewcommand{\baselinestretch}{1.5}
\usepackage[singlelinecheck=false]{caption}
\usepackage{tabularx}
\usepackage{tikz}

\usepackage{multicol}
\usepackage{multirow}
\usepackage{bigstrut}

\newtheorem{theorem}{Theorem}
\newtheorem{assump}{Assumption}
\newtheorem{assump*}{Assumption}[section]
\newtheorem{lemma}{Lemma}

\theoremstyle{definition}
\newtheorem{definition}{Definition}

\newtheorem{remark}{Remark}

\def\equationautorefname~#1\null{(#1)\null} 

\newtheorem{myexp}{Example}

\def\Pr{\mathrm{Pr}}
\def\Var{\mathrm{Var}}
\def\Cov{\mathrm{Cov}}

\newcommand{\argmin}{\operatornamewithlimits{argmin}}

\newcommand{\E}{{\bf E}}

\newcommand{\N}{\mathbf{N}}

\title{Regression Discontinuity Design with Spillovers\footnote{We thank Dmitry
Arkhangelsky, Andrii Babii, Ivan Canay, Elena Dal Torrione, Peter Hull, Simon Lee, Matt Masten, Michael Pollman, Jack Porter, Shuyang Sheng, Kevin Song, Valentyn Verdier, Basit Zafar for helpful comments and discussions.}
}
\author{ \parbox{0.45\linewidth}{\centering
\normalsize \setstretch{1.2}
Eric Auerbach \\
Department of Economics \\
Northwestern University \\
\url{eric.auerbach@northwestern.edu}}
\parbox{0.45\linewidth}{\centering
\normalsize \setstretch{1.2}
Yong Cai \\
Department of Economics \\
University of Wisconsin-Madison\\
\url{yong.cai@wisc.edu}}
\\
\\
\parbox{0.45\linewidth}{\centering
\normalsize \setstretch{1.2}
Ahnaf Rafi \\
Department of Economics \\
University of Virginia \\
\url{ahnafrafi@virginia.edu}}
}

 \date{\parbox{\linewidth}{\centering%
 \hfill \\
  \today \endgraf}}

\begin{document}
\maketitle

\begin{abstract} \setstretch{1}\noindent
	This paper studies regression discontinuity designs (RDD) when linear-in-means spill-overs occur between units that are close in their running variable. We show that the RDD estimand depends on the ratio of two terms: (1) the radius over which spillovers occur and (2) the choice of bandwidth used for the local linear regression. RDD estimates direct treatment effect when radius is of larger order than the bandwidth and total treatment effect when radius is of smaller order than the bandwidth. When the two are of similar order, the RDD estimand need not have a causal interpretation. To recover direct and spillover effects in the intermediate regime, we propose to incorporate estimated spillover terms into local linear regression. Our estimator is consistent and asymptotically normal and we provide bias-aware confidence intervals for direct treatment effects and spillovers. In the setting of \cite{gonzalez2021cell}, we detect endogenous spillovers in voter fraud during the 2009 Afghan Presidential election. We also clarify when the donut-hole design addresses spillovers in RDD. \\
\end{abstract}

\section{Introduction}
Regression discontinuity design (RDD) is a popular method for causal inference and policy evaluation, particularly in settings where experimental manipulation is not possible. 
However, in the continuity-based framework of \citet{hahn2001identification}, identification of treatment effect at the cutoff relies on the Stable Unit Treatment Value Assumption (SUTVA). This requires each unit’s potential outcomes to be unaffected by the treatment status of other units and may be unrealistic: in many economic environments, the variables of interest are equilibrium objects shaped by strategic interactions, making SUTVA violations natural. Although RDD is known to identify treatment effects local to the cutoff under SUTVA, this need not hold once SUTVA is violated.

Following \cite{manski1993identification}, the literature on social interactions typically emphasizes two types of SUTVA violations: exogenous spillovers (also known as contextual effects) and endogenous spillovers. Exogenous spillovers occur when  the outcome of an agent depends directly on the treatment status of their neighbors. On the other hand, endogenous spillovers arise when the outcome of an agent depends on the outcome of their neighbors. For a concrete example, consider \cite{gonzalez2021cell}, which studies the effect of cell phone coverage on voter fraud during the 2009 Afghan presidential election. Their spatial RDD compares polling stations on either side of the boundary of cell phone coverage areas and finds that cell phone coverage reduced fraud. However, the author is also explicitly concerned about potential spillovers. For example, suppose coverage at a polling station reduces fraud by allowing voters to report suspicious behavior to the electoral commission. Voters at non-covered stations could potentially walk to covered stations to report fraud, giving rise to exogenous spillovers. Alternatively, corrupt politicians might strategically allocate fraud to areas with less monitoring. In this way, higher fraud in one polling station will reduce the need to commit fraud in neighboring polling stations. This is a form of endogenous spillovers. Because spillovers appear plausible in many settings, it is important to understand their effects on the estimand of RDD.


This paper studies RDD in the presence of spillovers. To do so, we extend the framework of \cite{hahn2001identification} to incorporate exogenous and endogenous spillovers that occur along the running variable. We make two important modeling assumptions. Firstly, we assume that the outcome of a given unit depends linearly on the mean treatment status and mean outcome of their neighbors. The linear-in-means specification characterizes a large literature on peer effects (see e.g. \citealt{manski1993identification, bramoulle2009identification, deGiorgi2010identification,goldsmith2013social,dePaula2024identifying}) and spatial autoregression models (e.g. \citealt{cliff1973spatial,kelejian1998generalized, kelejian2010specification,lee2004asymptotic, lee2007gmm}).  Secondly, we assume that the running variable which defines the RDD also determines the social interactions in the linear-in-means model. Specifically, units with similar values of the running variables are also assumed to be neighbors. When the running variable is geographical coordinates, this model captures interaction between units that are close in space. In the classic \cite{thistlethwaite1960regression}, which studies the effect of scholarships on academic achievements, the running variable is test scores. Our model then captures the idea that students with similar test scores exert peer effects on one another, for example by forming study groups. 

We make two main contributions. Firstly, we show that the estimand of RDD depends on the ratio of two terms: (1) the radius over which spillovers occur and (2) the bandwidth used for local linear regression. Specifically, RDD estimates direct treatment effect at the boundary when the radius is of larger order than the bandwidth. This is the effect on a unit at the boundary when we switch their treatment status from control to treatment, keeping treatment assignment fixed for all other units. When radius is of smaller order than the bandwidth, RDD instead estimates total treatment effect, which is the effect on the unit at the boundary when we switch the treatment status of the entire population from control into treatment. In the regime where radius is of similar order as the bandwidth, the RDD estimand is a mixture of the above effects and lacks a clear interpretation. We argue, however, that this is the more realistic regime wherein the asymptotic approximation captures more features of the finite sample distribution.

As such, our second contribution is a proposal for recovering direct and spillover effects in the intermediate regime. We do so by incorporating estimated spillover terms into local linear regression, in a procedure we call the local spillover regression (LSR). We show that LSR, which is essentially the local analog of peer effects regressions, leads to consistent estimators which are asymptotically normally distributed. We also adapt the method of \cite{armstrong2020simple} to construct bias-aware confidence intervals for direct and spillover effects.

Simulation results show that LSR performs well relative to local linear regression with naively chosen bandwidths. The adaptive bandwidth rules of \cite{calonico2014robust} and \cite{armstrong2020simple} lead to estimators that are more robust to spillovers, but our method can achieve lower MSE and better coverage particularly in settings with intermediate amounts of spillovers. We apply LSR to the setting of \cite{gonzalez2021cell} and find evidence of positive endogenous spillovers in electoral fraud. The implied social multiplier may be relevant for cost-benefit analysis for fraud deterrence policies and may provide motivation for network-based interventions. Along the way, we also discuss the estimands of the popular RD donut and highlight settings under which it recovers either direct or total treatment effects.

This paper contributes to the vast literature on RDD (see \citealt{cattaneo2022regression} for a recent review). Empirical researchers have long been concerned with spillovers in RDD, with papers such as \cite{jardim2022boundary} arguing against the use of spatial discontinuity design for policy evaluation. However, to our knowledge, prior theoretical work on this issue is limited to \cite{aronow2017rdspill}, which conducts their analysis under the local randomization framework of \cite{cattaneo2015randomization}, finding that RDD always recovers a weighted average of direct treatment effect. We consider RDD under the continuity-based approach of \cite{hahn2001identification} and find that the estimand can exhibit more complex behavior, particularly when neighborhoods are determined by the running variable.
Since the first version of this paper was posted, \cite{torrione2024regression} and \cite{borusyak2024rdaggregation} have also studied RDD in the presence of spillovers. \cite{torrione2024regression} takes the continuity-based approach, assuming that the running variable and the variable determining neighborhood structure have a continuous joint density. They find that RDD recovers a weighted average of direct treatment effect, drawing a connection to multi-score RDD. Our paper focuses on the case where neighborhoods are completely determined by the running variable and is unique in having spillovers that appear in the asymptotic approximation, provided that radius and bandwidth are scaled appropriately. This paper also differs from \cite{aronow2017rdspill} and \cite{torrione2024regression} in considering endogenous spillovers, which arises in many settings of interest to economists. Finally, we note that the aforementioned papers, as well as ours, focus on spillovers between units within an RD. \cite{borusyak2024rdaggregation} studies the aggregation of multiple RDDs and considers effects of spillovers across designs.

This paper joins a burgeoning body of work that considers violations of SUTVA under various research designs, such as experiments (e.g. \citealt{hudgens2008toward,aronow2017estimating,savje2021average, hu2022average, leung2022causal,li2022random, gao2023causal, vazquez2023experiment, auerbach2025local}), differences-in-differences (e.g. \citealt{clarke2017estimating,butts2021difference, xu2023difference}), synthetic control (\citealt{cao2019estimation}), instrumental variables (\citealt{sobel2006randomized, vazquez2023iv}) and other observational settings (\citealt{forastiere2020causal}). RDD poses unique challenges relative to these other settings because it concerns parameters that are local to the cutoff and estimation is nonparametric. Additionally, we consider endogenous spillovers, which has received relatively less attention in this literature, though exceptions include, in the experiment context, \cite{leung2022causal,munro2021treatment, li2023experimenting, munro2023efficient} and \cite{faridani2024linear}. In the presence of endogenous spillovers, a unit's outcome may depend on the treatment status of the entire population, even if spillovers are assumed to have a small radius. The interaction of this dependence with the discontinuity of the observed outcome function poses novel technical challenges that we address.

The rest of the paper is organized as follows. Section \ref{section--setup} describes our econometric framework. Section \ref{section--estimands} characterizes the estimand of local linear regression under spillovers. Section \ref{section--estimands--donut_hole} in particular considers the use of donut-hole RDs. Section \ref{section--localspilloverreg} presents the local spillovers regression, our proposal for recovering direct treatment and spillover effects. Section \ref{section--application} applies our method to the setting of \cite{gonzalez2021cell}. Section \ref{section--conclusion} concludes. Proofs are contained in the Appendix. Simulation results and the auxiliary lemmas used in the proofs are available in the Supplemental Appendix.

The remainder of this paper uses the following notation. We write $A_n \gg B_n$ if $A_n/B_n \to \infty$, $A_n \propto B_n$ if $A_n/B_n \to c$ where $0 < c < \infty $, and $A_n \ll B_n$ if $A_n/B_n \to 0$. Let $\boldsymbol{\iota}$ and $\boldsymbol{0}$ be constant functions that take values 1 and 0 on $[-1,1]$ respectively.

\section{Econometric Framework}\label{section--setup}

In this section, we introduce the econometric framework for studying RDD with spillovers. Section \ref{section--setup--model} presents our model for potential outcomes. Section \ref{section--setup--parameters} defines the parameters of interest. Section \ref{section--setup--neighborhood} discusses our assumed neighborhood structure.
Section \ref{section--setup--sampling} presents the sampling process.

\subsection{Potential Outcomes}\label{section--setup--model}

Consider a continuum of agents that are indexed by their coordinates $z  \in \mathcal{Z} = [-1,1]$. For convenience, we will assume that agents are uniformly distributed according to $F = \text{Uniform}(\mathcal{Z})$, so that their density with respect to the Lebesgue measure is $f(z) = \frac{1}{2}$. Our results extend easily to any density bounded away from $0$ at the cutoff, although our expressions for the bias of RDD need not apply in the more general setting.

We work in the usual potential outcomes framework with binary treatment, where the observed outcome at $z$ satisfies
\begin{equation*}
	Y_d(z) = d(z) Y^+_d(z) + (1-d(z))Y^-_d(z)~.
\end{equation*}
Here, $d: \mathcal{Z} \to \{0,1\}$ is the treatment assignment function. $Y^+_d(z)$ and $Y^-_d(z)$ are the potential outcomes under treatment and control respectively. Because of the treatment assignment rule in RDD, defined below, we will denote quantities related to treatment with ``$+$" and those related to control with ``$-$".

Potential outcomes are indexed by $d$ because as a result of spillovers, they may depend on the entire treatment assignment function. Let every agent $z$ have the set of relevant neighbors $R(z) \subset \mathcal{Z}$ with  measure $|R(z)|$. These are agents whose realized outcomes affect $z$. Let potential outcomes be defined as follows:
\begin{assump}[Potential Outcomes]  \label{assump-model}
	For a given treatment assignment function $d: \mathcal{Z} \to [0,1]$ and neighborhood structure $R: \mathcal{Z} \to \mathcal{P}(\mathcal{Z})$, the potential outcome of agent $z$ under treatment $d$ is:
	\begin{align*}
		Y_d^+(z) = m^+(z) + \delta(z)\mu_d(z) + \gamma(z)\nu_d(z) \\
		Y_d^-(z) =	m^-(z) + \delta(z)\mu_d(z) + \gamma(z)\nu_d(z)
	\end{align*}
	where
	\begin{align*}
		\mu_d(z) & = \frac{1}{|R(z)|}\int_{R(z)}  Y_d(u) f(u) \, du \\
		\nu_d(z) & = \frac{1}{|R(z)|}\int_{R(z)}  d(u) f(u) \, du~.
	\end{align*}
	Furthermore, let $m^+(z)$ and $m^-(z)$ be Lipschitz continuous on $\mathcal{Z} = [-1, 1]$ with Lipschitz constant $C$. Let $\delta(z)$ be continuously differentiable with $|\frac{d\delta}{dz}| < C_\delta$, $\sup_{z \in \mathcal{Z}} |\delta(z)| \leq \bar{\delta}$, $\sup_{z_1,z_2 \in \mathcal{Z}} |\delta(z_1) - \delta(z_2)| \leq \bar{\delta}$, $\bar{\delta}<1$. Finally, let $\gamma(z)$ be continuously differentiable with $|\frac{d\gamma}{dz}| < C_\gamma$.
\end{assump}

The above model extends the continuity-based framework of \cite{hahn2001identification} to include two sources of spillovers, both of which are linear-in-means. The exogenous spillover term is $\gamma(z)\nu_d(z)$, where $\nu_d(z)$ is the mean treatment status of $z$'s neighbors, and $\gamma(z)$ is the effect of this term on $z$'s outcome. The endogenous spillover term is $\delta(z)\mu_d(z)$, where $\mu_d(z)$ is the mean outcome of $z$'s neighbors, and $\delta(z)$ is the effect of this term. Spillovers are assumed to enter the treated and control potential outcomes in the same way. As will become clear in Section \ref{section--setup--parameters}, this leads to a particularly simple formula for the parameters of interest.

The linear-in-means assumption is potentially restrictive. However, it is the workhorse model in the peer effects (e.g. \citealt{manski1993identification, bramoulle2009identification, deGiorgi2010identification,goldsmith2013social,dePaula2024identifying}) and spatial autoregression (e.g. \citealt{cliff1973spatial,kelejian1998generalized, kelejian2010specification,lee2004asymptotic, lee2007gmm}) literatures. We consider it to be a reasonable first step for analyzing endogenous spillovers, a task which is challenging in many settings. The linear-in-means assumption is in fact stronger than necessary for our characterization of the estimands. In particular, the qualitative results in Section \ref{section--estimands} does not require linearity in the effects of spillovers, and allows neighbors to have different weights in treated and control outcomes. The main requirement being that neighborhoods have approximately bounded support. However, as will become clear in Section \ref{section--localspilloverreg}, the specific form of spillovers needs to be assumed in order to recover target parameters in our preferred asymptotic regime. We therefore focus on the above model. Nonetheless, it is more general than the standard peer effects model in that the effects of spillovers, $\delta$ and $\gamma$, is allowed to vary with $z$.

The remaining conditions in Assumption \ref{assump-model} concerns continuity of the functions $m^+, m^-, \delta$ and $\gamma$ as well as the boundedness of $\delta$. Continuity of $m^+$ and $m^-$ around the cutoff is the main identifying assumption in RDD. Standard RDD requires only that $m^+(z)$ and $m^-(z)$ are continuous at $z = 0$. However, due to spillovers, continuity of potential outcomes also requires continuity of the two functions around the boundaries of $R_n(0)$ as well as the continuity of $\gamma$ and $\delta$. Since $R_n(0)$ is fixed under some of the regimes we consider, we assume continuity over $\mathcal{Z}$ for simplicity. We also slightly strengthen the condition to Lipschitz continuity. To ensure that $Y^+_d$ and $Y^-_d$ are well-defined, we require $\sup_{z \in \mathcal{Z}} |\delta(z)| < 1$, as well as for $\sup_{z_1,z_2 \in \mathcal{Z}} |\delta(z_1) - \delta(z_2)| \leq 1$.

\subsection{Parameters of Interest}\label{section--setup--parameters}

In a setting with spillovers, any two treatment assignment functions $d$ and $d'$ can give rise to a treatment effect
$$Y_d(z) - Y_{d'}(z)$$ that is potentially of interest. Following the literature on causal inference with spillovers (see e.g. {\citealt{hudgens2008toward}), we focus on the following two parameters:

\begin{definition}[Treatment Effects] \label{definition--parameters}\hfill
\begin{itemize}
\item The direct treatment effect at $z = 0$ is
	$$\tau_\text{DIR} := Y_d^+(0) - Y_d^-(0) = m^+(0) - m^-(0)~.$$
\item  The total treatment effect at $z = 0$ is
\begin{align*}
	\tau_{\text{TOT}} &:=  Y_{\boldsymbol{\iota}}^+(0) - Y_{\boldsymbol{0}}^-(0) = m^+(0) - m^-(0) + \delta(0)\left(\mu_{\boldsymbol{\iota}}(0) - \mu_{\boldsymbol{0}}(0)\right) + \gamma(0) ~.
\end{align*}
\end{itemize}
\end{definition}

The direct treatment effect on a given agent is the effect of treatment on their outcomes, keeping the treatment assignment of all other agents unchanged at some $d$. This parameter is useful for thinking about selective rollout of some policy, under which it is plausible to consider only the direct treatment effect, treating the spillovers as fixed.
Here, we abuse notation in using $d$ to refer to two different treatment assign functions, although they are identical on $\mathcal{Z} \setminus {0}$. In principle, direct treatment effect depends on $d$. In RDD, it is natural to focus on:
\begin{definition}[RDD Treatment Assignment]  \label{definition-rdd}
$  	d(z) = \mathbf{1}\left\{z\geq 0\right\}.$
\end{definition}

The above treatment assignment function is the natural reference point for evaluating direct treatment effects in RDD. However, under Assumption \ref{assump-model}, the agent at $z = 0$ experiences the same spillover whether or not they are treated for two reasons. Firstly, $\delta(z)$ and $\gamma(z)$ are assumed to be the same regardless of treatment status. Secondly, the agent $z = 0$ is infinitesimal, so that changing their treatment status does not affect $\mu_d(z)$ or $\nu_d(z)$ for any $z \in \mathcal{Z}$. Implicitly, this is a model involving dense network asymptotics. The result of these two factors is that $\tau_\text{DIR} = m^+(0) - m^-(0)$ regardless of $d$. Additionally, $\tau_\text{DIR}$ is equal to local average treatment effect in the standard setting with no spillovers.

With infinitesimal agents, direct treatment effects are easy to define since we can change the treatment status of one individual but still keep spillovers fixed.  Nonetheless, there is a coherent notion of spillovers in this model: if two treatment functions $d$ and $d'$ differ on a set of positive measure, then $\mu_d(z)$ and $\nu_d(z)$ need not be equal to $\mu_{d'}(z)$ and $\nu_{d'}(z)$. This point is also evident in the definition of our next parameter.

The total treatment effect on a given agent is the effect on their outcome when we switch the treatment status of the entire population from control to treatment. This corresponds to the effect on $z = 0$ from a large-scale rollout of treatment. In our notation, the subscripts ${\boldsymbol{\iota}}$ and ${\boldsymbol{0}}$ denote treatment assignments in which everyone and no one is treated respectively. As before, we will focus on the agent at the cutoff, $z = 0$. With total treatment effect, we are evaluating potential outcomes under different treatment assignment functions. Consequently, spillovers no longer cancel out. We remark that $\nu_{\boldsymbol{\iota}}(z) = 1$ and $\nu_{\boldsymbol{0}}(z) = 0$ so that exogenous spillovers at $z = 0$ is $\gamma(0)$.

%

\subsection{Neighborhood Structure}\label{section--setup--neighborhood}

In the presence of spillovers, the neighborhood structure is critical in determining the estimands of RDD. We focus on the case where spillovers occur along the running variable:
\begin{assump}[Neighborhood]\label{assump--neighborhood}
\begin{equation*}
	R_n(z) = \{u \, : \, \lVert z-u \rVert \leq r_n\}~.
\end{equation*}
\end{assump}
Here, $\lVert \cdot \rVert$ is taken to be the Euclidean distance. As such, $z$ is affected only by units whose running variable takes value within $r_n$ of $z$. All other units exert no effect. Together with an i.i.d sampling assumption introduced in the next section, the above assumption implies that the exposure map between sampled units is a random geometric graph (see e.g. \citealt{penrose2003random}). This is a well-studied model that is commonly used for spillovers and interference (see e.g. \citealt{leung2020treatment}).

When neighborhoods are defined by the running variable, nearby units that are comparable in their conditional means ($m^+, m^-$) can experience spillovers that are not comparable. Consequently, RDD may not recover any meaningful treatment effect parameters, as our analysis in Section \ref{section--estimands} shows. This fundamental tension is absent when neighborhoods are defined by another variable, say $w$, that is suitably continuous with respect to $z$. 
We will refer to such a $w$ as a background variable. Suppose for now that units are defined by their values of $(z,w)$ and that neighborhoods are defined as
\begin{equation*}
		R_n(z,w) = \{(\tilde{z}, \tilde{w}) \, : \, |\tilde{w}-w| \leq r_n\}~.
\end{equation*}
Additionally, suppose $z$ and $w$ have a joint density in a neighborhood containing the line $z = 0$. Then, local to the cutoff, we have that $z$ is uncorrelated $w$, so that neighborhoods, and in turn spillovers, are uncorrelated to the RDD treatment assignment. The result is that RDD always recovers the direct treatment effect, provided that the correlation between $z$ and $w$ remain fixed in the asymptotics.\footnote{The first version of this paper presented the above intuition for spillovers on background variables (\citealt[Section 3.3.3]{auerbach2024regression}). \cite{torrione2024regression} proves this result in a related setting, drawing a connection to multi-score RDD. Also see the discussion after Theorem \ref{theorem--consistency}.} 

In light of the above discussion, spillovers in RDD may appear pathological. However, there are many settings in which neighborhoods appear strongly correlated with the running variable. In spatial RDDs, it is often plausible that nearby units interact. For example, \cite{jardim2022boundary} studies the effect of minimum wage policies on labor market outcomes at the census tract level. However, it is clear that census tracts compete in the same labor market as all other census tracts within a reasonable commute. \cite{gonzalez2021cell} provides another example based on cell phone coverage and electoral fraud which we take up in greater detail in Section \ref{section--application}. \cite{leung2022rate} provides many more examples. The same is true with test score RDDs, such as the classic \cite{thistlethwaite1960regression} which studies the effect of merit scholarships on academic achievements. Here, scholarships are assigned based on test score cutoffs and it is plausible that students who are close in test scores interact. This could happen because of homophily, because students take classes and therefore socialize with those of similar abilities, or because they compete for the same set of career opportunities. In instances like the above, our model is a useful starting point for understanding the effect of spillovers.



\subsection{Sampling}\label{section--setup--sampling}

Having defined a sequence of models for generating potential outcomes, we close this section by considering sampling. We will assume that our data takes the following form:
\begin{assump}\label{assump-sampling}
Suppose we observe the i.i.d. sample $\{\left(Y_i, Z_i\right)\}_{i=1}^n$, where $Z_i \sim F$,
\begin{equation*}
	Y_i = Y_d(Z_i) + \varepsilon_i \quad , \quad E[\varepsilon_i \lvert Z_i] = 0
\end{equation*}
and $d(z) = \mathbf{1}\{z \geq 0\}$ is the RDD treatment assignment function. Moreover, suppose that $E[\varepsilon_i^2 \lvert Z_i = z] \leq \bar{\sigma}^2$ for all $z \in \mathcal{Z}$ and that $\mathbb{E}[\varepsilon^2 \mid Z = z]$ is continuous in a neighborhood of $0$.
\end{assump}

In words, the data-generating process operates as follows. A continuum of agents interact and their outcomes are determined by the values of their running variables as well as spillovers. The econometrician then samples agents randomly and observes their outcomes, possibly with an error that has $0$ conditional mean. We also assume that the error has conditional variance that is continuous in $z$ -- a standard assumption. Our framework resembles \cite{dePaula2018identifying} in defining interactions to occur in the population.


\section{Local Linear Regression in the Presence of Spillovers}\label{section--estimands}

In this section, we show that the local linear regression estimator may not recover meaningful treatment effect parameters in the presence of spillovers, necessitating proposals such as that in Section \ref{section--localspilloverreg}. Section \ref{section--estimands--llr} defines the local linear regression estimator and Section \ref{section--estimands--main_result} describes its estimands in the presence of spillovers. Our results suggest that the ratio of $r_n/h_n$ is a key modeling assumption and we argue for using $r_n \propto h_n$ in Section \ref{section--estimands--choice_of_framework}. Finally, Section \ref{section--estimands--donut_hole} considers the use of donut-hole designs as a solution to spillovers.

\subsection{The Local Linear Regression Estimator}\label{section--estimands--llr}

The local linear regression estimator for the local average treatment (equivalently, $\tau_\text{DIR}$) is defined as follows:

\begin{definition}[Local Linear Regression Estimator]\label{definition--LLR}
	Let $K$ be a bounded second-order kernel so that $\int_{-1}^{1} K(u) du = 1$ and $\int_{-1}^{1} uK(u) du = 0$. Furthermore, assume that for all $u$, $0 \leq K(u) \leq \bar{K}$, and that $K(u) = 0$ if $u < -1$ or $u > 1$. Let $h_n$ be a sequence of bandwidths such that $h_n \to 0$ and $nh_n \to \infty$. Then, define
	\begin{align*}
		\left(\hat{\beta}^+_0,  \hat{\beta}^+_1\right) & = \arg \min_{b \in \mathbb{R}^2} \sum_{Z_i \geq 0} K\left(\frac{Z_i}{h_n}\right)\left( Y_i - {b}_0 - {b}_1Z_i  \right)^2 \\
		\left(\hat{\beta}^-_0,  \hat{\beta}^-_1\right) & = \arg \min_{b \in \mathbb{R}^2} \sum_{Z_i < 0} K\left(\frac{Z_i}{h_n}\right)\left( Y_i - b_0 - {b}_1Z_i  \right)^2
	\end{align*}
	The local linear estimator for the local average treatment effect is $\hat{\tau}_{\text{RDD}} := \hat{\beta}^+_0 - \hat{\beta}^-_0$.
\end{definition}
Second order kernels are commonly used in local linear regression. For a discussion on the properties of higher order kernels, see e.g. \cite{wand1995kernel}. We additionally assume that $K$ has finite support and is bounded and non-negative. 

In the absence of spillovers, it is standard practice to estimate the local average treatment effect using $\hat{\tau}_{\text{RDD}}$ (see e.g. \citealt{hahn2001identification, cattaneo2022regression}). In this setting, $\hat{\tau}_{\text{RDD}}$ is consistent for $\tau_\text{DIR}$, though inference is complicated by the presence of an asymptotic bias of order $h_n^2$. Various solutions are available for the problem of inference (see e.g. \citealt{calonico2014robust, armstrong2018optimal}). In sum, the properties of RDD are relatively well-understood in settings without spillovers.

\subsection{Estimands of RDD}\label{section--estimands--main_result}

Our first result characterizes the estimands of RDD when spillovers occur along the running variable.

\begin{theorem}\label{theorem--consistency}
Suppose Assumptions \ref{assump-model}, \ref{assump--neighborhood} and \ref{assump-sampling} hold.
\begin{itemize}
	\item[(a)] If $r_n \gg h_n$, then $\hat{\tau}_{\text{RDD}}  \overset{p}{\to} \tau_\text{DIR}$
	\item[(b)] If $r_n \ll h_n$, then $\hat{\tau}_{\text{RDD}}  \overset{p}{\to} \tau_\text{TOT}$
	\item[(c)] If $r_n/h_n \to c$ where $0 < c < \infty$, then $\hat{\tau}_{\text{RDD}}  \overset{p}{\to} \tau_*$
\end{itemize}
\end{theorem}

When spillovers occur along the running variable, the estimand of RDD exhibits phase transition, with the phases depending on the ratio of $r_n$ to $h_n$. When $r_n \gg h_n$, RDD estimates direct treatment effect. On the other hand, when $r_n \ll h_n$, RDD estimates total treatment effect. In the intermediate regime where $r_n \propto h_n$, it estimates a parameter $\tau_*$, which is displayed in full in Equation \eqref{equation--tau_star}. $\tau_*$ depends on the $c$, the limiting ratio of $r_n$ and $h_n$, even though this is suppressed in the notation. Generally speaking, $\tau_*$ does not have a clear interpretation. In particular, it is not equal to either $\tau_\text{DIR}$ or $\tau_\text{TOT}$ unless either $\delta(0) = \gamma(0) = 0$ (i.e. no spillovers) or $\tau_\text{DIR} = \gamma(0) = 0$ (i.e. treatment has 0 effect), in which case $\tau_\text{DIR} = \tau_\text{TOT} = \tau_*$. It may be close to $0$ even when $\tau_\text{DIR}$ and $\tau_\text{TOT}$ are large in magnitude -- a common concern expressed by empirical papers regarding spillovers. 

The above result is intuitive. The local linear regression estimator essentially treats units within the bandwidth as being comparable. If $r_n$ is large relative to $h_n$, units in the bandwidth have neighborhoods that overlap almost completely so that they experience similar spillovers. As such, units on different sides of the cutoff differ only in their treatment status and $\hat{\tau}_\text{RDD}$ estimates direct treatment effect. Conversely, when $r_n$ is small relative to $h_n$, units to the left of the cutoff essentially only has control neighbors, while units to the right of the boundary only has treated neighbors. Comparing these units therefore reveals total treatment effect. When $r_n \propto h_n$, RDD compares a mix of units, some of which have similar neighborhoods, some of which do not. The resulting estimand is therefore a complicated interpolation of direct and spillover effects.

\begin{remark}
	Although the exact form of $\tau_*$ depends on the structure of the spillovers, changing assumptions about the functional form or the relative weights of neighbors will not qualitatively affect the results in Theorem \ref{theorem--consistency}. In particular (a) and (b) will continue to hold as long as spillovers occur over approximately bounded neighborhood. Similarly, when $r_n \propto h_n$, the limit of the $\hat{\tau}_{\text{RDD}}$ will be combination of terms with no clear interpretation.
\end{remark}

\subsubsection*{Relation to Existing Work}\label{section--estimands--relation}

Theorem \ref{theorem--consistency}, in particular cases (b) and (c), stands in stark contrast to existing work on spillovers within an RDD. \cite{aronow2017rdspill} studies spillovers under the local randomization framework of \cite{cattaneo2015randomization} and find that RDD always recovers a weighted average of direct treatment effect. Under local randomization, RDD is exactly an experiment under some neighborhood of the cutoff. Consequently, treatment assignment is orthogonal to spillovers, reproducing the results seen in the literature on experiments with spillovers (see e.g. \citealt{hudgens2008toward, aronow2017estimating, leung2022causal}). \cite{torrione2024regression} provides a similar analysis under the continuity-based approach of \cite{hahn2001identification}. They assume that the running variable and the variable determining neighborhood structure have a continuous joint density, ruling out the case we consider. Given continuous joint density, they find that treatment assignment is approximately orthogonal to spillovers. As a consequence, RDD recovers a weighted average of direct treatment effect even under the continuity framework.

Our paper studies RDD under a continuity-based framework and we focus on the case where neighborhoods are completely determined by the running variable. In this setting, the extent of orthogonality between treatment assignment and spillovers depends on the ratio $r_n/h_n$. Our approach -- particularly the $r_n \propto h_n$ case -- leads to an asymptotic model in which spillovers can affect the estimands of RDD. This is not possible in the framework of \cite{aronow2017rdspill} and \cite{torrione2024regression}. Our model therefore provides a basis for using data to speak to the effects of spillovers, a task which we take up in Section \ref{section--localspilloverreg}. 
Finally, we note that this paper is also unique relative to the above two papers in considering endogenous spillovers.

\subsection{Choice of Asymptotic Framework}\label{section--estimands--choice_of_framework}

Theorem \ref{theorem--consistency} shows that how we model the relative rates between $r_n$ and $h_n$ affects the interpretation of $\hat{\tau}_\text{RDD}$. We argue that the $r_n \propto h_n$ regime is the natural choice when researchers are concerned about spillovers.

In any given application, spillovers occur over some radius that is a finite ratio $c$ of the bandwidth. The regimes $r_n \gg h_n$ or $r_n \ll h_n$ are therefore unrealistic edge cases that take $c = 0$ or $c = \infty$. Consequently, they assert that $\hat{\tau}_\text{RDD}$ must either recover direct treatment effect or total treatment effect. There is no room for the data to speak to the amount of direct effects relative to spillovers.

In contrast, the estimand of RDD depends explicitly on the ratio of $c$ under the regime $r_n \propto h_n$, which is a better approximation to how spillovers manifest in practice. It leads to an asymptotic model that preserves more features of the finite sample distribution and therefore provides greater scope for reasoning about spillovers in the data. For this reason, $r_n \propto h_n$ is our preferred framework for analyzing RDD with spillovers. We do not claim $r_n$ to be a quantity that changes with sample size, only that taking it to $0$ leads to a useful approximation. Such an approach is also taken in traditional analysis of boundary bias for kernel methods (see e.g. Section 5.5 in \citealt{wand1995kernel}). Here, the target point is modeled as drifting towards the boundary at rate $h_n$ to capture the notion that these two points are close. 

In our preferred regime, RDD is potentially inconsistent for either treatment effects parameter. Section \ref{section--localspilloverreg} proposes a method for recovering both direct treatment effects and spillovers.

\subsection{Donut-Hole Designs}\label{section--estimands--donut_hole}

In the remainder of this section, we informally discuss donut-hole designs as viewed through the lens of our model. Often suggested as a solution for estimating $\tau_\text{DIR}$ in the presence of spillovers, donut-hole designs involve excluding observations close to the cutoff from estimation. The idea is that these units are contaminated by spillovers, so that removing them should lead to the recovery of $\tau_\text{DIR} = m^+(0) - m^-(0)$. This approach does not work under our model, but we state an alternative model of spillovers under which donut-hole RD can be justified, provided that there are no endogenous spillovers.

To define the donut-hole RD, let $h_n^o > 0$ be such that $h_n^o < h_n$.  Then donut-hole RD estimates $\hat{\beta}^+$ by local linear regression on observations for which $Z_i \in [h_n^o, h_n]$. Similarly for  $\hat{\beta}^-$ and observations for which $Z_i \in [-h_n, -h_n^o]$. Let the estimand of the donut-hole RD be denoted $\tau_o$.

Under our linear-in-means model, suppose $h_n^o = r_n$. That is, we exclude all observations that have neighbors on the other side of the cutoff. Suppose there are no endogenous spillovers ($\delta(0) = 0$). Then $\tau_o = \tau_\text{TOT}$ and not $\tau_\text{DIR}$. In our model, agents closest to the cutoff experience the most similar spillovers, so that comparing them leads to the best estimate of $\tau_\text{DIR}$. However, after excluding units in the donut hole $[-h_n^o, h_n^o]$, the remaining units only have neighbors who have the same treatment status as them so that RDD estimates $\tau_\text{TOT}$. In this sense, a donut-hole RD is exactly the wrong thing to do if researchers are interested in $\tau_\text{DIR}$.

However, suppose $\gamma(z) = \check{\gamma}(z) \mathbf{1}\{z \leq 0\}$. In other words, only control units experience spillovers based on how many of their neighbors are treated. This is reasonable, for example, in the case of information treatment, where information can diffuse from the treated to control units but not vice versa. In this case, if $h_n^o = r_n$ and $\delta(0) = 0$, the intuition in the above paragraph still applies so that $\tau_o$ estimates total treatment effect. However, in this model, total treatment effect is also equal to direct treatment effect under the reference treatment assignment function $d = 0$. Note that direct treatment effect is now dependent on $d$, since spillovers enter the treated and control potential outcomes asymmetrically. Donut-hole RDs can therefore be used to recover $\tau_\text{DIR}$ under this model.

Finally, we note that once $\delta(0) \neq 0$, donut-hole RDs do not recover $\tau_\text{DIR}$ or $\tau_\text{TOT}$ in either model. This is because under endogenous spillovers, the outcome of a given unit is affected by all other units in the domain: a given unit $z$ has an outcome that depends on the outcomes of $[z-r_n, z+r_n]$. In turn, the outcome for $z+r_n$ depends on the outcomes on $[z, z+2r_n]$, and so on. As such, it is not possible to isolate units that are ``contaminated" by spillovers using a donut design.

\section{Local Spillover Regression}\label{section--localspilloverreg}

As we argued in Section \ref{section--estimands--choice_of_framework}, our preferred asymptotic regime features $r_n \propto h_n$, in which the RDD estimand need not have a causal interpretation. In order to disentangle direct treatment effect from spillovers, Section \ref{section--localspilloverreg--estimation} proposes a local analog of the peer effects regression that we term the local spillover regression (LSR). We discuss estimation consistency and inference in Section \ref{section--localspilloverreg--theory}. LSR requires researchers to specify a radius $r_n$. We consider choosing $r_n$ as well as $h_n$ in Section \ref{section--localspilloverreg--choice}.

\subsection{Proposed Method}\label{section--localspilloverreg--estimation}

Since spillovers are the cause of inconsistency, our proposal is to estimate $\mu_d(z)$ and $\nu_d(z)$, and then to incorporate them into the local linear regression.

Suppose for now that $r_n$ is known. For a given $Z_i$, define its observed neighbors to be:
\begin{equation*}
	\tilde{R}(Z_i) = \{j \neq i \, : \, \lVert Z_j - Z_i \rVert \leq r_n\}~.
\end{equation*}
To be consistent with our econometric model, $\lVert \cdot \rVert$ is taken to be the Euclidean distance. We then estimate the mean outcome of $Z_i$'s neighbors using a leave-$i$-out estimator:
\begin{equation*}
	\tilde{\mu}_d(Z_i) = \frac{1}{|\tilde{R}(Z_i)|} \sum_{j \in \tilde{R}(Z_i)} Y_j~.
\end{equation*}
Additionally, we define
\begin{equation*}
		\tilde{\mu}_d(0) = \frac{1}{|\tilde{R}(0)|} \sum_{j \in \hat{R}(0)} Y_j \quad , \quad \tilde{R}(0) = \{j \neq i \, : \, \lVert Z_j  \rVert \leq r_n\}~.
\end{equation*}
where the only difference is that $z = 0$ is not observed. We focus on the case when the distribution of $Z$ is uniform, so that $\nu_d(Z_i)$ known. When $f$ is not uniform, it is straightforward to form the analogous leave-$i$-out estimator for $\nu_d(Z_i)$.

We propose to estimate $\tau_\text{DIR}$, $\gamma(0)$ and $\delta(0)$ using the following:
\begin{definition}[Local Spillover Regression Estimator]\label{definition-localspill}
	For a given kernel $K$ and bandwidths $h_n$, let
	\begin{equation*}\label{equation--localspill}
		\tilde{\beta} = \argmin_{b} \sum_{i=1}^n K\left(\frac{Z_i}{h_n}\right) \left(Y_i - \tilde{X}_i'b\right) ^2
	\end{equation*}
	where $D_i = \mathbf{1}\{Z_i \geq 0\}$,
	\begin{align*}
			{X}_i = \begin{pmatrix}
				1 \\
				Z_i \\
				D_i \\
				Z_iD_i \\
				{\mu}_d(Z_i) - {\mu}_d(0) \\
				Z_i({\mu}_d(Z_i) - {\mu}_d(0))\\
				{\nu}_d(Z_i) - {\nu}_d(0) \\
				Z_i({\nu}_d(Z_i) - {\nu}_d(0))\\
			\end{pmatrix} \quad \mbox{and} \quad
		\tilde{X}_i = \begin{pmatrix}
			1 \\
			Z_i \\
			D_i \\
			Z_iD_i \\
			\tilde{\mu}_d(Z_i) - \tilde{\mu}_d(0) \\
			Z_i(\tilde{\mu}_d(Z_i) - \tilde{\mu}_d(0))\\
			{\nu}_d(Z_i) - {\nu}_d(0) \\
			Z_i({\nu}_d(Z_i) - {\nu}_d(0))\\
		\end{pmatrix}~.
	\end{align*}
\end{definition}

$\tilde{X}_i$ is our estimate of the unobserved $X_i$. The estimator $\tilde{\beta}$ targets the following:
\begin{equation*}
	\beta = \begin{pmatrix}
					m^-(0) + \delta(0)\mu_d(0) + \gamma(0)\nu_d(0) \\
					(m^-)'(0) + \delta'(0)\mu_d(0) + \gamma'(0)\nu_d(0) \\
					{m^+(0) - m^-(0)}   \\
					(m^+)'(0) - (m^-)'(0)   \\
					{\delta(0)} \\
					\delta'(0)\\
					{\gamma(0)} \\
					\gamma'(0)
				\end{pmatrix} ~.
\end{equation*}
The first four components of $\tilde{X}_i$ are terms coming from the standard LLR. They are obtained by rewriting the two separate local regressions in Definition \ref{definition--LLR} as a single regression. As such, the coefficient of $D_i$ targets $\tau_\text{DIR} = {m^+(0) - m^-(0)}$. The remaining components address spillovers.
$\tilde{\mu}_d(Z_i) - \tilde{\mu}_d(0)$ and ${\nu}_d(Z_i) - {\nu}_d(0)$ are our estimated mean outcome and mean treatment status in the neighborhood of $Z_i$. As such, their coefficients target $\delta(0)$ and $\gamma(0)$ respectively. $Z_i(\tilde{\mu}_d(Z_i) - \tilde{\mu}_d(0))$ and $Z_i({\nu}_d(Z_i) - {\nu}_d(0))$ are not needed for consistency but they reduce the asymptotic bias when $\delta(z)$ and $\gamma(z)$ are not constant, analogous to how introducing $Z$ reduces the bias of LLR relative to the Nadaraya-Watson estimator.

In view of the above discussion, we will also use the notation $\tilde{\tau}_\text{DIR} := \tilde{\beta}_3$, $\tilde{\delta}(0) := \tilde{\beta}_5$, $\tilde{\gamma}(0) := \tilde{\beta}_7$. When $r_n \to 0$, $\tau_\text{TOT} \to \frac{\tau_\text{DIR} + \gamma(0)}{1-\delta(0)}$. As such, we also define $\tilde{\tau}_\text{TOT} :=\frac{\tilde{\tau}_\text{DIR} + \tilde{\gamma}(0)}{1-\tilde{\delta}(0)}$.

\subsection{Theoretical Properties of LSR}\label{section--localspilloverreg--theory}

Section \ref{section--localspilloverreg--theory--consistency} presents the consistency properties of LSR. Section \ref{section--localspilloverreg--theory--inference} provides the asymptotic distribution for $\tilde{\beta}$. Section \ref{section--localspilloverreg--theory--bias_aware} covers confidence intervals based on the bias-aware approach of \cite{armstrong2020simple}.

\subsubsection{Consistency}\label{section--localspilloverreg--theory--consistency}

This section presents results on the consistency of LSR.  We first discuss consistency of $\tau_\text{DIR}$, which obtains under fairly unrestrictive conditions. We then turn to $\gamma(0)$ and $\delta(0)$, which are more challenging to estimate and will require more assumptions.

For direct treatment effect,
\begin{theorem}\label{theorem--spilloverreg_direct}
Suppose Assumptions \ref{assump-model}, \ref{assump--neighborhood} and \ref{assump-sampling} hold. Additionally, suppose $n \to \infty$, $h_n \to 0$, $nh_n \to \infty$ and $r_n \propto h_n$. Then $\tilde{\tau}_\text{DIR} \overset{p}{\to} \tau_\text{DIR}$.
\end{theorem}
The main condition for consistency of $\tilde{\tau}_\text{DIR}$ is the Lipschitz continuity of $m^+$, $m^-$, $\delta$ and $\gamma$. This is a slight strengthening of the conditions needed for LLR, which typically assumes continuity of $m^+$ and $m^-$ at the cutoff. Including estimated spillover terms is therefore a simple and effective method to ensure that the target parameter is $\tau_\text{DIR}$. 

Recovering spillover effects requires more assumptions:
\begin{assump}\label{assumption--identification_spillreg}
Suppose the following hold:
\begin{enumerate}[\quad (a)]
	\item Suppose $r_n/h_n \to c < 1$.
	\item $\delta(0) \neq 0$ and $\tau_\text{DIR} \neq 0$ or $\gamma(0) \neq 0$.
\end{enumerate}
\end{assump}

Then, we have the following:
\begin{theorem} \label{theorem--spilloverreg_indirect}
Suppose Assumptions \ref{assump-model}, \ref{assump--neighborhood}, \ref{assump-sampling}  and \ref{assumption--identification_spillreg} hold. Additionally, suppose $n \to \infty$, $h_n \to 0$, $nh_n \to \infty$ and $r_n \propto h_n$. Then $\tilde{\beta} \overset{p}{\to} \beta$. In particular,
\begin{equation*}
	\tilde{\delta}(0) \overset{p}{\to} \delta(0) \quad , \quad \tilde{\gamma}(0) \overset{p}{\to} \gamma(0) \quad , \quad \tilde{\tau}_\text{TOT} \overset{p}{\to} \tilde{\tau}_\text{TOT}~.
\end{equation*}
\end{theorem}
The additional assumptions here address various sources of non-identification in the model. Condition (a) is specific to $z \in \mathbb{R}$. In this case, when $c > 1$, $\nu_d(z) = z$ under the assumption of uniform $f$. Having $c < 1$ is therefore necessary for the identification of $\gamma(0)$. This collinearity does not arise when units are in e.g. $\mathbb{R}^2$. Condition (b) is needed to ensure that $\mu_d(z)$ is not collinear with $\nu_d(z)$. Our approximation result in Lemma \ref{lemma--uniform_approx} shows that locally,
\begin{align*}
	\mu_d(z) - \mu_d(0) \approx (\nu_d(z) - \nu_d(0)) \cdot \big( & \tau_\text{DIR} \cdot (\text{first order term smoothing due to $\delta(0)$}) \\
	& \gamma(0) \cdot (\text{second order term smoothing due to $\delta(0)$})  \big)
\end{align*}
If $\delta(0) = 0$, the first and second order terms are constant, so that $\mu_d(z) - \mu_d(0) = (\nu_d(z) - \nu_d(0))$. Meanwhile, if both $ \tau_\text{DIR}$ and $\gamma(0)$, then $\mu_d(z) - \mu_d(0) \approx 0$. Intuitively, to learn about $\delta(0)$, we need treatment to create some effect ($\tau_\text{DIR}$ or $ \gamma(0) \neq 0$) that decays in $z$.

Condition (b) is relatively strong but it is reminiscent of the assumption that $\tau \gamma + \delta \neq 0$ common in the peer effects literature (see e.g. Proposition 1 in \citealt{bramoulle2009identification}). Finally, since the collinearities pertain to some combination of $z$, $\mu_d(z)$ and $\nu_d(z)$, $\tilde{\tau}_\text{DIR}$ is consistent for $\tau_\text{DIR}$ even when these conditions are not satisfied.

\subsubsection{Asymptotic Distribution}\label{section--localspilloverreg--theory--inference}

We next state a central limit theorem for $\tilde{\beta}$. For ease of exposition, we will maintain the full set of assumptions in Theorem \ref{theorem--spilloverreg_indirect}. We focus on the case when the bandwidth is assumed to have the MSE-optimal rate of $n^{-1/5}$ and introduce the following assumption:
\begin{assump}\label{assumption--inference}
	Suppose that
		\begin{enumerate}[(a)]
			\item $h_n, r_n \propto n^{-1/5}$.
			\item For some $s > 0$,
			\begin{equation*}
				\sup_{z \in (-s, s)} \mathbb{E}[|\varepsilon|^3 \mid Z = z] < \infty~.
			\end{equation*}
		\end{enumerate}
\end{assump}
Then,

\begin{theorem}
\label{thm--spilloverreg-inference}
Suppose Assumptions \ref{assump-model}, \ref{assump--neighborhood}, \ref{assump-sampling}  \ref{assumption--identification_spillreg} and \ref{assumption--inference} hold.
Then, as \(n \to \infty\), we have that
\begin{equation}
  \mathbf{V}^{-1/2} \sqrt{n} \left( \widetilde{\beta} - \beta - \mathbf{B}_{n}
  \right)
  \overset{d}{\to} \N \left( 0,\mathbb{I}_{8} \right)~,
  \label{eqn--spilloverreg-inference-std-gauss}
\end{equation}
where \(\mathbf{B}_{n} = \widetilde{\mathbf{Q}}^{-1}
\widetilde{\mathbf{T}}_{\rho}\) is a bias term,
\(\widetilde{\mathbf{T}}_{\rho}\) is defined in
\eqref{eqn--betatilde-breadown-1} in the appendix,
\(\mathbf{V}^{- 1 / 2} = \left( \mathbf{V}^{1 / 2} \right)^{- 1}\),
\(\mathbf{V}^{1 / 2} = \widetilde{\mathbf{Q}}^{- 1} \bm{\Omega}^{1 / 2}\), and
\begin{equation}
  \begin{split}
    \widetilde{\mathbf{Q}} =
    & \, \frac{1}{n} \sum_{i = 1}^{n} K \left( \frac{Z_{i}}{h} \right)
    \widetilde{X}_{i} \widetilde{X}_{i}^{\prime}, \\
    \bm{\Omega} =
    & \, \mathrm{Var} \left[ K \left( \frac{Z_{i}}{h} \right) X_{i}
    \varepsilon_{i} + \eta_{X} \left( \xi_{i} \right) \right].
  \end{split}
  \label{eqn--Omega-def}
\end{equation}
In the above expression,
\begin{equation}
  \begin{split}
    \eta_{X} \left( \xi_{1} \right) =
    & \, \E \left[ K \left( \frac{Z_{0}}{h} \right) \left( \delta (0) +
    \delta^{\prime} (0) Z_{0} \right) X \left( Z_{0} \right) \left( Y_{1} -
    \mu_{d} (0) \right) \frac{R \left( \frac{Z_{1}}{r} \right)}{\pi_{r} (0)}
    \middle| \xi_{1} \right] \\
    & - \E \left[ K \left( \frac{Z_{0}}{h} \right) \left( \delta (0) +
    \delta^{\prime} (0) Z_{0} \right) X \left( Z_{0} \right) \left( Y_{1} -
    \mu_{d} \left( Z_{0} \right) \right) \frac{R \left( \frac{Z_{1} - Z_{0}}{r}
    \right)}{\pi_{r} \left( Z_{0} \right)} \middle| \xi_{1} \right].
  \end{split}
  \label{eqn--etaX-def}
\end{equation}
where \(\xi_{i} \sim P\) independently.
Furthermore, \(R (z) = \mathbf{1}_{[-1,1]} (z)\) and \(\pi_{r} (z) =
\E \left[ R \left( \frac{Z_{1} - z}{r} \right) \right]\).
\end{theorem}


When $h_n \propto n^{-1/5}$, there is an asymptotic bias term $\mathbf{B}_{n}$
that we address below. \(\widetilde{\mathbf{Q}}\) is the regression design
matrix.
The matrix \(\bm{\Omega}\) in the definition of \(\mathbf{V}\) has two
components.
The term \(K (Z_{i} / h) X_{i} \varepsilon_{i}\) captures the usual source of
estimation error in weighted regression if the ``ideal'' regressors \(X_{i}\)
were known.
In this case, the only source of estimation error would be due to the unobserved
error term \(\varepsilon_{i}\).
The feasible regressors \(\widetilde{X}_{i}\), used in place of \(X_{i}\),
contain an additional source of error: estimation of the unknown
neighborhood average outcomes \(\mu_{d} (Z_{i})\).
The term \(\eta_{X} \left( \xi_{i} \right)\) captures this additional source of
estimation error, i.e. the difference \(\widetilde{X}_{i} - X_{i}\).
It is straightforward to see that the sample analogue of $\bm{\Omega}$ combined
with \(\widetilde{\mathbf{Q}}\) is a consistently estimates \(\mathbf{V}\).
Let the plug-in estimator be denoted $\widetilde{\mathbf{V}}_{n}$.

\subsubsection{Bias-Aware Confidence Intervals}\label{section--localspilloverreg--theory--bias_aware}

The bias term in Theorem \ref{thm--spilloverreg-inference}, $\mathbf{B}_{n}$, is asymptotically non-negligible when $h_n \propto n^{-1/5}$. To form valid confidence intervals, we follow the bias-aware approach of \cite{armstrong2020simple}. This entails restricting the classes of functions to which $m^+, m^-, \delta$ and $\gamma$ can potentially belong, so that an upper bound for $\mathbf{B}_{n}$ can be obtained. Confidence intervals can then be appropriately widened to ensure coverage under this worst-case bias.

To that end, define the Taylor class  of order 2:
\begin{equation*}
	\mathcal{F}_{T,2}(M) = \left\{  f: \; \left\lvert f(z) - f(0)- f'(0) \cdot z\right\rvert  \leq M \cdot \frac{z^2}{2} \; , \; z \in \mathcal{Z}  \right\}
\end{equation*}
This is the class of functions whose first order Taylor approximation has error less than $Mz^2/2$. We can loosely think of the tuning parameter $M$ as the upper bound for the second derivative of $f$ at $0$. We will now assume that:

\begin{assump}\label{assumption--taylor_class}
	Suppose
	\begin{equation*}
		m^+, m^- \in \mathcal{F}_{T,2}(M_m) \quad , \quad \delta \in \mathcal{F}_{T,2}(M_\delta) \quad , \quad \gamma \in \mathcal{F}_{T,2}(M_\gamma) ~.
	\end{equation*}
\end{assump}

It is then straightforward to see that $|\mathbf{B}_{n,j}| \leq \overline{\mathbf{B}}_{n,j} $, where
\begin{align}\label{equation--max_bias}
	\overline{\mathbf{B}}_{n,j} = \sum^n_{i=1} |w(Z_i)| \cdot  \frac{Z_i^2}{2}\left( M_m + M_\delta \lvert \mu_d(Z_i) - \mu_d(0)\rvert + M_\gamma \lvert \nu_d(Z_i) - \nu_d(0)\rvert  \right)
\end{align}
and $w_j(Z_i) = e_j' \hat{Q}_n^{-1} \frac{1}{h_n}K\left(\frac{Z_i}{h_n}\right) \tilde{X}_i $. Since $\mu_d(Z_i)$ and $\mu_d(0)$ are unobserved, we replace them with $\tilde{\mu}_d(Z_i)$ and $\tilde{\mu}_d(0)$. Denote the resulting estimator $\widetilde{\mathbf{B}}_{n,j}$. We can then form asymptotically valid confidence intervals as follows:

\begin{theorem}
Suppose Assumptions \ref{assump-model}, \ref{assump--neighborhood},
\ref{assump-sampling}, \ref{assumption--identification_spillreg},
\ref{assumption--inference} and \ref{assumption--taylor_class} hold. Let $ t_j =
\frac{\sqrt{n} \widetilde{\mathbf{B}}_{n,j}}{\sqrt{e_j'\widetilde{\mathbf{V}}_n e_j}}$ where
$e_j$ be an $8 \times 1$ vector with 1 in the j-th component and 0 everywhere
else. Denote by $c_{1-\alpha}$ the $1-\alpha$ quantile of the folded normal
distribution $\left\lvert N(t_j, 1)\right\rvert$ and let
	\begin{equation*}
		C_j = \left[ \tilde{\beta}_j - c_{1-\alpha}\sqrt{\frac{e'_j\widetilde{\mathbf{V}}_ne_j}{n}}  \; , \; \tilde{\beta}_j + c_{1-\alpha}\sqrt{\frac{e'_j\widetilde{\mathbf{V}}_ne_j}{n}}  \right]~.
	\end{equation*}
	Then,
	\begin{equation*}
		\liminf_{n \to \infty} \mathbb{P}(\beta_j \in C_j) \geq 1-\alpha~.
	\end{equation*}
\end{theorem}

The above result is an immediate consequence of Theorem \ref{thm--spilloverreg-inference} and the fact that $|\mathbf{B}_{n,j}| \leq \overline{\mathbf{B}}_{n,j}$ under Assumption \ref{assumption--taylor_class}. It yields asymptotically valid $1-\alpha$ confidence intervals for $\tau_\text{DIR}$, $\delta(0)$ and $\gamma(0)$. It is then straightforward to obtain valid confidence interval for $\tau_\text{TOT}$ by combining these confidence intervals together with a Bonferroni correction.

\subsection{Choice of $r_n$ and $h_n$}\label{section--localspilloverreg--choice}

Results in the previous section are predicated on researcher's choices of $r_n$ and $h_n$. This section presents heuristics for choosing these parameters.

\subsubsection{Choice of $r_n$}\label{section--localspilloverreg--choice_r}

In principle, researchers should choose $r_n$ based on their beliefs about likely sources of spillovers. In practice, however, such a choice may not be obvious. Researchers may instead consider choosing $r_n$ via cross-validation as follows.

Let the fold number $L$ be specified by the user and suppose we are given some initial bandwidth $h^0_n$. Partition the data set $D$ into $L$ equal-sized subsets $D_1, ..., D_L$ and let $\tilde{\beta}^{(l)}(r)$ be the LSR estimator computed on $D \setminus D_l$ under radius $r$. Define the out-of-sample MSE as:
\begin{align*}
	\text{MSE}^{(l)}({r}) = \sum_{i \in D_l} K\left(\frac{Z_i}{h^0_n}\right)\left(Y_i - \tilde{X}_i'\tilde{\beta}^{(l)}(r)\right)^2 \bigg/ \sum_{i \in D_l} K\left(\frac{Z_i}{h^0_n}\right)
\end{align*}
We can then choose
\begin{equation*}
	\tilde{r}_n = \argmin_{r} \sum_{l \in L} \text{MSE}^{(l)}({r})~.
\end{equation*}
Intuitively, if spillovers exist, getting the true radius correct should lead to good out-of-sample fit. It would be difficult to do better than the true radius in MSE since this would require the misspecified model to overfit on out-of-sample $\varepsilon_i$'s, which is independent of the training sample.

Finally, to choose the initial bandwidth $h_0$, we might consider a rule of thumb such as
\begin{equation}\label{equation--initial_h}
	h^0_n = n^{-1/5} \cdot \max_{i} \, \lVert Z_i \rVert
\end{equation}
Here, the $ \max \, \lVert Z_i \rVert$ scaling ensures that $h^0_n$ is invariant to the units of measurement of the running variable.

The simulations in Section \ref{section--simulations} explores the finite sample performance of the above method. We find that cross-validation can increase the MSE of various parameters by around 4 times relative to the oracle procedure in which $r_n$ is known. This appears to be a reasonable trade-off when researchers have only weak priors about spillovers in a given setting.


\subsubsection{Choice of $h_n$}\label{section--localspilloverreg--choice_h}

The rule of thumb in Equation \eqref{equation--initial_h} leads to a reasonable sequence of bandwidths, but it does not take MSE into consideration. Here, we present an intuitive adaptation of the procedure in \cite{armstrong2018optimal}, which aims to reduce finite sample version of the worst-case MSE.

Given an initial bandwidth $h^0_n$ and a radius that is either pre-specified or selected via cross validation, we can perform LSR and obtain an initial estimate of the asymptotic variance based on the sample analog of $V_n$ in Theorem \ref{thm--spilloverreg-inference}. Call this estimate $\widetilde{\mathbf{V}}^0_n$. For any given bandwidth $h$, we can also obtain an estimate for the maximum bias for $\tilde{\beta}_j$ using the sample analog of $\overline{\mathbf{B}}_{j,n}$ in \eqref{equation--max_bias}. Call this term $\widetilde{\mathbf{B}}_{j,0}(h)$. We can then choose bandwidth by solving:
\begin{equation*}
	h^*_n = \argmin_{h} \;  \widetilde{\mathbf{B}}_{j,0}^2 + \frac{h^0_n \left[\widetilde{\mathbf{V}}^0_n\right]_{jj}}{h \cdot n}
\end{equation*}
While any $\tilde{\beta}_j$ can potentially work with this procedure, we recommend the use of $\tilde{\beta}_3 = \tilde{\tau}_\text{DIR}$. This is because $\tilde{\tau}_\text{DIR}$ targets a parameter of immediate interest and it is consistent under weaker conditions than $\tilde{\delta}(0)$ and $\tilde{\gamma}(0)$.

\section{Empirical Demonstration}\label{section--application}

In this section, we revisit \cite{gonzalez2021cell}, which studies the effect of cell phone coverage on fraud in the 2009 Afghan presidential election. The analysis is conducted at the polling station level and the outcomes of interest are (1) indicator for whether fraud has likely occurred and (2) likely share of fraudulent votes. Using a spatial regression discontinuity design, the author finds that cell phone coverage reduces the probability that fraud occurs by 8 percentage points and decreases the share of fraudulent votes by 4 percentage points. \cite{gonzalez2021cell} is concerned with spillovers, highlighting for example that voters may walk from a non-covered to a covered area to report fraud. To test for spillovers, the author invokes donut-hole type reasoning and compares the fraud level at non-covered polling stations between 0--2 km of the boundary to those that are 2--4 km, 4--6 km or 6--8 km away. The idea is that spillovers should lead to spikes in fraud levels at polling stations closer to the boundary. They find that the differences in outcomes are not statistically significant and conclude that there is likely no spillovers.

In the spirit of the above exercise, we apply our method in the setting of \cite{gonzalez2021cell} as a robustness check. We first replicate the paper's result using either the method of \cite{armstrong2020simple} (AK) or \cite{calonico2014robust} (CCT). All tuning parameters are selected using the default procedures in the authors' respective \texttt{R} packages. Relative to the original specification, we do not include spatial fixed effects since they fall outside the scope of conventional theory. We then proceed with local spillover regression (LSR) taking the bandwidth choices of AK and CCT as given. Following \cite{gonzalez2021cell}, we consider circular neighborhoods that are between 2 to 8 km in radius. In principle, we can compute $\nu_i$ given the exact boundaries of cell phone coverage. Since that is not available, we estimate $\nu_i$ using the number of treated polling station in a given station's neighborhood, although we treat it as known when conducting inference. With the running variable measured in kilometers, we chose $M_m$ to be 0.01, which implies that the local linear approximation to the conditional probability of fraud has an error up to 1 (i.e. the entire support of the variable) at the 10 km radius. Similarly, we chose $M_\delta$ and $M_\gamma$ to be 0.01. Finally, we consider the case in which we select the spillover radius by cross-validation and choose bandwidth using our rule-of-thumb (LSR-CV).

Table \ref{tab:application_any_fraud} presents the results when using the indicator for fraud as outcome and the sample is the full set of polling stations. This is comparable to Panel A Column (1) in Table 2 of \cite{gonzalez2021cell}. Across the various bandwidth and radius choices, we find $\delta(0)$ to be positive and significant at the 5\% level. The estimates are also relatively stable across parameter values hovering between 0.6 to 0.75. Positive exogenous spillovers would arise, for example, if fraud is driven by local norms of corruption (\citealt{fisman2007corruption}; \citealt{barr2010corruption} and references therein). On the other hand, we do not find conclusive evidence of exogenous spillovers. These results continue to hold with data-driven choices of radius and bandwidth. Our cross-validation procedure suggests that spillovers occur across a radius of 6.1 km, which is within the plausible range of radii considered by \cite{gonzalez2021cell}. Results when using fraud share as outcome are similar.

The above exercise suggests that spillovers can be a threat to the identification of treatment effect parameters in RDD. In this case, our method can be useful as a robustness check. Additionally, the parameters for endogenous and exogenous spillovers may also be of interest since they can shed light on how social interactions mediate a variable of interest.

\begin{table}[htbp]
  \centering
    \begin{tabular}{ccccccc}
   \toprule\midrule
    Bandwidth & Eff. Obs & Method & Radius & $\tau_{DIR}$ & $\delta(0)$ & $\gamma(0)$ \bigstrut[b]\\
    \hline
    \multirow{10}[10]{*}{10.8} & \multirow{10}[10]{*}{1249} & \multirow{2}[2]{*}{AK} & \multirow{2}[2]{*}{-} & -0.028 &       &  \bigstrut[t]\\
          &       &       &       & [-0.093, 0.038] &       &  \bigstrut[b]\\
\cline{3-7}          &       & \multirow{2}[2]{*}{LSR} & \multirow{2}[2]{*}{2} & -0.077 & 0.585 & 0.103 \bigstrut[t]\\
          &       &       &       & [-0.372, 0.219] & [0.226, 0.944] & [-0.182, 0.388] \bigstrut[b]\\
\cline{3-7}          &       & \multirow{2}[2]{*}{LSR} & \multirow{2}[2]{*}{4} & -0.018 & 0.640 & -0.011 \bigstrut[t]\\
          &       &       &       & [-0.326, 0.291] & [0.240, 1.039] & [-0.309, 0.286] \bigstrut[b]\\
\cline{3-7}          &       & \multirow{2}[2]{*}{LSR} & \multirow{2}[2]{*}{6} & -0.021 & 0.719 & 0.000 \bigstrut[t]\\
          &       &       &       & [-0.333, 0.290] & [0.309, 1.129] & [-0.276, 0.277] \bigstrut[b]\\
\cline{3-7}          &       & \multirow{2}[2]{*}{LSR} & \multirow{2}[2]{*}{8} & -0.029 & 0.742 & -0.032 \bigstrut[t]\\
          &       &       &       & [-0.344, 0.286] & [0.307, 1.178] & [-0.330, 0.267] \bigstrut[b]\\
    \hline
    \multirow{10}[10]{*}{6.7} & \multirow{10}[10]{*}{1026} & \multirow{2}[2]{*}{CCT} & \multirow{2}[2]{*}{-} & 0.014 &       &  \bigstrut[t]\\
          &       &       &       & [-0.060, 0.088] &       &  \bigstrut[b]\\
\cline{3-7}          &       & \multirow{2}[2]{*}{LSR} & \multirow{2}[2]{*}{2} & -0.061 & 0.586 & 0.119 \bigstrut[t]\\
          &       &       &       & [-0.244, 0.123] & [0.347, 0.825] & [-0.043, 0.281] \bigstrut[b]\\
\cline{3-7}          &       & \multirow{2}[2]{*}{LSR} & \multirow{2}[2]{*}{4} & -0.008 & 0.642 & 0.002 \bigstrut[t]\\
          &       &       &       & [-0.191, 0.174] & [0.378, 0.906] & [-0.158, 0.162] \bigstrut[b]\\
\cline{3-7}          &       & \multirow{2}[2]{*}{LSR} & \multirow{2}[2]{*}{6} & -0.009 & 0.722 & 0.009 \bigstrut[t]\\
          &       &       &       & [-0.180, 0.162] & [0.471, 0.973] & [-0.141, 0.159] \bigstrut[b]\\
\cline{3-7}          &       & \multirow{2}[2]{*}{LSR} & \multirow{2}[2]{*}{8} & -0.024 & 0.752 & -0.025 \bigstrut[t]\\
          &       &       &       & [-0.191, 0.143] & [0.500, 1.004] & [-0.186, 0.136] \bigstrut[b]\\
    \hline
    \multirow{2}[2]{*}{3.2} & \multirow{2}[2]{*}{591} & \multirow{2}[2]{*}{LSR-CV} & \multirow{2}[2]{*}{6.10} & -0.004 & 0.744 & 0.023 \bigstrut[t]\\
          &       &       &       & [-0.103, 0.095] & [0.611, 0.876] & [-0.067, 0.112] \bigstrut[b]\\
    \midrule\bottomrule
    \end{tabular}%
     \caption{Comparison of AK, CCT and LSR in the setting of \cite{gonzalez2021cell}. Outcome is indicator for whether any fraud has occurred. This analysis uses the full sample of polling stations. Square brackets contain 95\% confidence intervals. Eff. Obs. is number of polling stations within the bandwidth.}
  \label{tab:application_any_fraud}%
\end{table}%

\section{Conclusion}\label{section--conclusion}

Regression Discontinuity Design allows researchers to obtain causal estimates of treatment effects at the cutoff. It relies on the assumption of no spillovers, which may not hold in practice. When neighborhoods are determined by the running variable, we find that the estimand of RDD is sensitive to the ratio of two terms: (1) radius over which spillovers occur and (2) the bandwidth used for local linear regression. In our preferred approximation, those two terms are of similar order so that an alternative to local linear regression is needed to recover direct treatment effects and spillovers. We propose the local spillover regression -- the local analog of peer effects regression -- and show that it can be a useful tool for addressing spillovers in RDD.

\renewcommand{\baselinestretch}{0}
\bibliographystyle{aer}
\bibliography{literature}
\renewcommand{\baselinestretch}{1.5}

\appendix

\section{Appendix}\label{section--proofs}

This Appendix provides proofs for the theorems in the main text. They draw on auxillary lemmas that are stated in Section \ref{section--auxillarylemmas} of the Supplemental Appendix.

\subsection{Proof of Theorem \ref{theorem--consistency}}

We consider the regimes case-by-case.  The proofs are similar across the cases except we use a different approximation for $\mu_d(xh_n) - \mu_d(0)$ depending on the regimes.

\subsubsection*{Case 1: $r_n \gg h_n$.}

We first derive the bias for $\hat{\beta}^+$. A similar argument follows for $\hat{\beta}^-$. The two together yield the bias for $\hat{\tau}_d$. By Assumption \ref{assump-model} and the standard expansion: {\small
\begin{align*}
	Y_i & = m^+(Z_i) + \delta(Z_i)\mu_d(Z_i) + \gamma(Z_i)\nu_d(Z_i) + \varepsilon_i \\
	& = m^+(0) + \delta(0)\mu_d(0) + \gamma(0)\nu_d(0) + m_z^+(0) \cdot Z_i + \frac{1}{2}m^+_{zz}(0) \cdot Z_i^2 \\
	& \qquad + \delta(0)\left(\mu_d(Z_i) - \mu_d(0)\right) + \delta_z(0) Z_i \mu_d(Z_i) + \frac{1}{2}\delta_{zz}(0) Z^2_i \mu_d(Z_i)  \\
	& \qquad + \gamma(0)\left(\nu_d(Z_i) - \nu_d(0)\right) + \gamma_z(0) Z_i \nu_d(Z_i) + \frac{1}{2}\gamma_{zz}(0) Z^2_i \nu_d(Z_i)  + \varepsilon_i + O(h_n^3)
\end{align*}}

\noindent In matrix form,
\begin{equation*}
	\mathbb{E}\left[ \hat{\beta}^+_0 \, \lvert \, \mathbf{Z} \right] = e_1^T \left(\mathbf{\tilde{Z}}^T\mathbf{W}\mathbf{\tilde{Z}}  \right)^{-1} \mathbf{\tilde{Z}} \mathbf{W} \mathbf{M}
\end{equation*}
where only observations with $Z_i \geq 0$ enter the following matrices: {\small
\begin{gather*}
	\mathbf{\tilde{Z}} = \begin{pmatrix}
		1 & Z_1 \\
		\vdots & \vdots \\
		1 & Z_{n_+}
	\end{pmatrix} \quad , \quad \mathbf{M} = \begin{pmatrix}
		m^+(Z_1) + \delta(Z_1)\mu_d(Z_1) + \gamma(Z_1)\nu_d(Z_1) \\
		\vdots \\
		m^+(Z_{n_+}) + \delta(Z_{n_+})\mu_d(Z_{n_+}) + \gamma(Z_{n_+}) \nu_d(Z_{n_+})
	\end{pmatrix}\\
	\mathbf{W} = \text{diag}\left(K_h(Z_1) \cdots K_h(Z_{n_+})  \right)~.
\end{gather*}}
As such, we can write
\begin{align*}
		\mathbf{E}\left[ \hat{\beta}^+_0 \, \lvert \, \mathbf{Z} \right] & =  m^+(0) + \delta(0)\mu_d(0) +  \gamma(0)\nu_d(0)   + A_n + B_n + R
\end{align*}
where $R = O_p(h_n^2/r_n)$ and {\footnotesize
\begin{align*}
		A_n & :=  \delta(0) \cdot e_1^T \left(\mathbf{\tilde{Z}}^T\mathbf{W}\mathbf{\tilde{Z}}  \right)^{-1}\mathbf{\tilde{Z}}\mathbf{W}\begin{pmatrix}
		\mu_d(Z_1) - \mu_d(0) \\
		\vdots \\
		\mu_d(Z_{n_+})  - \mu_d(0)
		\end{pmatrix} \quad , \quad 
		B_n & :=  \gamma(0) \cdot e_1^T \left(\mathbf{\tilde{Z}}^T\mathbf{W}\mathbf{\tilde{Z}}  \right)^{-1}\mathbf{\tilde{Z}}\mathbf{W}\begin{pmatrix}
		\nu_d(Z_1) - \nu_d(0) \\
		\vdots \\
		\nu_d(Z_{n_+})  - \nu_d(0)
		\end{pmatrix}~.
\end{align*}}

We first analyze $A_n$. Note that the denominator is standard. From \cite{wand1995kernel}, {\small
\begin{equation}\label{equation--consistency_large_r_denom} \hspace{-10mm}
\left(\frac{1}{n_+}\mathbf{\tilde{Z}}^T\mathbf{W}\mathbf{\tilde{Z}}  \right)^{-1} =  \left[h_n  \gamma_{2,1} \gamma_{0,1} - h_n \gamma_{1,1}^2 + o_p(h_n^2)\right]^{-1} \begin{pmatrix}
	h_n^2 \gamma_{2,1} + o_p(h_n^2) & -h_n\gamma_{1,1} + o_p(h_n) \\
	* & \gamma_{0,1} + o_p(h_n^2)
\end{pmatrix}
\end{equation}}
where have normalized $f(0)$ to $1$. For the numerator,{\small
\begin{equation*}
	\frac{1}{n_+} \mathbf{\tilde{Z}}\mathbf{W}\begin{pmatrix}
			\mu_d(Z_1) - \mu_d(0) \\
			\vdots \\
			\mu_d(Z_{n_+})  - \mu_d(0)
			\end{pmatrix} = \begin{pmatrix}
			\frac{1}{n_+} \sum_{i=1}^{n_+} K_h(Z_i)\left(\mu_d(Z_i ) - \mu_d(0)\right) \\
			\frac{1}{n_+} \sum_{i=1}^{n_+} K_h(Z_i) Z_i \left(\mu_d(Z_i ) - \mu_d(0)\right)
			\end{pmatrix}
\end{equation*}}

\noindent We evaluate the expectation of each term with respect to $\mathbf{Z}$. First, {\small
\begin{align}\label{equation--consistency_large_r_num1} \hspace{-10mm}
		\mathbb{E}\left[ \frac{1}{n_+} \sum_{i=1}^{n_+} K_h(Z_i)\left(\mu_d(Z_i ) - \mu_d(0)\right)  \right] 
		& = \frac{h^2_n}{r_n}  \left( \Gamma(r_n)  \int_0^1 K(x)  dF(x) + o(1)\right)= \frac{h^2_n}{r_n}  \left( \Gamma(r_n) \cdot \gamma_{0,1} +o(1) \right) ~,
\end{align}}
where the first equality follows from the usual change of variables argument and last equality follows from Lemma \ref{lemma--uniform_approx}, which also provides the exact form of $\Gamma(r_n)$. By a similar set of manipulations, {\small
\begin{align}\label{equation--consistency_large_r_num2} \hspace{-15mm}
	& \mathbb{E}\left[ \frac{1}{n_+} \sum_{i=1}^{n_+} K_h(Z_i)Z_i\left(\mu_d(Z_i ) - \mu_d(0)\right)  \right] = \frac{h^2_n}{r_n} \left(  \Gamma(r_n)  \int_0^1 x K(x)  dF(x) + o(1) \right)= \frac{h^2_n}{r_n} \left(  \Gamma(r_n) \cdot \gamma_{1,1} + o(1) \right)~.
\end{align}}

\noindent Combining equations \eqref{equation--consistency_large_r_denom}, \eqref{equation--consistency_large_r_num1} and \eqref{equation--consistency_large_r_num2}, we have that
$
A_n = O_p\left(\frac{h_n^2}{r_n}\right)
$.
Finally, observe that $\frac{1}{2}m^+_{zz}(0)Z^2_i$ contributes to the standard $O_p(h_n^2)$ bias term in RDD. It is also straightforward to show that the residual corresponding to $\delta_z(0) Z_i \mu_d(Z_i) $ and $\delta_z(0) Z^2_i \mu_d(Z_i)$ are $h_n$ times smaller than the above display.

Next, consider $B_n$. The denominator is the same as for $A_n$ and it's limit is given by Equation \eqref{equation--consistency_large_r_denom}. To analyze the numerator, first observe that {\small
\begin{align}\label{equation--nu--formula}
	\nu_d(Z_i) - \nu_d(0) = \begin{cases}
	-\frac{1}{2} & \mbox{ if } Z_i \leq -r_n \\
	\frac{Z_i}{2r_n} & \mbox{ if } -r_n \leq Z_i \leq Z_i \\
	\frac{1}{2} & \mbox{ if } Z_i \geq r_n
	\end{cases}
\end{align}}

\noindent As such, if $h_n \ll r_n$, $\nu_d(Z_i) - \nu_d(0) = O_p(h_n)$ conditional on $Z_i \in [-h_n, h_n]$. It is therefore immediate that analogs of Equations \ref{equation--consistency_large_r_num1} and \ref{equation--consistency_large_r_num2} hold for $B_n$. Therefore, we also have that
$
B_n = O_p\left(\frac{h_n^2}{r_n}\right)
$.
It remains to note that the residual error from approximating $\gamma(Z_i)$ at $\gamma(0)$ is $O_p(h_n^2)$, as in the case with $\delta(\cdot)$ above.

Conclude that
$
	\hat{\beta}^+_0 \overset{p}{\to} m^+(0) + \delta(0) \mu_d(0) + \gamma(0)\nu_d(0)
$.
The stated properties on $\hat{\tau}_{\text{RDD}}$ follow immediately by a similar set of manipulations for $\hat{\beta}_0^-$

\subsubsection*{Case 2: $r_n /h_n \to 0$}

Again, start by considering $A_n$. For $Z_i \geq 0$, the change of variables $x = z/h_n$ yields:
\begin{align}\label{equation--consistency_small_r_num1}
		\mathbb{E}\left[ \frac{1}{n_+} \sum_{i=1}^{n_+} K_h(Z_i)\left(\mu_d(Z_i ) - \mu_d(0)\right)  \right] 
		& = h_n \int_0^1 K(x) \left(\mu_d(xh_n)- \mu_d(0)\right) dF(x) \nonumber \\ 
		& = h_n \left(\frac{\tau_d + \delta(0)\gamma(0)}{2(1-\delta(0))} \cdot  \gamma_{0,1} + o(1)\right)
\end{align}
where the first equality follows from Lemma \ref{lemma--uniform_approx} and the fact that the integrand is bounded. And similarly,
\begin{align}\label{equation--consistency_small_r_num2}
		& \mathbb{E}\left[ \frac{1}{n_+} \sum_{i=1}^{n_+} K_h(Z_i)Z_i\left(\mu_d(Z_i ) - \mu_d(0)\right)  \right]  = h^2_n \left(\frac{\tau_d + \delta(0)\gamma(0)}{2(1-\delta(0))} \gamma_{1,1} + o(1)\right)
\end{align}
Combining equations \eqref{equation--consistency_large_r_denom}, \eqref{equation--consistency_small_r_num1} and \eqref{equation--consistency_small_r_num2} gives us
$
	A_n = \delta(0) \cdot \frac{\tau_d + \delta(0)\gamma(0)}{2(1-\delta(0))} + O_p(h_n)
$.

Next, consider $B_n$.  From Equation \ref{equation--nu--formula}, we have that $|Z_i| \in [w_n, 1]$, $\nu_d(Z_i) - \nu_d(0) = \text{sgn}(Z_i)/2$. On $[-w_n, w_n]$, the term is bounded. As such, we can write:
\begin{align*}
	\nu_d(Z_i) - \nu_d(0) = \frac{1}{2}\text{sgn}(Z_i) + O(w_n)
\end{align*}
uniformly over $Z_i \in \mathcal{Z}$. By similar manipulations as for $A_n$, but using the approximation above,
$
	B_n = \frac{\gamma(0)}{2} + O_p(h_n)
$.
As such,
\begin{align*}
	\hat{\beta}^+_0 &= m^+(0) + \delta(0)\mu_d(0) + \gamma(0)\nu_d(0) + \delta(0) \cdot \frac{\tau_d + \gamma(0)}{2(1-\delta(0))} + \frac{\gamma(0)}{2} + O_p(h_n)
\end{align*}
Similarly for $\hat{\beta}^1$. Therefore $\hat{\tau}_{\text{RDD}} \overset{p}{\to} \frac{\tau_d + \gamma(0)}{1-\delta(0)}$ \hfill \qed

\subsubsection*{Case 3: $r_n = \frac{1}{2} ch_n$}
As before, using Lemma \ref{lemma--uniform_approx} yields:
\begin{align}\label{equation--consistency_constant_r_num1}
		& \mathbb{E}\left[ \frac{1}{n_+} \sum_{i=1}^{n_+} K_h(Z_i)\left(\mu_d(Z_i ) - \mu_d(0)\right)  \right]
		& = h_n \left(\tau_d \Lambda^+_{0,1,1}+ \gamma(0) \tilde{\Lambda}^+_{0,1,1} + o(1) \right)
\end{align}
\vspace{-10mm}
\begin{align}\label{equation--consistency_constant_r_num2}
		& \mathbb{E}\left[ \frac{1}{n_+} \sum_{i=1}^{n_+} K_h(Z_i)Z_i\left(\mu_d(Z_i ) - \mu_d(0)\right)  \right]
		& = h^2_n \left(\tau_d  \Lambda^+_{1,1,1} + \gamma(0) \tilde{\Lambda}^+_{1,1,1} + o(1) \right)
\end{align}
Combining equations \eqref{equation--consistency_large_r_denom}, \eqref{equation--consistency_constant_r_num1} and \eqref{equation--consistency_constant_r_num2} gives us{\small
\begin{align*}
	A_n & = \frac{\tau_d}{\gamma_{2,1}\gamma_{0,1} - \gamma_{1,1}^2} \cdot \left( \gamma_{2,1}\Lambda^+_{0,1,1} - \gamma_{1,1}\Lambda^+_{1,1,1}\right)
	 + \frac{\gamma(0)}{\gamma_{2,1}\gamma_{0,1} - \gamma_{1,1}^2} \cdot \left( \gamma_{2,1}\tilde{\Lambda}^+_{0,1,1} - \gamma_{1,1}\tilde{\Lambda}^+_{1,1,1}\right) + O_p(h_n)~.
\end{align*}}
Next, for $B_n$:
\begin{align}\label{equation--consistency_constant_r_num1_nu}
		\mathbb{E}\left[ \frac{1}{n_+} \sum_{i=1}^{n_+} K_h(Z_i)\left(\nu_d(Z_i ) - \nu_d(0)\right)  \right]
		& = h_n \left(\Gamma^+_{0,1,1} + o(1)\right)
\end{align}
\vspace{-10mm}
\begin{align}\label{equation--consistency_constant_r_num2_nu}
		& \mathbb{E}\left[ \frac{1}{n_+} \sum_{i=1}^{n_+} K_h(Z_i)Z_i\left(\nu_d(Z_i ) - \nu_d(0)\right)  \right] = h^2_n \left(\Gamma^+_{1,1,1} + o(1)\right)~.
\end{align}
By Equations \eqref{equation--consistency_large_r_denom}, \eqref{equation--consistency_constant_r_num1_nu} and \eqref{equation--consistency_constant_r_num2_nu}, we have that
$
	B_n = \frac{\gamma_{2,1}\Gamma_{0,1,1}^+ - \gamma_{1,1} \Gamma^+_{1,1,1}}{\gamma_{2,1}\gamma_{0,1} - \gamma_{1,1}^2} + O_p(h_n)
$.
Noting that the usual bias term coming from $m_{zz}$ is smaller than the above, we therefore have that: {\small
\begin{align*}
	\hat{\beta}^+_0 & = m^+(0) + \delta(0)\mu_d(0) + \gamma(0)\nu_d(0) \\
	& \qquad + \frac{\delta(0)\tau_d}{\gamma_{2,1}\gamma_{0,1} - \gamma_{1,1}^2} \cdot \left( \gamma_{2,1}\Lambda^+_{0,1,1} - \gamma_{1,1}\Lambda^+_{1,1,1}\right) \\
	& \qquad + \frac{\delta(0)\gamma(0)}{\gamma_{2,1}\gamma_{0,1} - \gamma_{1,1}^2} \cdot \left( \gamma_{2,1}\tilde{\Lambda}^+_{0,1,1} - \gamma_{1,1}\tilde{\Lambda}^+_{1,1,1}\right) \\
	& \qquad + \gamma(0) \cdot \frac{\gamma_{2,1}\Gamma_{0,1,1}^+ - \gamma_{1,1} \Gamma^+_{1,1,1}}{\gamma_{2,1}\gamma_{0,1} - \gamma_{1,1}^2} + O_p(h_n)~.
\end{align*}}
Similarly for $\hat{\beta}^-$. 
As such, $	\hat{\tau}_{\text{RDD}} = \tau_* + O_p(h_n)$,  where {\small
\begin{align}\label{equation--tau_star}
	\begin{split}
		\tau_* - \tau_d& = \frac{\delta(0)\tau_d}{\gamma_{2,1}\gamma_{0,1} - \gamma_{1,1}^2} \cdot \left( \gamma_{2,1}\left(\Lambda^+_{0,1,1} - 	\Lambda^-_{0,1,1}\right) - \gamma_{1,1}\left(\Lambda^+_{1,1,1} - \Lambda^-_{1,1,1}\right)\right) \\
		& \qquad + \frac{\delta(0)\gamma(0)}{\gamma_{2,1}\gamma_{0,1} - \gamma_{1,1}^2} \cdot \left( \gamma_{2,1}\left(\tilde{\Lambda}^+_{0,1,1} - \tilde{\Lambda}^-_{0,1,1}\right) - \gamma_{1,1}\left(\tilde{\Lambda}^+_{1,1,1} - \tilde{\Lambda}^-_{1,1,1}\right)\right)\\
		& \qquad +\frac{\gamma(0)}{\gamma_{2,1}\gamma_{0,1} - \gamma_{1,1}^2} \left(\gamma_{2,1}\left(\Gamma^+_{0,1,1} - \Gamma^+_{0,1,1}\right) - \gamma_{1,1}\left(\Gamma^-_{1,1,1} - \Gamma^-_{1,1,1}\right) \right)~.
	\end{split}
\end{align}}
and $\Lambda, \tilde{\Lambda}, \Gamma$ are as in Definition \ref{definition--integrals}.
\hfill\qed

\subsection{Proof of Theorems \ref{theorem--spilloverreg_direct} and \ref{theorem--spilloverreg_indirect}}
\label{prf--thm--spilloverreg}

We prove Theorem 3. Theorem 2 follows from the usual Frisch-Waugh-Lovell argument after observing that the projection residuals of $Y_i$ and $D_i$ on $\tilde{\mu}(Z_i) - \tilde{\mu}(0)$, $\tilde{\nu}(Z_i) - \tilde{\nu}(0)$ and $Z_i$ are well-defined even if they are collinear. The proof for Theorem 3 largely follows standard arguments as in Chapter 5 of \cite{wand1995kernel}. There are two complications: (1) $\tilde{X}_i$ is measured with error. We use a concentration inequality in Lemma \ref{lemma--conc_mu} to show that this error is neglible. (2) We require a uniform approximation for $\mu(Z_i)-\mu(0)$. This is provided in Lemma \ref{lemma--uniform_approx}.

Under our assumptions, {\small
\begin{equation*}
  \widetilde{\beta} - \beta = \left( \frac{1}{n} \sum_{i = 1}^{n} K_{h} \left(
  Z_{i} \right) \widetilde{X}_{i} \widetilde{X}_{i}^{\prime} \right)^{- 1}
  \frac{1}{n} \sum_{i = 1}^{n} K_{h} \left( Z_{i} \right) \widetilde{X}_{i}
  \left\{ \left( X_{i} - \widetilde{X}_{i} \right)^{\prime} \beta + \rho_{i} +
  \varepsilon_{i} \right\},
\end{equation*}}

\noindent which can be rewritten as $\widetilde{\beta} - \beta =
   \widetilde{\mathbf{Q}}^{- 1} \left\{ \mathbf{B}_{n}+
  \widetilde{\mathbf{T}}_{X} + \widetilde{\mathbf{T}}_{\varepsilon}
  \right\}$, where {\small
  \begin{equation*}  \label{eqn--betatilde-breadown-1}
  \begin{array}{ll}
 \widetilde{\mathbf{Q}} = \frac{1}{n} \sum_{i = 1}^{n} K \left( Z_{i} / h_n \right) \widetilde{X}_{i}
  \widetilde{X}_{i}^{\prime} 
  & \mathbf{B}_{n} =
     \frac{1}{n} \sum_{i = 1}^{n} K \left( Z_{i} / h_n \right) \widetilde{X}_{i}
    \rho_{i} \\
   \widetilde{\mathbf{T}}_{X} =
      \frac{1}{n} \sum_{i = 1}^{n} K \left( Z_{i} / h_n \right) \widetilde{X}_{i}
     \left(X_{i} - \widetilde{X}_{i} \right)^{\prime} \beta \qquad \qquad& 
  \widetilde{\mathbf{T}}_{\varepsilon} =
  \frac{1}{n} \sum_{i = 1}^{n} K \left( Z_{i} / h_n \right) \widetilde{X}_{i}
  \varepsilon_{i}.     
  \end{array}
  \end{equation*}
We start with the terms in $\widetilde{\mathbf{Q}}$. First note that for $1 \leq i, j \leq 4$, $\widetilde{\mathbf{Q}}_{ij}$ are terms associated with the usual local linear regression. Standard arguments and noting that $f'(0) = 0$ give us:
{\scriptsize
\begin{equation*}
	\frac{1}{h_n}\widetilde{\mathbf{Q}}_{1:4, 1:4} = f(0) \cdot \begin{pmatrix}
	   1 + o_p(1)&  o_p(h_n^2)&   \int_{0}^{1} K(x) dx + o_p(1) &  h_n\left( \int_{0}^{1} x K(x) dx + o_p(1)  \right) \\
	 *&  h_n^2 \left(\int_{-1}^{1} x^2 K(x) dx  + o_p(1) \right) & h_n \left( \int_{0}^{1} xK(x) dx + o_p(1) \right) &  h_n^2 \left(\int_{0}^{1} x^2 K(x) dx  + o_p(1) \right) \\
	*& *&  \int_{0}^{1} K(x) dx + o_p(1) & h_n\left(   \int_{0}^{1} x K(x) dx + o_p(1)\right)  \\
	*& *& *& h_n^2 \left(  \int_{0}^{1} x^2 K(x) dx  + o_p(1)\right)  \\
	\end{pmatrix}
\end{equation*}}

Next, consider
{\small
\begin{align*}\hspace{-20mm}
	\frac{1}{h_n} \widetilde{\mathbf{Q}}_{1,5}  
	& =  \frac{1}{nh_n} \sum_{i = 1}^{n} K \left( Z_{i} / h \right)\left({\mu}(Z_i) - {\mu}(0)\right) +  \frac{1}{nh_n} \sum_{i = 1}^{n} K \left( Z_{i} / h_n \right)\left(\tilde{\mu}(Z_i) - \mu(Z_i)\right) +   \frac{1}{nh_n} \sum_{i = 1}^{n} K \left( Z_{i} / h \right)\left(\tilde{\mu}(0) - \mu(0)\right)~.
\end{align*}}
By the Lemma \ref{lemma--conc_mu},
\begin{align*}
	\frac{1}{nh_n} \sum_{i = 1}^{n} K \left( Z_{i} / h_n \right)\left(\tilde{\mu}(Z_i) - \mu(Z_i)\right) 
	& = o(1)  \cdot (f(0) +  o_p(1)) \quad \quad \text{w.p.a. 1}
\end{align*}
Next, denote the uniform approximation in the $r_n \propto h_n$ regime in Lemma \ref{lemma--uniform_approx} by $u(x)$. Using $u(x)$ and applying the standard arguments yields:
\begin{align*}
	 \frac{1}{nh_n} \sum_{i = 1}^{n} K \left( Z_{i} / h_n  \right)\left({\mu}(Z_i) - {\mu}(0)\right)   = \int_{-1}^{1}  u(x) K(x) dx \cdot f(0)  +  o_p(1)
\end{align*}
As such, we conclude that
$
	\frac{1}{h_n} \widetilde{\mathbf{Q}}_{1,5}  = \int_{-1}^{1}  u(x) K(x) dx \cdot f(0)  +  o_p(1)
$.
Similar arguments apply with the remaining components of $\widetilde{X}_i$. Note that $\nu(xh_n) - \nu(0) = \max\left\{-1, \min\left\{1, \frac{x}{c}\right\} \right\} =: v(x)$ is exact. This yields:
{\scriptsize
\begin{align*}
	& \hspace{-17mm} \frac{1}{h_n} \widetilde{\mathbf{Q}}_{1:8, 5:8} = f(0) \cdot \\
	&\hspace{-17mm} \begin{pmatrix}
		 \int_{-1}^{1}  u(x) K(x) dx  +  o_p(1) &  h_n \left(\int_{-1}^{1}  xu(x) K(x) dx   +  o_p(1) \right)
		 & \int_{-1}^{1}  v(x) K(x) dx  +  o_p(1)  & h_n \left(\int_{-1}^{1}  xv(x) K(x) dx  +  o_p(1) \right) \\
		h_n \left(\int_{-1}^{1}  x u(x) K(x) dx +  o_p(1) \right)&  h_n^2 \left(\int_{-1}^{1}  x^2u(x) K(x) dx   +  o_p(1) \right)
		& h_n \left(\int_{-1}^{1}  x v(x) K(x) dx  +  o_p(1) \right) & h_n^2 \left(\int_{-1}^{1}  x^2v(x) K(x) dx  +  o_p(1) \right) \\
		\int_{0}^{1}  u(x) K(x) dx  +  o_p(1) &  h_n \left(\int_{0}^{1}  xu(x) K(x) dx   +  o_p(1) \right)
		& \int_{0}^{1}  v(x) K(x) dx  +  o_p(1)  & h_n \left(\int_{0}^{1}  xv(x) K(x) dx  +  o_p(1) \right) \\
		h_n \left(\int_{0}^{1}  x u(x) K(x) dx +  o_p(1) \right)&  h_n^2 \left(\int_{0}^{1}  x^2u(x) K(x) dx   +  o_p(1) \right)
		& h_n \left(\int_{0}^{1}  x v(x) K(x) dx  +  o_p(1) \right) & h_n^2 \left(\int_{0}^{1}  x^2v(x) K(x) dx  +  o_p(1) \right) \\
		 \int_{-1}^{1}  u^2(x) K(x) dx  +  o_p(1) &  h_n \left(\int_{-1}^{1}  xu^2(x) K(x) dx   +  o_p(1) \right)
		 & \int_{-1}^{1}  u(x) v(x) K(x) dx  +  o_p(1)  & h_n \left(\int_{-1}^{1}  xu(x) v(x) K(x) dx  +  o_p(1) \right) \\
		* &  h_n^2 \left(\int_{-1}^{1}  x^2u^2(x) K(x) dx   +  o_p(1) \right)
		& h_n \left(\int_{-1}^{1}  x u(x) v(x) K(x) dx  +  o_p(1) \right) & h_n^2 \left(\int_{-1}^{1}  x^2u(x)v(x) K(x) dx  +  o_p(1) \right) \\
		*&  *
		 & \int_{-1}^{1}   v^2(x) K(x) dx  +  o_p(1)  & h_n \left(\int_{-1}^{1}  x v^2(x) K(x) dx  +  o_p(1) \right) \\
		* &  * & * & h_n^2 \left(\int_{-1}^{1}  x^2v^2(x) K(x) dx  +  o_p(1) \right) \\
	\end{pmatrix}
\end{align*}}

\noindent Invertibility of $\widetilde{\mathbf{Q}}$ follows from observing that $v(x)$ is piecewise linear (but not linear) while $u(x)$ is an infinite sum of piecewise polynomials given our assumed identification conditions.

Next, consider $\mathbf{B}_{n}$. Since $m^+, m^-, \delta$ and $\gamma$ are all twice continuously differentiable, $\rho_i = O(h_n^2)$. Write:{\small
\begin{align*}
	 \frac{1}{h_n}\mathbf{B}_{n}  = \frac{1}{nh_n} \sum_{i = 1}^{n} K \left( Z_{i} / h_n \right) \widetilde{X}_{i} \rho_i
	  =  \frac{1}{nh_n} \sum_{i = 1}^{n} K \left( Z_{i} / h_n \right) {X}_{i} \rho_i +  \frac{1}{nh_n} \sum_{i = 1}^{n} K \left( Z_{i} / h_n \right) \left(\tilde{X}_{i} -X_i \right)\rho_i
\end{align*}}
Now, {\small
\begin{align*}
	 \frac{1}{nh_n} \sum_{i = 1}^{n} K \left( Z_{i} / h_n \right) \left(\tilde{X}_{i} -X_i \right)\rho_i & = O(h_n^2) \cdot \frac{1}{nh_n} \sum_{i = 1}^{n} K \left( Z_{i} / h_n \right) \left(\tilde{X}_{i} -X_i \right) \\
	 & = O(h_n^2) \cdot o_p(1) \cdot \frac{1}{nh_n} \sum_{i = 1}^{n} K \left( Z_{i} / h_n \right) \iota = o_p(h_n^2)
\end{align*}}

\noindent Meanwhile, by the preceding analysis for $\widetilde{\mathbf{Q}}$, {\footnotesize
\begin{align*}
	 \frac{1}{nh_n} \sum_{i = 1}^{n} K \left( Z_{i} / h_n \right) {X}_{i} \rho_i = O(h_n^2) \cdot \begin{pmatrix}
	 	1 + o_p(1) \\
	 	o_p(h_n^2) \\
	 	\int_{0}^{1} K(x) dx + o_p(1) \\
	 	h_n\left( \int_{0}^{1} x K(x) dx + o_p(1)  \right)\\
 		\int_{-1}^{1}  u(x) K(x) dx  +  o_p(1) \\
   		h_n \left(\int_{-1}^{1}  xu(x) K(x) dx   +  o_p(1) \right) \\
   		\int_{-1}^{1}  v(x) K(x) dx  +  o_p(1)  \\
   		h_n \left(\int_{-1}^{1}  xv(x) K(x) dx  +  o_p(1) \right)
	 \end{pmatrix}
	  \quad \Rightarrow \quad
	  \frac{1}{h_n}\mathbf{B}_{n}  = \begin{pmatrix}
	 	 	o_p(h_n) \\
	 	 	o_p(h^2_n) \\
	 	 	o_p(h_n) \\
	 	 	o_p(h^2_n) \\
	 	 	o_p(h_n) \\
	 	 	o_p(h^2_n) \\
	 	 	o_p(h_n) \\
	 	 	o_p(h^2_n)
	 	 \end{pmatrix}~.
\end{align*}}

Next, consider $\widetilde{\mathbf{T}}_X$. By Lemma \ref{lemma--conc_mu}, we again have that:
{\footnotesize 
\begin{align*}
	  \frac{1}{h_n}\widetilde{\mathbf{T}}_{X} & 
	   = \frac{1}{n} \sum_{i = 1}^{n} K \left( Z_{i} / h_n \right) {X}_{i}
	  	  \left(X_{i} - \widetilde{X}_{i} \right)^{\prime} \beta + \frac{1}{n} \sum_{i = 1}^{n} K \left( Z_{i} / h_n \right)  \left(X_{i} - \widetilde{X}_{i}  \right)   \left(X_{i} - \widetilde{X}_{i} \right)^{\prime} \beta \\
	  	&= \begin{pmatrix}
	  	 	 	o_p(h_n) &
	  	 	 	o_p(h^2_n) &
	  	 	 	o_p(h_n) &
	  	 	 	o_p(h^2_n) &
	  	 	 	o_p(h_n) &
	  	 	 	o_p(h^2_n) &
	  	 	 	o_p(h_n) &
	  	 	 	o_p(h^2_n)
	  	 	 \end{pmatrix}' ~.
\end{align*}}

Finally, consider $\widetilde{\mathbf{T}}_\varepsilon$:
\begin{align*}
	 \frac{1}{h_n}\widetilde{\mathbf{T}}_{\varepsilon} 
	  & = \frac{1}{nh_n} \sum_{i = 1}^{n} K \left( Z_{i} / h_n \right) {X}_{i} \varepsilon_{i} +  \frac{1}{nh_n} \sum_{i = 1}^{n} K \left( Z_{i} / h_n \right) \left(X_i - \widetilde{X}_{i}\right) \varepsilon_{i}
\end{align*}
Standard arguments based on the continuous differentiability of $\sigma^2(z)$ yields:
{\small
\begin{align*}
	 \hspace{-20mm} \frac{1}{nh_n} \sum_{i = 1}^{n} K \left( Z_{i} / h_n \right) {X}_{i} \varepsilon_{i}  & = O_p\left(\frac{1}{\sqrt{nh_n}}\right) \cdot \begin{pmatrix}
	 	O_p(1) &
	 	O_p(h_n) &
	 	O_p(1) &
	 	O_p(h_n) &
	 	O_p(1) &
	 	O_p(h_n) &
	 	O_p(1) &
	 	O_p(h_n)
	 \end{pmatrix}'\\
	 & = 
	 \begin{pmatrix}
  	 	 	o_p(h_n) &
  	 	 	o_p(h^2_n) &
  	 	 	o_p(h_n) &
  	 	 	o_p(h^2_n) &
  	 	 	o_p(h_n) &
  	 	 	o_p(h^2_n) &
  	 	 	o_p(h_n) &
  	 	 	o_p(h^2_n)
 	 \end{pmatrix}' \quad \mbox{ since } h_n \gg n^{-1/3}~.
\end{align*}}

By another application of Lemma \ref{lemma--conc_mu},
{\small
\begin{align*}\hspace{-10mm}
	\frac{1}{nh_n} \sum_{i = 1}^{n} K \left( Z_{i} / h_n \right) \left(X_i - \widetilde{X}_{i}\right) \varepsilon_{i} & = o(h_n) \cdot \begin{pmatrix}
		0 &
		0 &
		0 &
		0 &
		O_p\left(\frac{1}{\sqrt{nh_n}}\right) &
		O_p\left(\frac{h_n}{\sqrt{nh_n}}\right) &
		0 &
		0
	\end{pmatrix}' \quad \mbox{ w.p.a. 1} \\
 	  \Rightarrow \qquad 
 	 \frac{1}{h_n}\widetilde{\mathbf{T}}_{\varepsilon} &=
 	 	 \begin{pmatrix}
 	   	 	 	o_p(h_n) &
 	   	 	 	o_p(h^2_n) &
 	   	 	 	o_p(h_n) &
 	   	 	 	o_p(h^2_n) &
 	   	 	 	o_p(h_n) &
 	   	 	 	o_p(h^2_n) &
 	   	 	 	o_p(h_n) &
 	   	 	 	o_p(h^2_n)
 	  	 \end{pmatrix}'~.
\end{align*}}

Given the above analysis, let $W_n = \text{diag}(1, h_n, 1, h_n, 1, h_n, 1, h_n)$ and write $\overline{\mathbf{Q}} = W_n \widetilde{\mathbf{Q}} W_n$. The entries of $\overline{\mathbf{Q}}$ converge to non-zero values in probability as $n \to \infty$ and it is asymptotically full-rank. Then,
$
	\widetilde{\beta} - \beta = 	W_n^{-1} \overline{\mathbf{Q}}^{-1} W_n^{-1} \left(\widetilde{\mathbf{T}}_{\rho}+\widetilde{\mathbf{T}}_{X}+\widetilde{\mathbf{T}}_{\varepsilon}\right)
$
and
\begin{equation*}
	W_n^{-1}\frac{1}{h_n}\mathbf{B}_{n} = o_p(h_n) \cdot \iota \quad \Rightarrow \quad \overline{\mathbf{Q}}^{-1} W_n^{-1}\frac{1}{h_n}\mathbf{B}_{n}= o_p(h_n) \cdot \iota \quad \Rightarrow \quad W_n^{-1} \overline{\mathbf{Q}}^{-1} W_n^{-1}\frac{1}{h_n}\mathbf{B}_{n}= o_p(1) \cdot \iota~.
\end{equation*}
By a similar argument for  $\widetilde{\mathbf{T}}_{X}$ and $\widetilde{\mathbf{T}}_{\varepsilon}$, we are done.

\subsection{Proof of Theorem \ref{thm--spilloverreg-inference}}

Henceforth, we denote \(h := h_{n}\) and \(r := r_{n}\) and drop
the subscript \(n\) unless there is risk of ambiguity.
Use \eqref{eqn--betatilde-breadown-1} and write
\begin{equation}
  \widetilde{\mathbf{T}} = \widetilde{\mathbf{Q}} \left( \widetilde{\beta} -
  \beta - \mathbf{B}_{n} \right) \quad \text{and} \quad
  \widetilde{\mathbf{T}}_{\ast} = \frac{1}{n} \sum_{i = 1}^{n} \left[ K
  \left(\frac{Z_{i}}{h} \right) X_{i} \varepsilon_{i} + \dot{\eta}_{X} \left(
  \xi_{i} \right) \right],
  \label{eqn--Ttilde-Ttilde-ast-def}
\end{equation}
where
\begin{equation*}
  \dot{\eta}_{X} \left( \xi_{i} \right) = \eta_{X} \left( \xi_{i} \right) - \E
  \left[ \eta_{X} \left( \xi_{1} \right) \right],
  \label{eqn--etaX-dot-def}
\end{equation*}
and \(\eta_{X} (\cdot)\) is given in \eqref{eqn--etaX-def}.
Note that \(\E \left[ \sqrt{n} \widetilde{\mathbf{T}}_{\ast} \right] =
\mathbf{0}\) and \(\Var \left[ \sqrt{n} \widetilde{\mathbf{T}}_{\ast} \right] =
\bm{\Omega}\).
The claim \eqref{eqn--spilloverreg-inference-std-gauss} is equivalent to
\(\bm{\Omega}^{- 1 / 2} \sqrt{n} \widetilde{\mathbf{T}}
\overset{\mathrm{d}}{\to} \N \left( 0, \mathbb{I}_{8} \right)\) and by Slutsky's
theorem, is implied by
\begin{equation}
  \bm{\Omega}^{- 1 / 2} \sqrt{n} \widetilde{\mathbf{T}}_{\ast}
  \overset{\mathrm{d}}{\to} \N \left( 0, \mathbb{I}_{8} \right) \quad \text{and}
  \quad \bm{\Omega}^{- 1 / 2} \sqrt{n} \left( \widetilde{\mathbf{T}} -
  \widetilde{\mathbf{T}}_{\ast} \right) = o_{\mathrm{p}} (1).
  \label{eqn--spilloverreg-inference-std-gauss-sufficient}
\end{equation}
The asymptotic normality result is due to
Lemma \ref{lem--spilloverreg-inference-std-gauss-sufficient-an}.
The asymptotic negligibility result is due to
Lemma \ref{lem--spilloverreg-inference-std-gauss-sufficient-op1}.
\qed

\clearpage
\newpage
\begin{center}
	\Large Supplemental Appendix
\end{center}

\section{Simulations}\label{section--simulations}

This section compares the performance of LSR to prevailing methods popular for RDD in the context of a simple DGP. We first show that LLR with a naively chosen bandwidth exhibits behavior best described by the $r_n \propto h_n$ case in Theorem \ref{theorem--consistency}. We then compare the performance of LSR with the methods of \cite{armstrong2020simple} (AK hereinafter) and \cite{calonico2014robust} (CCT hereinafter) which involve more sophisticated rules for bandwidth choice. 

Consider the data-generating process described in Section \ref{section--setup} with the following values:
\vspace{-3mm}
\begin{gather*}
	m^+(z) = 2 \quad , \quad m^-(z) = 1 \quad, \quad \tau_\text{DTE} = 1 \\
	\delta(z) = \delta \quad , \quad \gamma(z) = \gamma
\end{gather*}

The researcher observes
\begin{equation*}
	Y_i = D_i Y^+_d(Z_i) + (1-D_i) Y^-_d(Z_i) + \varepsilon_i \quad , \quad \varepsilon_i \sim \text{N}(0,\sigma^2)
\end{equation*}
where $\sigma = 0.2$. The values of $n$, $r_n$, $\delta$ and $\gamma$ will vary depending on the set up. We estimate LLR using a triangular kernel. LLR has no bias in this setting. We set $M_m = M_\delta = M_\gamma = 0.5$ for conservative inference. The following results are based on 1,000 draws for each set of parameter values.

\subsection{Accuracy of the Preferred Regime}

We first present simulations pertaining to Theorem \ref{theorem--consistency}. Figure \ref{fig:llr_transition_naive} plots the mean of $\hat{\tau}_\text{RDD}$ under various values of $\delta$ and $\gamma$ when $n = 2000$ and bandwidth  is chosen based on the rule-of-thumb in \eqref{equation--initial_h} ($\approx 0.22$). We see that as $c = r_n/h_n$ decreases to $0$, the LLR estimate goes from $\tau_\text{DIR} = 1$ to $\tau_\text{TOT} = \frac{\tau_\text{DIR}+\gamma}{1-\delta}$. This is predicted by the $r_n \propto h_n$ case in our Theorem \ref{theorem--consistency}. Approximations under the regime $r_n \gg h_n$ or $r_n \ll h_n$ would predict either end point but would miss the transition. As such, our preferred regime $r_n \propto h_n$ leads to a more accurate approximation.

In practice, LLR is implemented based on either AK or CCT, each of which entail its own rules for bandwidth choice. Figures \ref{fig:llr_transition_ak} and \ref{fig:llr_transition_cct} plots the same diagrams when bandwidths are chosen based on the recommendations of AK and CCT respectively. We see that both methods are more robust to spillovers than naive LLR in that the estimates are very close to $\tau_{\text{DIR}}$. Spillovers only distort these estimators when $r_n/h_n  \leq 0.2$. These plots do not give the full picture in that they omit MSE and coverage. We will return to these points below. Interestingly, distortion under AK seems to be increasing in the amount of spillovers. On the other hand, CCT seems to do worse under intermediate values of spillovers, with little distortion when spillovers are large.

The above simulations suggest that an alternative is needed to simple LLR under spillovers. AK and CCT appear more robust, but to our knowledge, this performance is not currently supported by theory, motivating consideration of LSR.

\begin{figure}[htbp]
\centering
\includegraphics[width=1\linewidth]{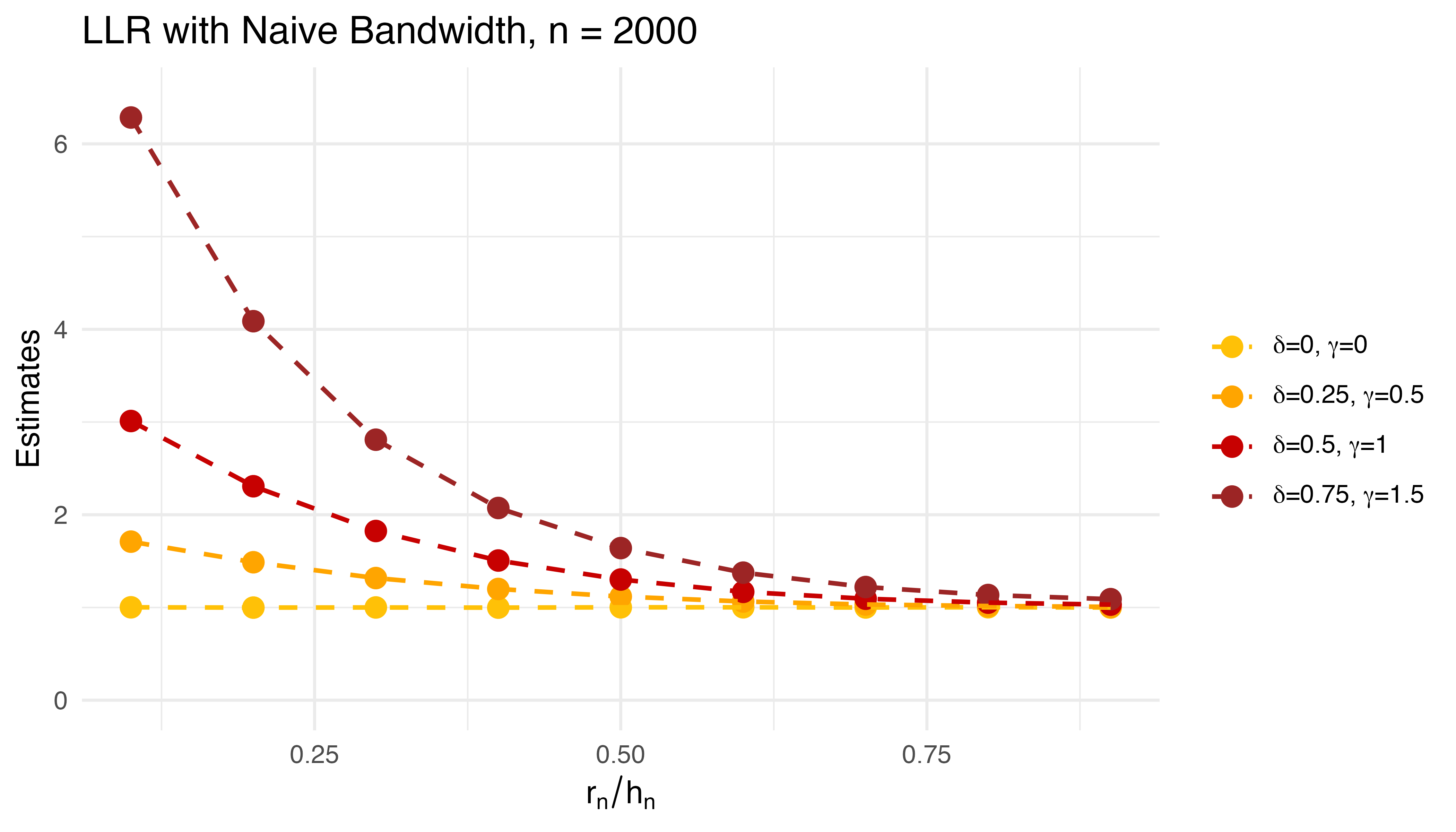}
\caption{Estimates of LLR when bandwidth is chosen based on the rule-of-thumb \eqref{equation--initial_h}. The behavior of $\hat{\tau}_\text{RDD}$ is best described by the $r_n \propto h_n$ case in Theorem \ref{theorem--consistency}.}
\label{fig:llr_transition_naive}
\end{figure}

\begin{figure}[htbp]
\includegraphics[width=1\linewidth]{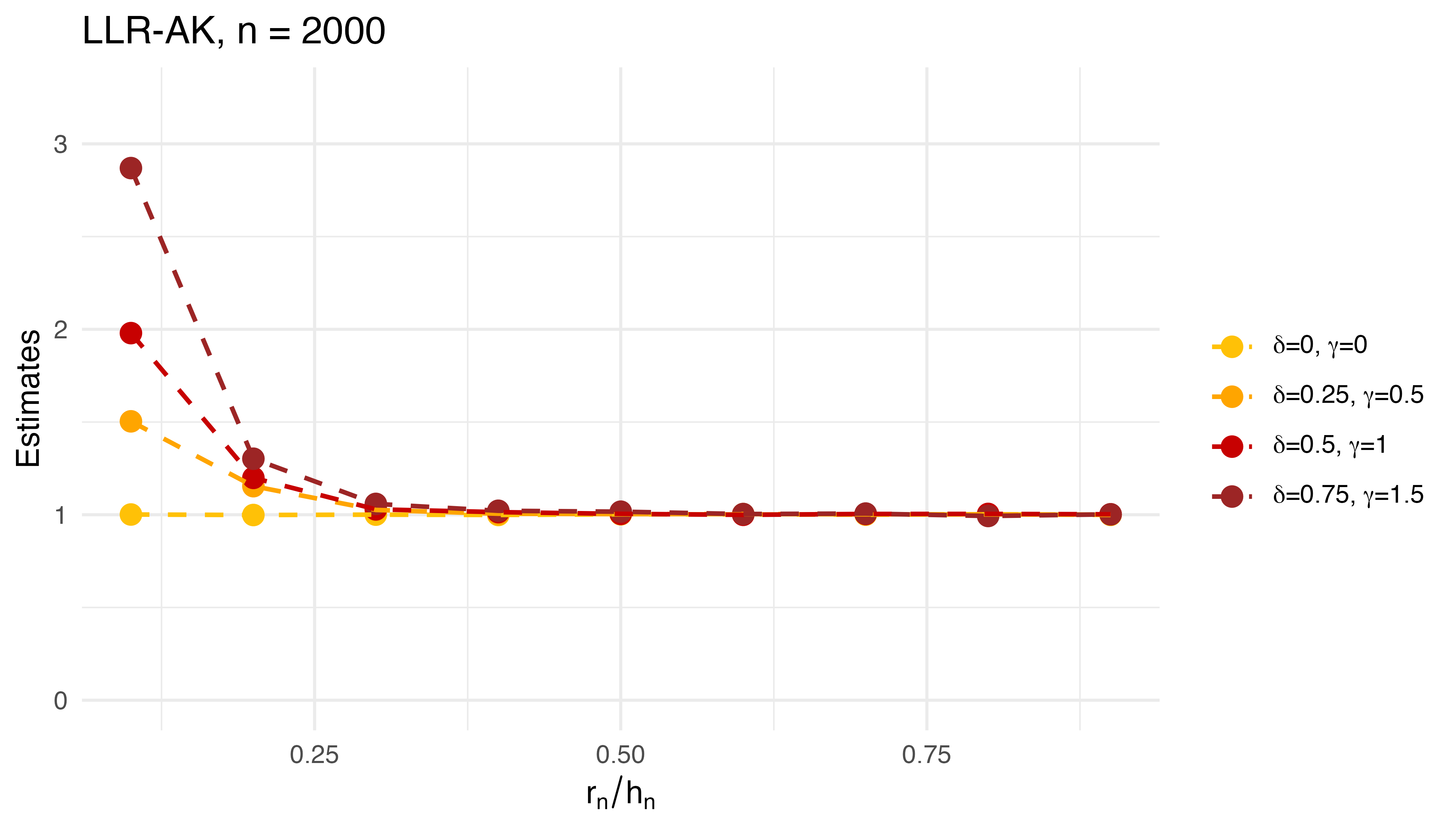}
\caption{Estimates of LLR when bandwidth is chosen based on \cite{armstrong2018optimal}, as implemented in the R package \texttt{RDHonest}.}
\label{fig:llr_transition_ak}
\vspace{3mm}
\includegraphics[width=1\linewidth]{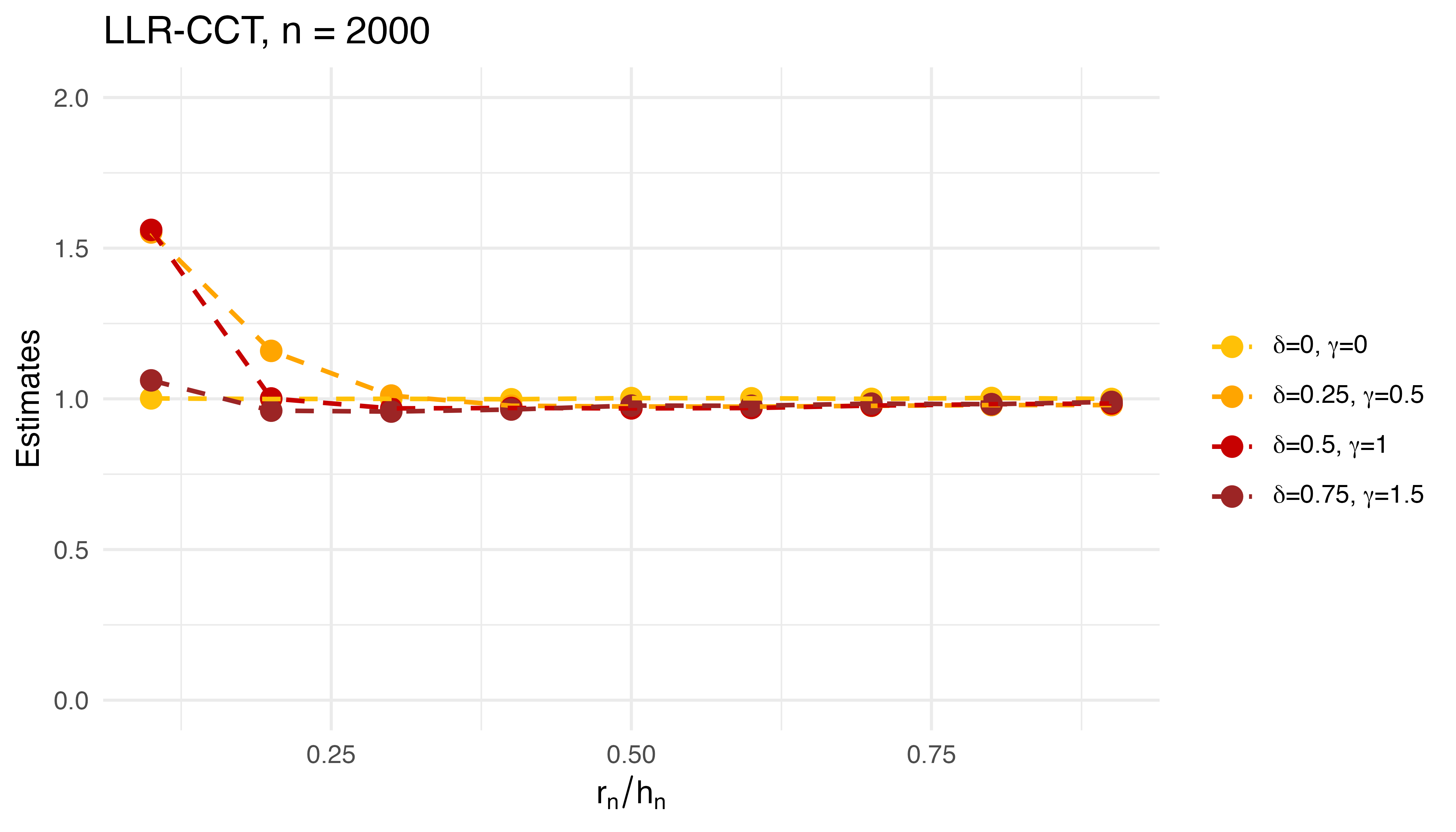}
\caption{Estimates of LLR when bandwidth is chosen based on \cite{calonico2014robust}, as implemented in the R package \texttt{rdrobust}.}
\label{fig:llr_transition_cct}
\end{figure}

\subsection{Local Spillover Regression}

This section investigates the performance of LSR for recovering the various parameters of interest. We first compare its MSE and coverage for $\tau_\text{DIR}$ to various implementations of LLR. We then consider estimation of $\delta$ and $\gamma$. The remainder of this section maintains $r_n = 0.05$ fixed, with $n \in\{500, 2000, 10000\}$. This sequence of models is better described by the asymptotic regime $r_n \gg h_n$, which favors LLR over LSR.

Table \ref{tab:sims_low_spill_tau} and \ref{tab:sims_high_spill_tau} present results on $\tau_\text{DIR}$. In the first panel in Table\ref{tab:sims_low_spill_tau}, $\delta = \gamma = 0$. LLR (with naive bandwidth given in \eqref{equation--initial_h}), AK and CCT all perform well. In this setting, $\tilde{\delta}(0)$ and $\tilde{\gamma}(0)$ are inconsistent. However, $\tilde{\tau}_\text{DIR}$ is consistent, although it has MSE that can be close to 5 times higher than methods designed for settings with no spillovers. Our method leads to conservative coverage even relative to AK. We conjecture that the conservatism can be reduced with data-driven choices of $M_m, M_\delta$ and $M_\gamma$, similar to the approach in AK.

Panel 2 in Table \ref{tab:sims_low_spill_tau} considers the case where $\delta = 0.25$ and $\gamma = 0.50$. As we explained at the start of this section, this sequence of models is not strictly comparable since $r_n$ is fixed as $n \to \infty$ and $h_n \to 0$. The asymptotics here favors LLR over LSR since $c = r_n/h_n \to \infty$ implies that $\hat{\tau}_\text{RDD} \overset{p}{\to} \tau_\text{DIR}$. As such, we see the largest gains of LSR over various implementations of LLR for smaller values of $n$. When $n = 500$, LSR has MSE that is 1/2 of AK and 1/3 of CCT. The two methods also have coverage of 60\% and 40\% respectively, whereas LSR has coverage that is consistently above 90\%. LSR-CV has relatively higher MSE and lower coverage but performance appears reasonable. We also see that it appears to deliver consistent estimates for $r_n$.

The picture remains similar when we consider Table \ref{tab:sims_high_spill_tau}. The top panel features intermediate levels of spillovers with $\delta = 0.5, \gamma = 1$ while the bottom panel features relatively high levels of spillovers with $\delta = 0.75, \gamma = 1.5$. Comparing across panels, we see that the MSE of AK increases as spillovers increase. However, CCT does the worst under small values of spillovers. Performance of LSR with known radius seems to suffer under high levels of spillovers, but LSR-CV surprisingly does better.

We next turn to results on $\tilde{\gamma}(0)$ and $\tilde{\delta}(0)$, presented in Table \ref{tab:sims_delta_gamma}. The top panel considers the case when $\delta = \gamma = 0$. Here, our estimators are not consistent  for $\gamma(0)$ and $\delta(0)$ and MSE indeed appears to diverge. However, it is encouraging that coverage is 1, even if our theory cannot speak to inference in this case. The remaining panels in Table \ref{tab:sims_delta_gamma} covers cases when $\delta$ and $\gamma$ are not $0$. We see that estimation of these parameters is more challenging that for $\tau_\text{DIR}$, although we achieve reasonable performance with sufficiently large sample size. The confidence intervals exhibit considerable over-coverage, possibly because of upward bias in our various plug-in estimators in addition to the built-in conservativeness of the bias-aware confidence intervals. Finally, we see that estimation of $\delta$ and $\gamma$ is much more sensitive to selection of the radius compared to $\tau_\text{DIR}$, although it can still be a compelling alternative when $r_n$ is unknown.

In sum, our simulations show that naive LLR -- and to a lesser extent AK and CCT -- exhibits the phase transition described in Theorem \ref{theorem--consistency}. LSR deliver more precise estimates of $\tau_\text{DIR}$, particularly for smaller sample sizes. It also yields estimates of $\delta(0)$ and $\gamma(0)$, which the other approaches cannot provide. These parameters are interesting in and of themselves, although their estimation requires more assumptions and larger sample sizes. Finally, cross-validation seems to be a reasonable approach to selecting radius, even though it is not without loss. Taken together, LSR and LSR-CV provides a useful set of tools for researchers who are concerned about spillovers in the RDDs .

\begin{table}[htbp]
  \centering
    \begin{tabular}{|c|c|c|c|c|c|c|c|}     \hline
          & $n$   & Method & Radius & Bandwidth & Eff. Obs. & MSE   & Coverage \bigstrut[b]\\
    \hline
    \multirow{15}[6]{*}{\begin{tabular}{@{}l@{}} $\delta = 0$, \\ $\gamma = 0$\end{tabular}} & \multirow{5}[2]{*}{500} & LLR   & -     & 0.289 & 72.1  & 0.005 & - \bigstrut[t]\\
          &       & AK    & -     & 0.175 & 43.8  & 0.011 & 0.959 \\
          &       & CCT   & -     & 0.303 & 75.7  & 0.009 & 0.932 \\
          &       & LSR   & 0.050 & 0.622 & 155.4 & 0.032 & 0.991 \\
          &       & LSR-CV & 0.105 & 0.645 & 161.2 & 0.032 & 0.996 \bigstrut[b]\\
\cline{2-8}          & \multirow{5}[2]{*}{2000} & LLR   & -     & 0.219 & 218.7 & 0.002 & - \bigstrut[t]\\
          &       & AK    & -     & 0.176 & 176.5 & 0.002 & 0.973 \\
          &       & CCT   & -     & 0.308 & 308.2 & 0.002 & 0.951 \\
          &       & LSR   & 0.050 & 0.489 & 489.2 & 0.008 & 0.997 \\
          &       & LSR-CV & 0.089 & 0.537 & 537.3 & 0.009 & 0.999 \bigstrut[b]\\
\cline{2-8}          & \multirow{5}[2]{*}{10000} & LLR   & -     & 0.158 & 792.4 & 0.000 & - \bigstrut[t]\\
          &       & AK    & -     & 0.176 & 878.3 & 0.000 & 0.971 \\
          &       & CCT   & -     & 0.309 & 1543.3 & 0.000 & 0.947 \\
          &       & LSR   & 0.050 & 0.330 & 1649.1 & 0.002 & 0.999 \\
          &       & LSR-CV & 0.092 & 0.386 & 1931.3 & 0.002 & 0.999 \bigstrut[b]\\
    \hline
    \multirow{15}[6]{*}{\begin{tabular}{@{}l@{}} $\delta = 0.25$, \\ $\gamma = 0.50$\end{tabular}} & \multirow{5}[2]{*}{500} & LLR   & -     & 0.289 & 72.1  & 0.305 & - \bigstrut[t]\\
          &       & AK    & -     & 0.122 & 30.4  & 0.062 & 0.628 \\
          &       & CCT   & -     & 0.164 & 41.0  & 0.098 & 0.383 \\
          &       & LSR   & 0.050 & 0.559 & 139.8 & 0.032 & 0.992 \\
          &       & LSR-CV & 0.079 & 0.579 & 144.7 & 0.061 & 0.975 \bigstrut[b]\\
\cline{2-8}          & \multirow{5}[2]{*}{2000} & LLR   & -     & 0.219 & 218.7 & 0.186 & - \bigstrut[t]\\
          &       & AK    & -     & 0.094 & 93.7  & 0.014 & 0.750 \\
          &       & CCT   & -     & 0.102 & 102.3 & 0.017 & 0.632 \\
          &       & LSR   & 0.050 & 0.396 & 396.5 & 0.006 & 0.985 \\
          &       & LSR-CV & 0.058 & 0.420 & 420.1 & 0.014 & 0.958 \bigstrut[b]\\
\cline{2-8}          & \multirow{5}[2]{*}{10000} & LLR   & -     & 0.158 & 792.4 & 0.087 & - \bigstrut[t]\\
          &       & AK    & -     & 0.070 & 348.7 & 0.002 & 0.922 \\
          &       & CCT   & -     & 0.060 & 301.9 & 0.002 & 0.937 \\
          &       & LSR   & 0.050 & 0.291 & 1457.4 & 0.001 & 0.982 \\
          &       & LSR-CV & 0.050 & 0.299 & 1494.5 & 0.002 & 0.947 \bigstrut[b]\\
    \hline
    \end{tabular}%
    \vspace{2mm}
  \caption{MSE and coverage for $\tau_\text{DIR}$ under zero to low spillovers. LLR uses the naive bandwidth \eqref{equation--initial_h} and we do not consider inference. AK is the implementation of \cite{armstrong2020simple}. CCT is the implementation of \cite{calonico2014robust}. LSR is local spillover regression when the true radius is known. LSR-CV selects the radius by cross-validation, as described in Section \ref{section--localspilloverreg--choice_r}. Both variants of LSR choose bandwidth using the procedure described in Section \ref{section--localspilloverreg--choice_h}.}
  \label{tab:sims_low_spill_tau}%
\end{table}%

\begin{table}[htbp]
  \centering

    \begin{tabular}{|c|c|c|c|c|c|c|c|}
    \hline
          & $n$   & Method & Radius & Bandwidth & Eff. Obs. & MSE   & Coverage \bigstrut\\
    \hline
    \multirow{15}[6]{*}{\begin{tabular}{@{}l@{}} $\delta = 0.50$, \\ $\gamma = 1.00$\end{tabular}} & \multirow{5}[2]{*}{500} & LLR   & -     & 0.289 & 72.1  & 2.281 & - \bigstrut[t]\\
          &       & AK    & -     & 0.094 & 23.6  & 0.107 & 0.659 \\
          &       & CCT   & -     & 0.083 & 20.6  & 0.064 & 0.840 \\
          &       & LSR   & 0.050 & 0.501 & 125.2 & 0.034 & 0.994 \\
          &       & LSR-CV & 0.057 & 0.499 & 124.7 & 0.061 & 0.966 \bigstrut[b]\\
\cline{2-8}          & \multirow{5}[2]{*}{2000} & LLR   & -     & 0.219 & 218.7 & 1.328 & - \bigstrut[t]\\
          &       & AK    & -     & 0.074 & 74.0  & 0.018 & 0.844 \\
          &       & CCT   & -     & 0.049 & 49.5  & 0.009 & 0.933 \\
          &       & LSR   & 0.050 & 0.418 & 417.6 & 0.008 & 0.999 \\
          &       & LSR-CV & 0.050 & 0.421 & 421.2 & 0.013 & 0.942 \bigstrut[b]\\
\cline{2-8}          & \multirow{5}[2]{*}{10000} & LLR   & -     & 0.158 & 792.4 & 0.583 & - \bigstrut[t]\\
          &       & AK    & -     & 0.053 & 266.9 & 0.002 & 0.958 \\
          &       & CCT   & -     & 0.031 & 155.8 & 0.003 & 0.925 \\
          &       & LSR   & 0.050 & 0.384 & 1918.0 & 0.002 & 0.999 \\
          &       & LSR-CV & 0.049 & 0.386 & 1929.1 & 0.002 & 0.996 \bigstrut[b]\\
    \hline
    \multirow{15}[6]{*}{\begin{tabular}{@{}l@{}} $\delta = 0.75$, \\ $\gamma = 1.50$\end{tabular}} & \multirow{5}[2]{*}{500} & LLR   & -     & 0.289 & 72.1  & 13.106 & - \bigstrut[t]\\
          &       & AK    & -     & 0.088 & 22.0  & 0.270 & 0.746 \\
          &       & CCT   & -     & 0.051 & 12.7  & 0.068 & 0.992 \\
          &       & LSR   & 0.050 & 0.653 & 163.4 & 0.273 & 1.000 \\
          &       & LSR-CV & 0.056 & 0.608 & 152.0 & 0.166 & 0.990 \bigstrut[b]\\
\cline{2-8}          & \multirow{5}[2]{*}{2000} & LLR   & -     & 0.219 & 218.7 & 7.059 & - \bigstrut[t]\\
          &       & AK    & -     & 0.066 & 65.7  & 0.035 & 0.946 \\
          &       & CCT   & -     & 0.029 & 29.2  & 0.017 & 0.931 \\
          &       & LSR   & 0.050 & 0.567 & 566.9 & 0.039 & 1.000 \\
          &       & LSR-CV & 0.049 & 0.563 & 563.4 & 0.036 & 1.000 \bigstrut[b]\\
\cline{2-8}          & \multirow{5}[2]{*}{10000} & LLR   & -     & 0.158 & 792.4 & 2.787 & - \bigstrut[t]\\
          &       & AK    & -     & 0.042 & 211.8 & 0.004 & 0.975 \\
          &       & CCT   & -     & 0.020 & 99.9  & 0.004 & 0.932 \\
          &       & LSR   & 0.050 & 0.422 & 2107.9 & 0.007 & 1.000 \\
          &       & LSR-CV & 0.049 & 0.424 & 2118.6 & 0.007 & 1.000 \bigstrut[b]\\
    \hline
    \end{tabular}%
      \vspace{2mm}
  \caption{MSE and coverage for $\tau_\text{DIR}$ under moderate to high spillovers. LLR uses the naive bandwidth \eqref{equation--initial_h} and we do not consider inference. AK is the implementation of \cite{armstrong2020simple}. CCT is the implementation of \cite{calonico2014robust}. LSR is local spillover regression when the true radius is known. LSR-CV selects the radius by cross-validation, as described in Section \ref{section--localspilloverreg--choice_r}. Both variants of LSR choose bandwidth using the procedure described in Section \ref{section--localspilloverreg--choice_h}.}
  \label{tab:sims_high_spill_tau}%
\end{table}%

\begin{table}[htbp]
  \centering \small
    \begin{tabular}{|c|c|c|c|c|c|c|c|c|c|}
    \hline
    \multirow{2}[4]{*}{} & \multirow{2}[4]{*}{$n$} & \multirow{2}[4]{*}{Method} & \multirow{2}[4]{*}{Radius} & \multirow{2}[4]{*}{Bdwth} & \multirow{2}[4]{*}{Eff. Obs.} & \multicolumn{2}{c|}{$\delta(0)$} & \multicolumn{2}{c|}{$\gamma(0)$} \bigstrut\\
\cline{7-10}          &       &       &       &       &       & MSE   & Cov.  & MSE   & Cov. \bigstrut\\
    \hline
    \multirow{6}[6]{*}{\begin{tabular}{@{}l@{}} $\delta = 0$, \\ $\gamma = 0$\end{tabular}} & \multirow{2}[2]{*}{500} & LSR   & 0.050 & 0.622 & 155.4 & 0.837 & 1.000 & 0.945 & 1.000 \bigstrut[t]\\
          &       & LSR-CV & 0.105 & 0.645 & 161.2 & 5.583 & 1.000 & 5.738 & 1.000 \bigstrut[b]\\
\cline{2-10}          & \multirow{2}[2]{*}{2000} & LSR   & 0.050 & 0.489 & 489.2 & 1.485 & 1.000 & 1.536 & 1.000 \bigstrut[t]\\
          &       & LSR-CV & 0.089 & 0.537 & 537.3 & 6.587 & 1.000 & 6.650 & 1.000 \bigstrut[b]\\
\cline{2-10}          & \multirow{2}[2]{*}{10000} & LSR   & 0.050 & 0.330 & 1649.1 & 2.865 & 1.000 & 2.887 & 1.000 \bigstrut[t]\\
          &       & LSR-CV & 0.092 & 0.386 & 1931.3 & 9.726 & 1.000 & 9.768 & 1.000 \bigstrut[b]\\
    \hline
    \multirow{6}[6]{*}{\begin{tabular}{@{}l@{}} $\delta = 0.25$, \\ $\gamma = 0.50$\end{tabular}} & \multirow{2}[2]{*}{500} & LSR   & 0.050 & 0.559 & 139.8 & 0.455 & 1.000 & 1.732 & 1.000 \bigstrut[t]\\
          &       & LSR-CV & 0.079 & 0.579 & 144.7 & 1.567 & 1.000 & 5.623 & 1.000 \bigstrut[b]\\
\cline{2-10}          & \multirow{2}[2]{*}{2000} & LSR   & 0.050 & 0.396 & 396.5 & 0.200 & 1.000 & 0.720 & 1.000 \bigstrut[t]\\
          &       & LSR-CV & 0.058 & 0.420 & 420.1 & 0.644 & 1.000 & 2.258 & 1.000 \bigstrut[b]\\
\cline{2-10}          & \multirow{2}[2]{*}{10000} & LSR   & 0.050 & 0.291 & 1457.4 & 0.040 & 1.000 & 0.134 & 1.000 \bigstrut[t]\\
          &       & LSR-CV & 0.050 & 0.299 & 1494.5 & 0.120 & 1.000 & 0.419 & 1.000 \bigstrut[b]\\
    \hline
    \multirow{6}[6]{*}{\begin{tabular}{@{}l@{}} $\delta = 0.50$, \\ $\gamma = 1.00$\end{tabular}} & \multirow{2}[2]{*}{500} & LSR   & 0.050 & 0.501 & 125.2 & 0.095 & 0.982 & 1.309 & 0.973 \bigstrut[t]\\
          &       & LSR-CV & 0.057 & 0.499 & 124.7 & 0.257 & 0.972 & 3.333 & 0.959 \bigstrut[b]\\
\cline{2-10}          & \multirow{2}[2]{*}{2000} & LSR   & 0.050 & 0.418 & 417.6 & 0.020 & 0.998 & 0.255 & 0.996 \bigstrut[t]\\
          &       & LSR-CV & 0.050 & 0.421 & 421.2 & 0.051 & 0.992 & 0.642 & 0.985 \bigstrut[b]\\
\cline{2-10}          & \multirow{2}[2]{*}{10000} & LSR   & 0.050 & 0.384 & 1918.0 & 0.003 & 0.997 & 0.037 & 0.997 \bigstrut[t]\\
          &       & LSR-CV & 0.049 & 0.386 & 1929.1 & 0.010 & 0.997 & 0.131 & 0.997 \bigstrut[b]\\
    \hline
    \multirow{6}[6]{*}{\begin{tabular}{@{}l@{}} $\delta = 0.75$, \\ $\gamma = 1.50$\end{tabular}} & \multirow{2}[2]{*}{500} & LSR   & 0.050 & 0.653 & 163.4 & 0.015 & 1.000 & 1.179 & 1.000 \bigstrut[t]\\
          &       & LSR-CV & 0.056 & 0.608 & 152.0 & 0.051 & 0.987 & 3.488 & 0.985 \bigstrut[b]\\
\cline{2-10}          & \multirow{2}[2]{*}{2000} & LSR   & 0.050 & 0.567 & 566.9 & 0.003 & 1.000 & 0.194 & 1.000 \bigstrut[t]\\
          &       & LSR-CV & 0.049 & 0.563 & 563.4 & 0.009 & 1.000 & 0.746 & 1.000 \bigstrut[b]\\
\cline{2-10}          & \multirow{2}[2]{*}{10000} & LSR   & 0.050 & 0.422 & 2107.9 & 0.001 & 1.000 & 0.046 & 1.000 \bigstrut[t]\\
          &       & LSR-CV & 0.049 & 0.424 & 2118.6 & 0.002 & 1.000 & 0.195 & 1.000 \bigstrut[b]\\
    \hline
    \end{tabular}%
    \vspace{2mm}
     \caption{MSE and coverage for $\delta(0)$ and $\gamma(0)$. LSR is local spillover regression when the true radius is known. LSR-CV selects the radius by cross-validation, as described in Section \ref{section--localspilloverreg--choice_r}. Both variants of LSR choose bandwidth using the procedure described in Section \ref{section--localspilloverreg--choice_h}.}
  \label{tab:sims_delta_gamma}%
\end{table}%

\section{Auxillary Lemmas}\label{section--auxillarylemmas}

This section states lemmas which are used in Appendix \ref{section--proofs}. Proofs of these lemmas can be found in \cite{auerbach2025regression}. We sometimes suppress the dependence of variables (e.g. $r_n, h_n$) on $n$ to reduce notational clutter.

\subsection{Preliminaries}

The following definitions and results will be useful. In particular, Lemma \ref{lemma--uniform_approx} is the technical core of the results in Sections \ref{section--estimands} and \ref{section--localspilloverreg}.

\begin{definition}
	Let the spillover operator $G_n$ be defined as
	\begin{equation*}
		(G_n \circ f)(z) = \frac{ \delta(z)}{|R_n(z)|} \int \mathbf{1}\left\{|u-z|<r_n\right\}f(z) \frac{1}{2}dz
	\end{equation*}
	where $|R_n(z)| = \int \mathbf{1}\left\{|u-z|<r_n\right\} \frac{1}{2}dz$.
\end{definition}

\begin{definition}
	For a function $g: [-1,1] \to \mathbf{R}$, let its sup-norm be denoted $\lVert g \rVert := \sup_{z \in [-1,1]} \left\lvert g(z) \right\rvert$.
\end{definition}

Let the ``structural" and ``reduced form" functions be, respectively,
\begin{definition}
\begin{align*}
	m(z) &:= \begin{cases}
		m^+(z) & \mbox{ if } z \geq 0 \\
		m^-(z) & \mbox{ if } z < 0
	\end{cases}~, \\
g_n(z) &:= \left(\left(I - G_n\right)^{-1}\circ m\right)(z)~.
\end{align*}
\end{definition}
It will also be useful to split $m(\cdot)$ into the part that is Lipschitz continuous and the part that is not:
\begin{definition}\label{definition--m_cont} Let $m(x) = {m}^c(x) + \mathbf{1}\{z \geq 0 \} \tau_d$ where
\begin{align*}
{m}^c(z) = \begin{cases}
	m^+(z) - \tau_d  & \mbox{ if } z \geq 0 \\
	m^-(z) & \mbox{ otherwise}
\end{cases}
\end{align*}
is Lipschitz continuous with Lipschitz constant $C$.
\end{definition}
\begin{definition} Let the difference in the spillover neighborhoods at $xh_n$ and $0$ be:
	\begin{align*}
		{R}^+(xh_n) &:= R(xh_n) \setminus R(0) \\
		{R}^-(xh_n) &:= R(0) \setminus R(xh_n)
	\end{align*}
\end{definition}

\begin{definition}\label{definition--tilde_d}
	Let $\tilde{d}$ be
	\begin{align*}
		\tilde{d}(a) = \begin{cases}
			0 & \mbox{ if } a \leq -1 \\
			\frac{a+1}{2} & \mbox{ if } -1 \leq a \geq 1 \\
			1 & \mbox{ if } a \geq -1 \\
		\end{cases}
	\end{align*}
\end{definition}
Observe that if $\delta(z) = 1$, $\tilde{d}(a/r_n) = \left(G_n \circ \mathbf{1}\{z \geq 0\}\right)(a)$.

\begin{definition}\label{definition--g_infty}
	Let $G_\infty$ be the pointwise limit of $G_n$ when $r_n \to 0$. That is,
	\begin{align*}
		\left(G_\infty \circ g\right)(z) = \lim_{n\to\infty} \frac{\delta(z)}{R_n(z)} \int_{R(z)} g(y) dy = \delta(z)g(z)~,
	\end{align*}
	where the last equality follows from the Lebesgue Differentiation Theorem. Furthermore, let $g_\infty(z) = \left(\left(I - G_\infty\right)^{-1} \circ g \right)(z) = \frac{g(z)}{1-\delta(z)}$.
\end{definition}

\begin{lemma}\label{lemma--shrinkingnorm}
	Suppose $\Vert f - g \rVert \leq \varepsilon$. Then $\left\lVert \left(G_n \circ f\right) - \left(G_n \circ g\right) \right\rVert \leq \bar{\delta} \varepsilon~.$ In particular, if $\lVert g(z) \rVert <\varepsilon$, then $\left\lVert G_n^k \circ g \right\rVert < \bar{\delta}^k \varepsilon$.
\end{lemma}

\begin{lemma}\label{lemma--discont_limt}
Under Assumptions \ref{assump-model} and \ref{assump--neighborhood}, suppose $r_n \to 0$. Then, there exists a function $\lambda$ such that for $a \neq 0$,
\begin{align*}
		\lvert ((I-G_n)^{-1} \circ \mathbf{1}\{z \geq 0\})(ar_n) -\lambda(a)\rvert \leq \frac{2C_\delta r_n}{a\left(1-\bar{\delta} \right)^2} + \frac{\bar{\delta}^{1/r_n-|a|}}{1-\bar{\delta}}~,
\end{align*}
where $ \lambda(a) > 0$ if $\delta(z) > 0$ for all $z \in \mathcal{Z}$. Furthermore, if $\delta(0) \neq 0$, then
\begin{multline*}
		\left\lvert \left((I-G_n)^{-1} \circ \gamma(z)\tilde{d}(z/r_n)\right)(ar_n) - \frac{\gamma(0)}{\delta(0)}(\lambda(a) - \mathbf{1}\{a \geq 0\})\right\rvert \leq \\
		\frac{2C_\delta r_n}{a\left(1-\bar{\delta} \right)^2} + \frac{\bar{\delta}^{1/r_n-|a|}}{1-\bar{\delta}}
		+ \frac{C_\gamma \bar{\delta}}{1-\bar{\delta}}\cdot ar_n~.
\end{multline*}
If $\delta(0) = 0$, then
\begin{equation*}
	\left((I-G_n)^{-1} \circ \gamma(z)\tilde{d}(z/r_n)\right)(ar_n) = \gamma(ar_n)\tilde{d}(a) = \gamma(0) \tilde{d}(a) + O(r_n)~.
\end{equation*}
\end{lemma}

In fact, we know more about the structure of $\lambda(a)$. Let $G_*$ be operator that smooths over $f \in \mathbf{R} \to \mathbf{R}$:
\begin{equation*}
	(G_* \circ f)(z) = \frac{1}{2} \int \mathbf{1}\left\{|u-z|<1\right\}f(z) \frac{1}{2}dz
\end{equation*}
Then $\lambda(a)$ is approximately $((I-\delta(0)G_*)^{-1} \circ \mathbf{1}\{z \geq 0\})(a)$.
Note that $\lambda$ depends on $\delta$, though this is suppressed in the notation. Given Lemma \ref{lemma--discont_limt}, define:
\begin{definition}\label{definition--integrals}
	Let $\tilde{R}^+(x) = \Big[\max\left\{1, \frac{2x}{c}-1\right\} \, , \, 1+\frac{2x}{c}\Big]$ and $\tilde{R}^-(x) = \Big[-1\, , \, \min\left\{1, \frac{2x}{c}-1\right\} \Big]$. Define:
	\begin{align*}
		\lambda^+(x) :=  \frac{1}{|\tilde{R}^+(x)|} \int_{\tilde{R}^+(x)} \lambda(a) \, dF(a) \quad &, \quad \lambda^-(x) :=  \frac{1}{|\tilde{R}^-(x)|} \int_{\tilde{R}^-(x)} \lambda(a) \, dF(a) \\
		\tilde{\lambda}^+(x) :=  \frac{1}{|\tilde{R}^+(x)|} \int_{\tilde{R}^+(x)} \lambda(a) - \mathbf{1}\{a  \geq 0\} \, dF(a) \quad &, \quad \tilde{\lambda}^-(x) :=  \frac{1}{|\tilde{R}^-(x)|} \int_{\tilde{R}^-(x)} \lambda(a)- \mathbf{1}\{a  \geq 0\}\, dF(a)~.
	\end{align*}
And:
\begin{align*}
	\Lambda^+_{p,q,s} := \int^1_0 x^p \left(\lambda^+(x)- \lambda^-(x)\right)^q K^s(x) ds \quad &, \quad \Lambda^-_{p,q,s} := \int^0_{-1}  x^p \left(\lambda^+(-x) - \lambda^-(-x)\right)^q K^s(x) ds \\
	\tilde{\Lambda}^+_{p,q,s} := \int^1_0 x^p \left(\tilde{\lambda}^+(x)-\lambda^-(x)\right)^q K^s(x) ds \quad &, \quad \tilde{\Lambda}^-_{p,q,s} := \int^0_{-1}  x^p \left(\tilde{\lambda}^+(-x) - \lambda^-(-x)\right)^q K^s(x) ds \\
	\Gamma^+_{p,q,s} := \int^1_0 x^p \left(\tilde{d}\left(\frac{x}{c}\right)\right)^q K^s(x) ds \quad &, \quad \Gamma^-_{p,q,s} := \int^0_{-1}  x^p \left(\tilde{d}\left(\frac{x}{c}\right)\right)^q K^s(x) ds ~.
\end{align*}
\end{definition}

\begin{lemma}\label{lemma--preserveLipschitz}
	Suppose $g$ is Lipschitz on $[-1,1]$ with Lipschitz constant $C$ and $  \left\lVert g \right\rVert \leq M~.$ Then $\left(G_n \circ g\right)$ is Lipschitz on $[-1,1]$ with Lipschitz constant $C_\delta \cdot M + \bar{\delta}C$. Moreover, $\left(I-G_n\right)^{-1} \circ g$ is Lipschitz on $[-1,1]$ with Lipschitz constant $$\frac{C_\delta \cdot M }{\left(1-\bar{\delta}\right)^2} + \frac{C}{1-\bar{\delta}}~.$$
\end{lemma}

\begin{lemma}\label{lemma--preserveOdd}
Suppose $g$ is an odd function. If $\delta(z)$ is an even function, then $\left(G_n \circ g\right)$ is an odd function.
\end{lemma}

\begin{lemma}\label{lemma--g_nLipschitz}
	$g_n(z) =  g_n^c(z) + \tau_d \mathbf{1}\{z \geq 0\}$ where $g_n^c(z)$ is Lipschitz with Lipschitz constant
	\begin{equation*}
	\frac{C_\delta \cdot (1+\bar{\delta})M }{\left(1-\bar{\delta}\right)^2} + \frac{C}{1-\bar{\delta}} + \frac{C\bar{\delta}}{2r_n\left(1-\bar{\delta}\right)}
	\end{equation*}
\end{lemma}

\begin{lemma}\label{lemma--Lipschitz_unif_conv}
	Let $g$ be a Lipschitz continuous function with Lipschitz constant $C$ and suppose $r_n \to 0$. Then:
	\begin{equation*}
		\sup_{z \in [-1/2,1/2]} \left\lvert \left(\left(I - G_n \right)^{-1} \circ g\right)(z) - \left(\left(I - G_\infty \right)^{-1} \circ g\right)(z) \right\rvert \leq \frac{Cr_n}{2}	\frac{\bar{\delta}}{\left(1-\bar{\delta}\right)^2}~.
	\end{equation*}
\end{lemma}

\begin{lemma}\label{lemma--delta0approx}
	Let $h_n$ be such that $h_n > Kr_n$.  Let $\tilde{G}_n$ be $G_n$ except with $\delta(z)$ set to $\delta(0)$. Then for all $x \in [-h_n, h_n]$, we have that
	\begin{equation*}
		\left(G^k_n\circ \mathbf{1}\left\{z \geq 0\right\}\right)(x) = 	\left(\tilde{G}^k_n\circ \mathbf{1}\left\{z \geq 0\right\}\right)(x) + \sum_{j=1}^{k} \left(G_n^{k-j} \circ R_{j} \right)(x)
	\end{equation*}
	where
	\begin{equation*}
		\sup_{x \in [-h_n, \, h_n]} \left\lvert \left(G_n^{k-j} \circ R_{j} \right)(x) \right\rvert \leq 2 \bar{\delta}^k C_\delta h_n~.
	\end{equation*}
\end{lemma}


\begin{lemma}\label{lemma-mu0_approx}
	Under Assumptions \ref{assump-model} and \ref{assump--neighborhood}, suppose $r_n \to0$. The following holds for all $\eta \in (0,1)$:
	\begin{equation*}
		\mu_d(0) = \frac{m^+(0)+m^-(0) + \gamma(0)}{2(1-\delta(0))} + O\left(\frac{C_\delta r_n^\eta}{\left(1-\bar{\delta}\right)^2} + \frac{\bar{\delta}^{1/r_n^\eta}}{1-\bar{\delta}}\right)~.
	\end{equation*}
\end{lemma}

\begin{lemma}\label{lemma--uniform_approx}
	Under Assumptions \ref{assump-model} and \ref{assump--neighborhood}, the following approximations are uniform in $x \in [0,1]$ as $n \to \infty$: {\small
	\begin{align*}
		\mu_d(xh_n) - \mu_d(0) = \begin{cases}
					\ \frac{xh_n}{r_n}\left(g_1(r) - g_1(-r)+ l_1(r) - l_1(-r) + o(1)   \right) & \mbox{ if } r_n = r \\
			\frac{xh_n}{r_n}\cdot \left(\tau_d\cdot \left(  \lambda(1) - \lambda(-1)  \right)  + \gamma(0)\cdot \left(  \left(\lambda(1) - 1\right) - \lambda(-1)  \right) + o(1) \right)& \mbox{ if } \frac{h_n}{r_n} \to 0 \\
			 \min\left\{1, \frac{|x|}{c} \right\} \cdot  \left(\tau_d\cdot \left(  \lambda^+(x) - \lambda^-(x)  \right)  + \gamma(0)\cdot \left(  \tilde{\lambda}^+(x) - \tilde{\lambda}^-(x)  \right) + o(1) \right)& \mbox{ if } r_n = \frac{1}{2}ch_n
		\end{cases}~.
	\end{align*}}
	If $r_n/h_n \to 0$, let $w_n$ be such that $w_n/r_n \to \infty$ and $w_n/h_n \to 0$. Then for $x \in[w_n, 1]$,
	\begin{equation*}
		\mu_d(xh_n) - \mu_d(0) =  \frac{\tau_d + \gamma(0)}{2\left(1-\delta(0)\right)} + O \left(h_n + \bar{\delta}^{w_n/r_n}\right)~.
	\end{equation*}
	The same statements hold for $x\in [-1,0]$ mutatis mutandis.
\end{lemma}

Finally, we define the (mixed) incomplete moments:
\begin{definition}[Incomplete Moments] The incomplete moment of order $p$ at $0$ is
\begin{equation*}
	\gamma_{p,s} := \int_{0}^1 x^p K^s(x) dx 
\end{equation*}
and the incomplete mixed moment of order $p$ at  $0$ is
\begin{multline*}
	\phi(p,q,r,s):= \int_{0}^1 x^{p} \left(\min\left\{ \frac{x}{c} , 1 \right\}  \left(\tau_d\cdot \left(  \lambda^+(x) - \lambda^-(x)  \right)  + \gamma(0)\cdot \left(  \left(\lambda^+(x) - 1\right) - \lambda^-(x)  \right) \right)\right)^q \\
	\cdot  \left( \min\left\{ \frac{x}{c} , \frac{1}{2} \right\} \right)^r K^s(x) \, dx 
\end{multline*}
\end{definition}
Incomplete moments are standard objects in the analysis of local polynomial regressions at the boundary (see e.g. \cite{wand1995kernel}).The incomplete mixed moment is so named since also includes terms that arise from the approximations of $\mu_d(Z_i)$ and $\nu_d(Z_i)$ around the cutoff. We suppress the dependence of $\phi$ on $c$ for convenience.

\begin{lemma}
\label{lem--spilloverreg-inference-std-gauss-sufficient-an}
Under the conditions of
Theorem \ref{thm--spilloverreg-inference}, \(\bm{\Omega}^{- 1 / 2} \sqrt{n}
\widetilde{\mathbf{T}}_{\ast} \overset{\mathrm{d}}{\to} \N \left( 0,
\mathbb{I}_{8} \right)\), where \(\widetilde{\mathbf{T}}_{\ast}\) is as defined
in \eqref{eqn--Ttilde-Ttilde-ast-def}.
\end{lemma}

\begin{lemma}
\label{lem--spilloverreg-inference-std-gauss-sufficient-op1}
Let the premises of Theorem \ref{thm--spilloverreg-inference} hold and
\(\widetilde{\mathbf{T}}\) and \(\widetilde{\mathbf{T}}_{\ast}\) be as defined
in \eqref{eqn--Ttilde-Ttilde-ast-def}.
Then \(\bm{\Omega}^{- 1 / 2} \sqrt{n} \left( \widetilde{\mathbf{T}} -
\widetilde{\mathbf{T}}_{\ast} \right) = o_{\mathrm{p}} (1)\).
\end{lemma}

\begin{lemma}
\label{lem--Tg-linear-rep}
\(T_{g}\) in \eqref{eqn--Tg-def} has the following representation:
\begin{equation}
  \begin{split}
    T_{g} =
    & \, T_{\ast, g} + U_{g} + R_{g}, \\
    \text{where} \quad T_{\ast, g} =
    & \, \frac{1}{n} \sum_{i = 1}^{n} \dot{\eta}_{g, 1} \left( \xi_{i}
    \right), \\
    U_{g} =
    & \, \frac{1}{n (n - 1)} \sum_{i = 1}^{n - 1} \sum_{j = i + 1}^{n}
    \dot{\eta}_{g, 2} \left( \xi_{i}, \xi_{j} \right) - T_{\ast, g}, \\
    R_{g} =
    & \, \frac{1}{n} \sum_{i = 1}^{n} K \left( Z_{i} / h \right) g \left(
    \xi_{i} \right) \left( \widehat{\mathrm{Rem}}_{i, r} \left( Z_{i} \right) -
    \widehat{\mathrm{Rem}}_{i, r} (0) \right).
  \end{split}
  \label{eqn--Tg-linear-rep}
\end{equation}
\end{lemma}

\begin{lemma}
\label{lem--Tg-linear-rep-Tast-rate-normality}
Let \(T_{\ast, g}, U_{g}, R_{g}\) be as in \eqref{eqn--Tg-linear-rep}.
Then \(\E \left[ T_{\ast, g} \right] = 0\), \(\E \left[ U_{g} \right] = 0\).
Furthermore,
\begin{equation}
  \begin{split}
    \E \left[ T_{\ast, g}^{2} \right] =
    & \, \frac{\E \left[ \dot{\eta}_{g 1} \left( \xi_{1} \right)^{2} \right]}{n}
    \\
    \E \left[ \dot{\eta}_{g 1} \left( \xi_{1} \right)^{2} \right]
    =
    & \, h \frac{h}{4 r} (1 + o (1)) \left( \int f (v r) \sigma^{2} (v r)
    \varphi_{\alpha, 1} (v; h, r)^{2} \mathrm{d} v + \int \varphi_{\alpha, 2}
    (v; h, r)^{2} f (v r) \; \mathrm{d} v \right)
    \\
    \E \left[ \left| \dot{\eta}_{g 1} \left( \xi_{1} \right) \right|^{3} \right]
    \leq
    & \,
    16 h \frac{h^{2}}{8 r^{2}} \left\{ \int f (v r) \rho_{3} (v r) \left|
    \varphi_{\alpha, 1} (v; h, r) \right|^{3} \mathrm{d} v + \int \left|
    \varphi_{\alpha, 2} (v; h, r) \right|^{3} f (v r) \; \mathrm{d} v \right\}
    \\
    & + o \left( h \frac{h^{2}}{r^{2}}  \right) + O \left( h^{3} \right),
  \end{split}
  \label{eqn--Tg-linear-rep-Tg-ast-rate}
\end{equation}
for \(\rho_{3} (z) \equiv \E \left[ \varepsilon_{1} \middle| Z_{1} = z \right]\)
and measurable bounded functions \(\varphi_{\alpha, j}\).
\end{lemma}

\begin{lemma}
\label{lem--Tg-linear-rep-rate-UR}
Let \(T_{\ast, g}, U_{g}, R_{g}\) be as in \eqref{eqn--Tg-linear-rep}.
For a bounded measurable function
\(C_{v} (\cdot)\) and \(m_{g^{2}} (z) \equiv \E \left[ g \left( \xi_{1}
\right)^{2} \middle| Z_{1} = z \right]\).
\begin{equation}
  \begin{split}
    \E \left[ U_{g}^{2} \right] =
    & \, \frac{2}{n (n - 1)} \mathrm{E} \left[ \mathrm{Var} \left[
    \dot{\eta}_{g, 2} \left( \xi_{1}, \xi_{2} \right) \middle| \xi_{1} \right]
    \right] - \mathrm{Var} \left[ \mathrm{E} \left[ \dot{\eta}_{g, 2} \left(
    \xi_{1}, \xi_{2} \right) \middle| \xi_{1} \right] \right] \\
    \leq
    & \, \frac{8}{n (n - 1)} \cdot \frac{h}{r} \int K (u)^{2} f (u h) m_{g^{2}}
    (u h) C_{v} (u; h, r) \; \mathrm{d} u, \\
    \text{so that } U_{g} =
    & \, o_{\mathrm{p}} \left( T_{\ast, g} \right) \quad
    \text{ if } n h \to \infty.
  \end{split}
  \label{eqn--Tg-linear-rep-Ug-rate}
\end{equation}
Furthermore, if \(n h \to \infty\) and \((\log n) / (n r) \to 0\), then
\begin{equation}
  R_{g} = O_{\mathrm{p}} \left( \sqrt{\E \left[ U_{g}^{2} \right]} \right).
  \label{eqn--Tg-linear-rep-Rg-rate}
\end{equation}
\end{lemma}

\clearpage
\newpage

\section{Proofs of Auxiliary Lemmas}

\subsection{Proof of Lemma \ref{lemma--shrinkingnorm}}

For any $z \in [-1,1]$,
\begin{equation*}
	 \left\lvert \left(G_n \circ f\right)(z) - \left(G_n \circ g\right)(z)\right\rvert \leq \frac{\left\lvert \delta(z)\right\rvert}{|R(z)|} \int_{R(z)} \left\lvert f(x) - g(x) \right\rvert dz \leq \bar{\delta}\varepsilon~.
\end{equation*}
Letting $g = \mathbf{0}$ and the bound for $G_n^k$ follows by induction. \hfill\qed

\subsection{Proof of Lemma \ref{lemma--discont_limt}}

We will prove that the sequence $\left(\left(I - G_n\right)^{-1} \circ \mathbf{1}\left\{ar_n\right\}\right)_{n=1}^\infty$ is Cauchy. Let $m,n$ be given, where $m>n$. We consider the Neumann series:
Writing $\left(I - G_n\right)^{-1}$ in terms of its Neumann series:
\begin{align*}
	\left(I - G_n\right)^{-1} = I + G_n + G^2_n + \cdots
\end{align*}
First observe that for all $a$ such that $|a| < 1/r_n$, we have that
\begin{equation*}
	\mathbf{1}\left\{z \geq 0 \right\}(ar_n) = 	\mathbf{1}\left\{z \geq  0\right\}(ar_m)~.
\end{equation*}

Next, suppose $\delta = 1$ is a constant function. Then,
\begin{equation*}
\left(G_n \circ \mathbf{1}\left\{ z \geq 0 \right\}\right)(ar_n) =  \left(G_m \circ \mathbf{1}\left\{ z \geq 0 \right\}\right)(ar_m)
\end{equation*}
for all $a$ such that $|a| < 1/r_n$. This is because the average of $G_n \circ \mathbf{1}\left\{ z \geq 0 \right\}$ in the $r_n$ neighborhood is around $ar_n$ is exactly the average of $G_m \circ \mathbf{1}\left\{ z \geq 0 \right\}$ in the $r_m$ neighborhood is around $ar_m$. This is true also for all $a$ such that $|a| < 1/r_n - 1$. This is because our last result established equality on $[1,1]$, and this guarantees equality of the local averages on $[-1+r_n, 1-r_n]$.

Now suppose $\delta(x)$ is changing. We can write
\begin{equation*}
	\left(G_n \circ \mathbf{1}\left\{ z \geq 0 \right\}\right)(ar_n) =  \left(G_m \circ \mathbf{1}\left\{ z \geq 0 \right\}\right)(ar_m) + R_1
\end{equation*}
where $\lVert R_1 \rVert \leq \lvert \delta(ar_n) - \delta(ar_m) \rvert \leq C_\delta |a| (r_n + r_m) $ by Lemma \ref{lemma--shrinkingnorm}.

Suppose for induction that for $|a| < 1/r_n - k$,
\begin{align*}
	\left(G_n^k \circ \mathbf{1}\{ z \geq 0 \} \right)(ar_n) = 	\left(G_m^k \circ \mathbf{1}\{ z \geq 0 \} \right)(ar_m) + \sum_{j=1}^{k-1} G_n^j R_{k-j}
\end{align*}
where $\lVert R_k \rVert \leq C_\delta  |a|  (r_n + r_m)$ for all $k$. Now,
\begin{align*}
	\left(G_n^{k+1} \circ \mathbf{1}\{ z \geq 0 \} \right)(ar_n) = 	G_n \circ \left(G_m^k \circ \mathbf{1}\{ z \geq 0 \} \right)(ar_m) + G_n \circ  \sum_{j=1}^{k-1} G_n^j R_{k-j}
\end{align*}
for $|a| < 1/r_n -(k+1)$.
Now for a fixed $a^*$, choose $K$ such that $(K+|a|)r_n \leq 1$. Then we have that:
\begin{align*}
	& \left\lvert (I-G_n)^{-1}\mathbf{1}\{z \geq 0\})(a^*r_n) - (I-G_m)^{-1}\mathbf{1}\{z \geq 0\})(a^*r_m) \right\rvert \\
	& = \left\lvert \sum_{j=0}^\infty  	\left(G_n^{k} \circ \mathbf{1}\{ z \geq 0 \} \right)(a^*r_n) - 	\left(G_m^{k} \circ \mathbf{1}\{ z \geq 0 \} \right)(a^*r_m)\right\rvert \\
	& \leq  \left\lvert \sum_{j=0}^K  	\left(G_n^{k} \circ \mathbf{1}\{ z \geq 0 \} \right)(a^*r_n) - 	\left(G_m^{k} \circ \mathbf{1}\{ z \geq 0 \} \right)(a^*r_m)\right\rvert \\
	& \qquad +  \left\lvert \sum_{j=K+1}^\infty  	\left(G_n^{k} \circ \mathbf{1}\{ z \geq 0 \} \right)(a^*r_n) - 	\left(G_m^{k} \circ \mathbf{1}\{ z \geq 0 \} \right)(a^*r_m)\right\rvert \\
	& \leq \sum_{j=1}^{K-1} \left\lVert k G_n^jR_{k-j}\right\rVert + \frac{\bar{\delta}^{K+1}}{1-\bar{\delta}} \\
	& \leq \frac{C_\delta (r_n+r_m)}{a^*\left(1-\bar{\delta} \right)^2} + \frac{\bar{\delta}^{1/r_n-|a^*|}}{1-\bar{\delta}}~.
\end{align*}
where the second to last inequality follows from our inductive arguments earlier as well as repeated application of Lemma \ref{lemma--shrinkingnorm}. Applying Lemma \ref{lemma--shrinkingnorm} again, and rewriting $K = \lfloor 1-/r_n - |a| \rfloor$ yields the final inequality. As such, the sequence is Cauchy for fixed $a^*$. The limit therefore exists with the final term above being the rate of convergence.

For the next part, first suppose $\delta(0) \neq 0$. Then, we can write:
\begin{align*}
	& \left((I-G_n)^{-1} \circ \gamma(z) \tilde{d}(z/r_n)\right)(ar_n) \\
	& \qquad = 	\frac{\gamma(0)}{\delta(0)}\left( (I-G_n)^{-1} \circ \left(G_n \circ \mathbf{1}\{z \geq 0\}\right)\right)(ar_n) \\
	& \qquad \qquad + \frac{\gamma(0)}{\delta(0)} \left((I-G_n)^{-1} \circ \left(\gamma(z) - \gamma(0)\right) \left(G_n \circ \mathbf{1}\{z \geq 0\}\right)\right)(ar_n) + R
\end{align*}
and
\begin{equation*}
	|R| \leq \frac{C_\delta (r_n+r_m)}{a^*\left(1-\bar{\delta} \right)^2} + \frac{\bar{\delta}^{1/r_n-|a^*|}}{1-\bar{\delta}} ~.
\end{equation*}
This is because as before, the error arises solely from approximating $\delta(z)$ with $\delta(0)$ on the relevant part of the domain.

By the first part of this lemma,
\begin{align*}
	\left\lvert \frac{\gamma(0)}{\delta(0)}\left((I-G_n)^{-1} \circ \left(G_n \circ \mathbf{1}\{z \geq 0\}\right)\right)(ar_n) - \frac{\gamma(0)}{\delta(0)} \left(\lambda(a) - \mathbf{1}\{a \geq 0\}\right) \right\rvert \leq \frac{2C_\delta r_n}{a\left(1-\bar{\delta} \right)^2} + \frac{\bar{\delta}^{1/r_n-|a|}}{1-\bar{\delta}}~.
\end{align*}
Next observe that
\begin{align*}
\left(\gamma(z) - \gamma(0)\right) \left(G_n \circ \mathbf{1}\{z \geq 0\}\right))(ar_n)  & =  \left(\gamma(ar_n) - \gamma(0)\right) \delta(ar_n)\tilde{d}(a)
\end{align*}
where $\tilde{d}$ is given in Definition \ref{definition--tilde_d}.

As such,
\begin{align*}
\left\lvert \left(\gamma(z) - \gamma(0)\right) \left(G_n \circ \mathbf{1}\{z \geq 0\}\right))(ar_n)  \right\rvert \leq C_\gamma \bar{\delta}\cdot ar_n~.
\end{align*}
By the Neumann expansion and repeatedly applying Lemma \ref{lemma--shrinkingnorm} as above,
\begin{align*}
	\left\lvert (I-G_n)^{-1} \circ \left(\gamma(z) - \gamma(0)\right) \left(G_n \circ \mathbf{1}\{z \geq 0\}\right))(ar_n)  \right\rvert \leq \frac{C_\gamma \bar{\delta}}{1-\bar{\delta}}\cdot ar_n
\end{align*}

Finally, when $\delta(0) = 0$, $G_n = \mathbf{0}$ and the result follows immediately.

\subsection{Proof of Lemma \ref{lemma--preserveLipschitz}}

Let $x,y \in [-1,1]$ be given.
\begin{align*}
	\left\lvert \left(G_n \circ g \right)(x) - \left(G_n \circ g\right)(y) \right\rvert & = \left\lvert \frac{\delta(x)}{\lvert R(x) \rvert} \int_{R(x)} g(z) dz - \frac{\delta(y)}{\lvert R(y) \rvert} \int_{R(y)} g(z) \, dz \right\rvert \\
	& = \left\lvert \frac{\delta(x)}{\lvert R(x) \rvert} \int_{R(x)} g(z) \, dz - \frac{\delta(y)}{\lvert R(x) \rvert} \int_{R(x)} g(z+x-y) \, dz \right\rvert \\
	& \leq \left\lvert \frac{\delta(x)-\delta(y)}{\lvert R(x) \rvert} \int_{R(x)} g(z) \, dz \right\rvert + \left\lvert\frac{\delta(y)}{\lvert R(x) \rvert} \int_{R(x)} g(z) - g(z+x-y) \, dz \right\rvert \\
	& \leq M \cdot C_\delta  \lvert x - y \rvert + \bar{\delta} C \lvert x - y\rvert
\end{align*}

Now suppose for induction $\left(G_n^k \circ g \right)$ has Lipschitz constant $k \bar{\delta}^{k-1} C_\delta \cdot M + \bar{\delta}^k C$. We know from the above that $\left(G_n^{k+1} \circ g \right)$ has Lipschitz constant
\begin{align*}
	\bar{\delta}^k C_\delta \cdot M + \bar{\delta}\left(k \bar{\delta}^{k-1} C_\delta \cdot M + \bar{\delta}^k C\right) = (k+1) \bar{\delta}^k C_\delta \cdot M + \bar{\delta}^{k+1} C~.
\end{align*}
Writing $\left(I - G_n\right)^{-1}$ in terms of its Neumann series:
\begin{align*}
	\left(I - G_n\right)^{-1} = I + G_n + G^2_n + \cdots
\end{align*}
gives us that
\begin{align*}
& \left\lvert	\left(\left(I - G_n\right)^{-1} \circ g\right)(x) - 	\left(\left(I - G_n\right)^{-1} \circ g\right)(y) \right\rvert \\
& \leq \sum_{k=0}^\infty 	\left\lvert \left(G^k_n \circ g \right)(x) - \left(G^k_n \circ g\right)(y) \right\rvert  \\
& \leq \lvert x - y \rvert \sum_{k=0}^\infty k \bar{\delta}^{k-1} C_\delta \cdot M + \bar{\delta}^k C \\
& \leq \lvert x - y \rvert \cdot \left(\frac{C_\delta \cdot M }{\left(1-\bar{\delta}\right)^2} + \frac{C}{1-\bar{\delta}}\right)
\end{align*}
\hfill \qed

%

\subsection{Proof of Lemma \ref{lemma--preserveOdd}}
	Suppose $R(z) =  [z-a, z+b]$. Then $R(-z) = [-z-b, -z+a]$. Since, we have that
	\begin{align*}
		\left(G_n \circ g\right)(z) &= \frac{\delta(z)}{|R(z)|} \int_{z-a}^{z+b} g(x) \, dF(x) \\
		& = -\frac{\delta(-z)}{|R(-z)|} \int_{-z+a}^{-z-b} g(-y) \, dF(y) \mbox{ by change of variable: } y = -x \\
		& = -\frac{\delta(-z)}{|R(-z)|} \int^{-z+a}_{-z-b} g(y) \, dF(y)   \mbox{ since } g(z) = -g(-z) \\
		& = \left(G_n \circ g\right)(-z)
	\end{align*}
\hfill\qed

\subsection{Proof of Lemma \ref{lemma--g_nLipschitz}}
Using Definition \ref{definition--m_cont}, write
\begin{equation*}
	g_n(z) = \left(\left(I-G_n\right)^{-1} \circ m^c\right)(z) + \left(\left(I-G_n\right)^{-1} \circ \mathbf{1}\{z \geq 0\}\tau_d \right)(z)
\end{equation*}
From Lemma \ref{lemma--preserveLipschitz}, $\left(\left(I-G_n\right)^{-1} \circ m^c\right)(z)$ is Lipschitz with Lipschitz constant $$\frac{C_\delta \cdot M }{\left(1-\bar{\delta}\right)^2} + \frac{C}{1-\bar{\delta}}~.$$ Next, by the Neumann expansion,
\begin{align*}
	\left(\left(I-G_n\right)^{-1} \circ \mathbf{1}\{z \geq 0\}\tau_d \right)(z) &=  \mathbf{1}\{z \geq 0\}\tau_d + \left(G_n \circ  \mathbf{1}\{z \geq 0\}\tau_d\right)(z) + \left(G^2_n \circ  \mathbf{1}\{z \geq 0\}\tau_d\right)(z) + \cdots \\
	& = \mathbf{1}\{z \geq 0\}\tau_d + 	\left(\left(I-G_n\right)^{-1} \circ \left(G_n\circ \mathbf{1}\{z \geq 0\}\tau_d \right)\right)(z)
\end{align*}
It is easy to see that $\left(G_n\circ \mathbf{1}\{z \geq 0\} \right)$ is Lipschitz continuous with Lipschitz constant $\bar{\delta}/2r_n$. We are done after applying Lemma \ref{lemma--preserveLipschitz} again. \hfill \qed

\subsection{Proof of Lemma \ref{lemma--Lipschitz_unif_conv}}
In this proof only, let $m^c$ be any Lipschitz continuous function with Lipschitz constant $C$. By Lipschitz continuity, no point in the $r_n$ neighborhood of $z$ can be larger than $m^c(z) + Cr_n$. Choose $z \in [-\frac{1}{2}, \frac{1}{2}]$. Then,
\begin{align*}
	\left(G_n \circ m^c\right)(z) \leq {\delta}(z)\left(m^c(z) + \frac{Cr_n}{2}\right)~.
\end{align*}
Similarly,
\begin{align*}
	\left(G_n^2 \circ m^c\right)(z) & \leq {\delta(z)} \left( \left(G_n \circ m^c \right)(z)  + \bar{\delta}\cdot \frac{Cr_n}{2}\right) \\
	& \leq \left(\delta(z)\right)^2 m^c(z) + \frac{Cr_n}{2} \cdot 2 \left(\bar{\delta}\right)^2
\end{align*}
where the first inequality follows from the Lemma \ref{lemma--preserveLipschitz} and the second inequality follows from substituting in our bound for $\left(G_n \circ m^c\right)(z)$. By induction,
\begin{align*}
	\left(G_n^k \circ m^c\right)(z) & \leq \left(\delta(z)\right)^k m^c(z) + \frac{Cr_n}{2} \cdot k \left(\bar{\delta}\right)^k
\end{align*}
This is true as long as for all $k \leq K := \lfloor {1}/{2r_n}\rfloor$. Summing the terms in the Neumann expansion, we can write
\begin{align*}
	\left(\left(I - G_n\right)^{-1} \circ m^c\right)(z) - \frac{m^c(z)}{1-\delta(z)} & \leq \frac{Cr_n}{2}  \cdot \sum_{K=1}^\infty k \bar{\delta}^k + 2 \cdot \frac{\bar{\delta}^{K+1}\cdot M}{1-\bar{\delta}} \\
	& = \frac{Cr_n}{2}   \cdot
	\frac{\bar{\delta}}{\left(1-\bar{\delta}\right)^2} + 2 \cdot \frac{\bar{\delta}^{1/2r_n}\cdot M}{1-\bar{\delta}}
\end{align*}
 where the second term above comes from bounding terms corresponding to $k \geq K+1$. An analogous lower bound gives us that
\begin{align*}
\left\lvert	\left(\left(I - G_n\right)^{-1} \circ m^c\right)(z) - \frac{m^c(z)}{1-\delta(z)} \right\rvert \leq  \frac{Cr_n}{2}   \cdot
	\frac{\bar{\delta}}{\left(1-\bar{\delta}\right)^2} + 2 \cdot \frac{\bar{\delta}^{1/2r_n}\cdot M}{1-\bar{\delta}}~.
\end{align*}
Note here that for $z \geq 0$,
\begin{equation*}
	\frac{m^c(z)}{1-\delta(z)}  = \left(\left(I - G_\infty\right)^{-1} \circ m^c\right)(z)~.
\end{equation*}
Putting the above together with a similar argument for $z < 0$ gives us that for $z \in [-\frac{1}{2},\frac{1}{2}]$,
\begin{align*}
\left\lvert	\left(\left(I - G_n\right)^{-1} \circ m^c\right)(z) -  \left(\left(I - G_\infty\right)^{-1} \circ m^c\right)(z) \right\rvert \leq \frac{Cr_n}{2}   \cdot
	\frac{\bar{\delta}}{\left(1-\bar{\delta}\right)^2}~.
\end{align*}
In other words, $\left(\left(I - G_n\right)^{-1} \circ m^c\right)$ converges uniformly to  $\left(\left(I - G_\infty\right)^{-1} \circ m^c\right)$ on $[-\frac{1}{2}, \frac{1}{2}]$. \hfill \qed

\subsection{Proof of Lemma \ref{lemma--delta0approx}}

The argument here is essentially the same as the inductive argument in Lemma \ref{lemma--discont_limt}. First note that for a given $h_n$, we have by Lipschitz continuity of $\delta(z)$ that for all $x \in [-2h_n, 2h_n]$, $\lvert \delta(x) - \delta(0) \rvert \leq 2C_\delta h_n$.

Next, observe that on ${x \in [-2h_n + r_n, 2h_n - r_n]}$
\begin{equation*}
	\left(G_n\circ \mathbf{1}\left\{z \geq 0\right\}\right)(x) = \left(\tilde{G}_n\circ \mathbf{1}\left\{z \geq 0\right\}\right)(x) + R_1(x)
\end{equation*}
where the requisite bound on $R_1$ holds by Lipschitz continuity of $\delta(z)$. We propagate this bound forward. On ${x \in [-2h_n+ 2r_n, 2h_n - 2r_n]}$,
\begin{align*}
	\left(G^2_n\circ \mathbf{1}\left\{z \geq 0\right\}\right)(x) & = \left(G_n \circ \left(\tilde{G}_n\circ \mathbf{1}\left\{z \geq 0\right\}\right)\right)(x) + \left(G_n \circ R_1\right)(x) \\
	& = 	\left(\tilde{G}^2_n\circ \mathbf{1}\left\{z \geq 0\right\}\right)(x) +R_2(x) +  \left(G_n \circ R_1\right)(x)~.
\end{align*}
The bounds on $G_n^{k-j}R_j$ follows from Lemma \ref{lemma--shrinkingnorm}. Note that the domain over which the property holds shrinks by $r_n$ in each step of the induction. Since $h_n > Kr_n$, however, we can perform the above inductive step $K$ times and still have the property hold for $[-h_n, h_n]$. Lemma \ref{lemma--delta0approx} therefore follows.

\subsection{Proof of Lemma \ref{lemma-mu0_approx}}

We decompose $\mu_d(0)$ into the parts corresponding to the structural function $m$ and the exogenous spillover:
\begin{equation*}
	\mu_d(0) = A_n + B_n~,
\end{equation*}
where
\begin{align*}
	A_n  = \frac{1}{|R_n(0)|} \int_{-r_n}^{r_n} g_n(x) \, dF(x) \quad , \quad B_n = \frac{1}{|R_n(0)|} \int_{-r_n}^{r_n} l_n(x) \; dF(x) ~.
\end{align*}

We start by evaluating the limit of $A_n$. Using Definition \ref{definition--m_cont} write:
\begin{align*}
	A_n & = \frac{1}{|R_n(0)|} \int_{-r_n}^{r_n} \left(\left(1-G_n\right)^{-1} \circ m \right) (x) \, dF(x) \\
	& = \frac{1}{|R_n(0)|} \int_{-r_n}^{r_n} \left(\left(1-G_n\right)^{-1} \circ m^c \right) (x) \, dF(x) + \frac{\tau_d}{|R_n(0)|} \int_{-r_n}^{r_n} \left(\left(1-G_n\right)^{-1} \circ \mathbf{1}\left\{z \geq 0\right\}\right) (x) \, dF(x)
\end{align*}
By Lemma \ref{lemma--Lipschitz_unif_conv},
\begin{equation*}
	\frac{1}{2r_n} \int_{-r_n}^{r_n} \left(\left(1-G_n\right)^{-1} \circ m^c \right) (x) \, dF(x) = \frac{m^-(0)}{1-\delta(0)} + O(r_n)
\end{equation*}

For the second term in $A_n$, we use the Neumann series to write:
\begin{align}\label{equation--lemma--mu0--discreteA}
 & \frac{1}{2r_n} \int_{-r_n}^{r_n} \left(\left(1-G_n\right)^{-1} \circ \mathbf{1}\left\{z \geq 0\right\}\right) (x) \, dF(x) \nonumber \\ \nonumber
 &  = \frac{1}{2r_n} \int_{-r_n}^{r_n} \sum_{k=0}^K \left(G^k_n \circ \mathbf{1}\left\{z \geq 0\right\}\right) (x) \, dF(x) +  \frac{1}{2r_n} \int_{-r_n}^{r_n} \sum_{k=K+1}^\infty \left(G^k_n \circ \mathbf{1}\left\{z \geq 0\right\}\right) (x) \, dF(x) \\ \nonumber
 &  \leq \frac{1}{2r_n} \int_{-r_n}^{r_n} \sum_{k=0}^K \left(\tilde{G}^k_n \circ \mathbf{1}\left\{z \geq 0\right\}\right) (x) \, dF(x) + \frac{C_\delta r_n^{1-\eta}}{(1-\bar{\delta}^2)} +  \frac{1}{2r_n} \int_{-r_n}^{r_n} \sum_{k=K+1}^\infty \left(G^k_n \circ \mathbf{1}\left\{z \geq 0\right\}\right) (x) \, dF(x) \\
 &  \leq \frac{1}{2r_n} \int_{-r_n}^{r_n} \sum_{k=0}^K \left(\tilde{G}^k_n \circ \mathbf{1}\left\{z \geq 0\right\}\right) (x) \, dF(x) + \frac{C_\delta r_n^{1-\eta}}{(1-\bar{\delta}^2)} + \frac{\bar{\delta}^{1/r^\eta_n}}{1-\bar{\delta}}
\end{align}
where the first inequality above follows from Lemma \ref{lemma--delta0approx} with $h_n$ in the lemma set to $r_n^{1-\eta}/2$ and $K$ set to $\lfloor 1/r_n^\eta \rfloor$.  The second inequality follows from Lemma \ref{lemma--shrinkingnorm} and the definition of $K$. A similar argument leads to an analogous lower bound.

Now observe that observe that $\mathbf{1}\left\{z \geq 0\right\} - \frac{1}{2}$ is an odd function. By repeated application of Lemma \ref{lemma--preserveOdd}, we have that $\tilde{G}^k_n \circ \left(\mathbf{1}\left\{z \geq 0\right\}  - \frac{1}{2}\right)$ is an odd function. Since odd functions integrate to $0$ on intervals symmetric around $0$,
\begin{align*}
	\frac{1}{2r_n} \int_{-r_n}^{r_n} \sum_{k=0}^K \left(\tilde{G}^k_n \circ \left(\mathbf{1}\left\{z \geq 0\right\}\right) - \frac{1}{2}\right) (x) \, dF(x) = 0~.
\end{align*}
In other words,
\begin{align}\label{equation--lemma--mu0--discreteB}
	\frac{1}{2r_n} \int_{-r_n}^{r_n} \sum_{k=0}^K \left(\tilde{G}^k_n \circ \mathbf{1}\left\{z \geq 0\right\}\right) (x) \, dF(x) & = 	\frac{1}{2r_n} \int_{-r_n}^{r_n} \sum_{k=0}^K \left(\tilde{G}^k_n \circ \frac{1}{2}\mathbf{I} \right) (x) \, dF(x) \nonumber \\
	& = \frac{1}{2} \sum_{k=0}^K \left(\delta(0)\right)^k
\end{align}
where $\mathbf{I}$ is the identity function. The final equality above follows from the fact that $G_n \circ \mathbf{I} = \mathbf{I}$. Conclude that
\begin{align*}
	A_n & = \frac{m^-(0)}{1-\delta(0)} + \frac{\tau_d}{2(1-\delta(0))} + O\left(\frac{C_\delta r_n^{1-\eta}}{(1-\bar{\delta}^2)} + \frac{\bar{\delta}^{1/r^\eta_n}}{1-\bar{\delta}} \right)\\
	 & = \frac{m^+(0) + m^-(0)}{2(1-\delta(0))}  + O\left(\frac{C_\delta r_n^{1-\eta}}{(1-\bar{\delta}^2)} + \frac{\bar{\delta}^{1/r^\eta_n}}{1-\bar{\delta}} \right)~.
\end{align*}
Next, consider $B_n$:
\begin{align*}
	B_n & =  \frac{1}{|R_n(0)|} \int_{-r_n}^{r_n} \left( \left(I - G_n\right)^{-1} \circ \left(\gamma(z) \cdot  \tilde{d}(z/r_n) \right)\right)(x) \; dF(x) \\
	& =  \frac{1}{|R_n(0)|} \int_{-r_n}^{r_n} \left( \left(I - G_n\right)^{-1} \circ \left(\left(\gamma(z) - \gamma(0)\right) \cdot \mathbf{1}\{z \geq 0\} \right)\right)(x) \; dF(x) \\
	& \qquad -  \frac{1}{|R_n(0)|} \int_{-r_n}^{r_n} \left( \left(I - G_n\right)^{-1} \circ \left(\left(\gamma(z) - \gamma(0)\right)\cdot \left( \mathbf{1}\{z \geq 0\} - 	\tilde{d}(z/r_n) \right) \right)\right)(x) \; dF(x) \\
	& \qquad + \frac{1}{|R_n(0)|} \int_{-r_n}^{r_n} \left( \left(I - G_n\right)^{-1} \circ \left(\gamma(0) \cdot  \tilde{d}(z/r_n)\right)\right)(x) \; dF(x)
\end{align*}
Observe that $\left(\gamma(z) - \gamma(0)\right)\mathbf{1}\{z \geq 0\}$ is Lipschitz continuous with Lipschitz constant $C_\gamma$. As such, by Lemma \ref{lemma--Lipschitz_unif_conv},
\begin{align*}
	 \frac{1}{|R_n(0)|} \int_{-r_n}^{r_n} \left( \left(I - G_n\right)^{-1} \circ \left(\left(\gamma(z) - \gamma(0)\right) \cdot \mathbf{1}\{z \geq 0\} \right)\right)(x) \; dF(x) = O(r_n)~.
\end{align*}
Furthermore, since $\left\lvert\gamma(z) - \gamma(0)\right\rvert \leq C_\gamma r_n$, we also have by Lemma \ref{lemma--shrinkingnorm} that
\begin{equation*}
	 \frac{1}{|R_n(0)|} \int_{-r_n}^{r_n} \left( \left(I - G_n\right)^{-1} \circ \left(\left(\gamma(z) - \gamma(0)\right)\cdot \left( \mathbf{1}\{z \geq 0\} - 	\tilde{d}(z) \right) \right)\right)(x) \; dF(x) = O(r_n)
\end{equation*}
Next, observe that when $\delta(0) \neq 0$, {\small
\begin{align*}
	& \frac{1}{|R_n(0)|} \int_{-r_n}^{r_n} \left( \left(I - G_n\right)^{-1} \circ  \tilde{d}(z/r_n) \right)(x) \; dF(x)\\
	& = \frac{1}{\delta(0)}\frac{1}{|R_n(0)|} \int_{-r_n}^{r_n} \left( \left(I - G_n\right)^{-1} \circ  \left(G_n \circ \mathbf{1}\{z \geq 0\}\right)\right)(x) \; dF(x) +  O\left(\frac{C_\delta r_n^{1-\eta}}{(1-\bar{\delta}^2)} + \frac{\bar{\delta}^{1/r^\eta_n}}{1-\bar{\delta}} \right)  \\
	& = \frac{1}{\delta(0)}\left(\frac{1}{|R_n(0)|} \int_{-r_n}^{r_n} \left( \left(I - G_n\right)^{-1} \circ  \mathbf{1}\{z \geq 0\}\right)(x) \; dF(x) - 	\frac{1}{|R_n(0)|} \int_{-r_n}^{r_n}  \mathbf{1}\{x \geq 0\} \; dF(x) \right)\\
	& \qquad + O\left(\frac{C_\delta r_n^{1-\eta}}{(1-\bar{\delta}^2)} + \frac{\bar{\delta}^{1/r^\eta_n}}{1-\bar{\delta}} \right) \\
	& = \frac{1}{\delta(0)}\left(\frac{1}{2(1-\delta(0))} -\frac{1}{2} \right)+ O\left(\frac{C_\delta r_n^{1-\eta}}{(1-\bar{\delta}^2)} + \frac{\bar{\delta}^{1/r^\eta_n}}{1-\bar{\delta}} \right)  \\
	& = \frac{1}{2(1-\delta(0))} + O\left(\frac{C_\delta r_n^{1-\eta}}{(1-\bar{\delta}^2)} + \frac{\bar{\delta}^{1/r^\eta_n}}{1-\bar{\delta}} \right)
\end{align*}}

\noindent where the first equality comes from approximating $\delta(z)$ with $\delta(0)$. The second to last equality above follows from the same arguments as in equations \eqref{equation--lemma--mu0--discreteA} and  \eqref{equation--lemma--mu0--discreteB}.  When $\delta(0) = 0$,
\begin{align*}
		& \frac{1}{|R_n(0)|} \int_{-r_n}^{r_n} \left( \left(I - G_n\right)^{-1} \circ  \tilde{d}(z/r_n) \right)(x) \; dF(x)\\
		& = \frac{1}{|R_n(0)|} \int_{-r_n}^{r_n} \left( \left(I - \tilde{G}_n\right)^{-1} \circ  \tilde{d}(z/r_n) \right)(x) \; dF(x) + O\left(\frac{C_\delta r_n^{1-\eta}}{(1-\bar{\delta}^2)} + \frac{\bar{\delta}^{1/r^\eta_n}}{1-\bar{\delta}} \right) \\
		& =  \frac{1}{|R_n(0)|} \int_{-r_n}^{r_n}   \tilde{d}(x/r_n) \; dF(x) + O\left(\frac{C_\delta r_n^{1-\eta}}{(1-\bar{\delta}^2)} + \frac{\bar{\delta}^{1/r^\eta_n}}{1-\bar{\delta}} \right) \\
		& = \frac{1}{2}  + O\left(\frac{C_\delta r_n^{1-\eta}}{(1-\bar{\delta}^2)} + \frac{\bar{\delta}^{1/r^\eta_n}}{1-\bar{\delta}} \right) ~.
\end{align*}
As such,
\begin{equation*}
	B_n = \frac{\gamma(0)}{2(1-\delta(0))} + O\left(\frac{C_\delta r_n^{1-\eta}}{(1-\bar{\delta}^2)} + \frac{\bar{\delta}^{1/r^\eta_n}}{1-\bar{\delta}} \right)~.
\end{equation*}
\hfill \qed

\subsection{Proof of Lemma \ref{lemma--uniform_approx}}

Write:
\begin{align*}
	\mu_d(xh_n) - \mu_d(0) & = A_n + B_n
\end{align*}
where
\begin{align*}
	A_n & = \frac{1}{|R_n(xh_n)|} \int_{R_n(xh_n)} g_n(z) \, dF(z) -  \frac{1}{|R_n(0)|} \int_{R_n(0)} g_n(z) \, dF(z) \\
		& = \frac{|R_n^+(xh_n)|}{|R_n(xh_n)|} \cdot \frac{1}{|R_n^+(xh_n)|} \int_{R^+_n(xh_n)} g_n(z) \, dF(z) \\
		& \qquad \qquad - \frac{|R_n^-(xh_n)|}{|R_n(xh_n)|} \cdot \frac{1}{|R_n^-(xh_n)|} \int_{R^-_n(xh_n)} g_n(z) \, dF(z)
\end{align*}
and
\begin{align*}
	B_n & = \frac{1}{|R_n(xh_n)|} \int_{R_n(xh_n)} l_n(z) \, dF(z) -  \frac{1}{|R_n(0)|} \int_{R_n(0)} l_n(z) \, dF(z) \\
		& = \frac{|R_n^+(xh_n)|}{|R_n(xh_n)|} \cdot \frac{1}{|R_n^+(xh_n)|} \int_{R^+_n(xh_n)} l_n(z) \, dF(z) \\
		& \qquad \qquad - \frac{|R_n^-(xh_n)|}{|R_n(xh_n)|} \cdot \frac{1}{|R_n^-(xh_n)|} \int_{R^-_n(xh_n)} l_n(z) \, dF(z)~.
\end{align*}
Here,
\begin{equation*}
	l_n(z) = \left[\left(I - G_n\right)^{-1} \circ \left(\gamma \cdot \left(G_n \circ \mathbf{1}\{u \geq 0\} \right)\right)\right](z)~.
\end{equation*}
$A_n$ and $B_n$ are the parts of $\mu_d$ coming from (1) the structural function $m$ and (2) the exogenous spillover $\nu_d = G_n \circ \mathbf{1}\{z \geq 0\}$ respectively. We analyze these term by considering cases separately.

\subsubsection*{Case 1: $r_n = r$ constant}

In this case, $G_n$, $g_n$ and $R_n$ do not change with $n$. Suppose first that $z \geq 0$. Then,
\begin{equation*}
	R_n^+(xh_n) = [r, r+xh_n] \quad , \quad 	R_n^-(xh_n) = [-r-xh_n, -r]
\end{equation*}
which shrink to $r$ and $-r$ respectively.

We start with $A_n$. By Lemma \ref{lemma--g_nLipschitz}, {\small
\begin{align*}
 \frac{1}{|R_n^+(xh_n)|} \int_{R^+_n(xh_n)} g_n(z) \, dF(z) & =   \frac{1}{|R_n^+(xh_n)|} \int_{R^+_n(xh_n)} g^c_n(z) \, dF(z) +  \frac{{\tau}_d}{|R_n^+(xh_n)|} \int_{R^+_n(xh_n)} \mathbf{1}\{z \geq 0\} \, dF(z) \\
 	& \leq g_n^c(r) + \tilde{C} xh_n +  \frac{\tau_d}{|R_n^+(xh_n)|} \int_{R^+_n(xh_n)} \mathbf{1}\{z \geq 0\} \, dF(z) \\
 	& = g_n^c(r) + \tilde{C} xh_n +  \tau_d\\
 	& = g_n(r) + \tilde{C} xh_n
\end{align*}}

\noindent where the inequality above follows from the fact that $g_n^c$ is Lipschitz, with $\tilde{C}$ being the relevant constant in Lemma \ref{lemma--g_nLipschitz}. Together with a similar argument for the lower bound, we have that
\begin{align*}
\left\lvert \frac{1}{|R_n^+(xh_n)|} \int_{R^+_n(xh_n)} g_n(z) \, dF(z) - g_n(r) \right\rvert & \leq \tilde{C} xh_n
\end{align*}
Furthermore, since $h_n/r_n \to 0$, we have that for $n$ large enough,
\begin{equation}\label{equation--overlap_size}
	\frac{|R_n^+(xh_n)|}{|R_n(xh_n)|}  = \frac{|R_n^-(xh_n)|}{|R_n(xh_n)|} = \frac{xh_n}{r_n}~.
\end{equation}
Hence,
\begin{align*}
	 A_n = \frac{xh_n}{r_n}\left(g_n(r) - g_n(-r) + O(h_n) \right)~.
\end{align*}
where the constants in $O(h_n)$ do not depend on $x$. Now, if $z < 0$, a similar argument to the one above gives us
\begin{align*}
	 A_n = \frac{xh_n}{r_n}\left(g_n(-r) - g_n(r) + O(h_n) \right)~.
\end{align*}
Next consider $B_n$. By Lemma \ref{lemma--preserveLipschitz}, $l_n(z)$ is Lipschitz continuous with Lipschitz constant
\begin{equation*}
	\tilde{C}_1 = \frac{C_\delta}{\left(1- \bar{\delta}\right)^2} + \frac{C_\gamma}{1-\bar{\delta}}
\end{equation*}
As such,
\begin{align*}
	 \frac{1}{|R_n^+(xh_n)|} \int_{R^+_n(xh_n)} l_n(z) \, dF(z) & \leq l_n(r) + \tilde{C}_1 xh_n~,
\end{align*}
with an analogous lower bound. We conclude that
\begin{equation*}
	B_n = \frac{xh_n}{r_n}\left(l_n(r) - l_n(-r) + O(h_n) \right)~.
\end{equation*}
when $z \geq 0$ and
\begin{equation*}
	B_n = \frac{xh_n}{r_n}\left(l_n(-r) - l_n(r) + O(h_n) \right)~.
\end{equation*}
when $z < 0$.

\subsubsection*{Case 2: $r_n \to 0$, $h_n/r_n \to 0$}

Start with $A_n$. Suppose again that $z \geq 0$. Since $h_n/r_n \to 0$, equation \eqref{equation--overlap_size} continues to hold. Write: {\small
\begin{align}\label{equation--mu_conv_decomp}
\begin{split}
	 \frac{1}{|R_n^+(xh_n)|} \int_{R^+_n(xh_n)} g_n(z) \, dF(z) & = \frac{1}{|R_n^+(xh_n)|} \int_{R^+_n(xh_n)}  \left(\left(I - G_n\right)^{-1} \circ m^c \right)(z)\, dF(z)  \\
	 & \quad  +  \frac{\tau_d}{|R_n^+(xh_n)|}  \int_{R^+_n(xh_n)}  \left(\left(I - G_n\right)^{-1} \circ \mathbf{1}\left\{z \geq 0\right\} \right)(z) \, dF(z)
\end{split}
\end{align}}

\noindent We start with first term above:
\begin{align}\label{equation--mu_conv_cont}
	& \frac{1}{|R_n^+(xh_n)|} \int_{R^+_n(xh_n)} \left(\left(I - G_n\right)^{-1} \circ m^c \right)(z) \, dF(z)  \nonumber \\
	 & \qquad = \frac{1}{|R_n^+(xh_n)|} \int_{R^+_n(xh_n)} \left(\left(I - G_\infty\right)^{-1} \circ m^c \right)(z) \, dF(z)  + O(r_n)  \mbox{ by Lemma \ref{lemma--Lipschitz_unif_conv}}\nonumber \\
	 & \qquad = \frac{m^c(r_n)}{1-\delta(r_n)} + O(h_n) + O(r_n) \mbox{ by Definition \ref{definition--g_infty} and the fact that $m^c$ is Lipschitz} \nonumber \\
	 & \qquad = \frac{m^c(0)}{1-\delta(0)} + O(h_n) + O(r_n)
\end{align}
where the last equality follows again from Lipschitz continuity of $m^c(z)$ and $\delta(z)$.

Next, consider
\begin{align*}
	 & \frac{\tau_d}{|R_n^+(xh_n)|} \int_{R^+_n(xh_n)} \left(\left(I - G_n\right)^{-1} \circ \mathbf{1}\{x \geq 0\} \right)(z) \, dF(z) \nonumber\\
	 & \qquad = \frac{\tau_d}{|R_n^+(xh_n)|} \int_{R^+_n(xh_n)} \lambda(z/r_n) \, dF(z) + O\left( \frac{2\bar{\delta}^{\lfloor1/r_n\rfloor-|a|}}{1-\bar{\delta}}\right)
\end{align*}
which follows by an application of Lemma \ref{lemma--discont_limt} with $|a| = 2$ since $h_n \ll r_n$. Next consider the change of variable $a = z/r_n$. This gives us
\begin{align}\label{equation--mu_conv_discont}
& \frac{\tau_d}{|R_n^+(xh_n)|} \int_{R^+_n(xh_n)} \left(\left(I - G_n\right)^{-1} \circ \mathbf{1}\{x \geq 0\} \right)(z) \, dF(z) \nonumber \\
	 & \qquad = \frac{\tau_d}{r_n(xh_n/r_n)} \int_{[1,1+2xh_n/r_n]} \lambda(a) \cdot r_n \, dF(a) + O\left( \frac{2\bar{\delta}^{\lfloor1/r_n\rfloor-|a|}}{1-\bar{\delta}}\right) \nonumber \\
	 & \qquad = \frac{\tau_d}{xh_n/r_n} \int_{[1,1+2xh_n/r_n]} \lambda(a) \, dF(a) + O\left( \frac{2\bar{\delta}^{\lfloor1/r_n\rfloor-|a|}}{1-\bar{\delta}}\right) \nonumber \\
	 & \qquad = \tau_d \cdot \lambda(1) + o(1)
\end{align}
where the last equality follows from the Lebesgue Differentiation Theorem. The domain of integration is a convex interval and trivially of bounded eccentricity.

Putting equations \eqref{equation--mu_conv_cont} and \eqref{equation--mu_conv_discont} together with equation \eqref{equation--overlap_size} yields:
\begin{equation*}
	A_n = \frac{xh_n}{r_n}\cdot \tau_d\cdot \left(  \lambda(1) - \lambda(-1) + o(1)   \right)~.
\end{equation*}

Finally, suppose $z < 0$. Note that analysis of the continuous part is identical. For the discontinuous part, the first term is now zero. The second term is expanded around $-r_n$, giving us
\begin{equation*}
	A_n = \frac{xh_n}{r_n}\cdot \tau_d\cdot \left(  \lambda(-1) - \lambda(1) + o(1)   \right)~.
\end{equation*}

Next, consider $B_n$. Using the second part of Lemma \ref{lemma--discont_limt}, and identical argument to the one in equation \eqref{equation--mu_conv_discont} gives us that
\begin{align}
	 \frac{1}{|R_n^+(xh_n)|} \int_{R^+_n(xh_n)} l_n(z) \, dF(z)  = \gamma(0) \cdot \left(\lambda(1)-\mathbf{1}\{1 \geq 0\}\right) + o(1)
\end{align}
Consequently, if $z \geq 0$,
\begin{equation*}
	B_n = \frac{xh_n}{r_n}\cdot \gamma(0)\cdot \left(  \left(\lambda(1) - 1\right) - \lambda(-1)  + o(1)   \right)~.
\end{equation*}
On the other hand, if $z \leq 0$,
\begin{equation*}
	B_n = \frac{xh_n}{r_n}\cdot \gamma(0)\cdot \left(  \lambda(-1) - \left(\lambda(1) -1\right)  + o(1)   \right)~.
\end{equation*}

\subsubsection*{Case 3: $r_n = \frac{1}{2}ch_n$}

Under this asymptotic regime,
\begin{equation}\label{equation--overlap_size_constant}
	\frac{|R_n^+(xh_n)|}{|R_n(xh_n)|}  = \frac{|R_n^-(xh_n)|}{|R_n(xh_n)|} = \min\left\{1, \frac{|x|}{c} \right\}~.
\end{equation}
To see this, note that when $|x|h_n > 2r_n = ch_n$, there is no overlap between $R_n(xh_n)$ and $R_n(0)$. When $|x| < c$, the area of $$R^+_n(xh_n) = R^-_n(xh_n)  = (c-|x|)h_n$$
while $R_n(xh_n) = 2r_n = ch_n$, yielding the expression above.

Suppose again that $z \geq 0$. We consider the same decomposition as in equation \eqref{equation--mu_conv_decomp}. Equation \eqref{equation--mu_conv_cont} continues to be valid since we did not make any assumption about $h_n/r_n$ in its derivation.

Now, by a change of variable $a = z/r_n$, we have that
\begin{align}\label{equation--mu_conv_discont_constant}
& \frac{\tau_d}{|R_n^+(xh_n)|} \int_{R^+_n(xh_n)} \left(\left(I - G_n\right)^{-1} \circ \mathbf{1}\{x \geq 0\} \right)(z) \, dF(z) \nonumber \\
	 & \qquad = \frac{\tau_d}{r_n|\tilde{R}(x)|} \int_{\tilde{R}(x)} \lambda(a) \cdot r_n \, dF(a) + O\left( \frac{2\bar{\delta}^{\lfloor1/r_n\rfloor-|a|}}{1-\bar{\delta}}\right) \nonumber \\
	 & \qquad =\tau_d \cdot \lambda^+(x)+ O\left( \frac{2\bar{\delta}^{\lfloor1/r_n\rfloor-|a|}}{1-\bar{\delta}}\right)
\end{align}
In first equality above, we applied Lemma \ref{lemma--discont_limt} and the fact that $$R^+_n(xh_n) = \left[\max\left\{r_n, xh_n-r_n\right\}, r_n+xh_n\right] = r_n\left[\max\left\{1, \frac{2x}{c}-1\right\}, 1+\frac{2x}{c}\right] = r_n \tilde{R}^+(x)~.$$
Since $|x| \leq 1$, the above bound is uniform in $x$ once we let $|a| = 1+\frac{2}{c}$.

By the same reasoning,
\begin{equation*}
	R_n^- = r_n\underbrace{\left[-1, \min\left\{\frac{2x}{c} - 1, 1 \right\} \right]}_{=: \tilde{R}^-(x)}
\end{equation*}
and
\begin{align}\label{equation--mu_conv_discont_constant_negative}
& \frac{\tau_d}{|R_n^-(xh_n)|} \int_{R^-_n(xh_n)} \left(\left(I - G_n\right)^{-1} \circ \mathbf{1}\{x \geq 0\} \right)(z) \, dF(z) \nonumber \\
	 & \qquad =\tau_d \cdot \lambda^-(x)+ O\left( \frac{2\bar{\delta}^{\lfloor1/r_n\rfloor-|a|}}{1-\bar{\delta}}\right)
\end{align}

Putting equations \eqref{equation--mu_conv_cont}, \eqref{equation--mu_conv_discont_constant} and \eqref{equation--mu_conv_discont_constant_negative} together with equation \eqref{equation--overlap_size_constant} yields:
\begin{equation*}
	A_n = \min\left\{1, \frac{|x|}{c} \right\} \cdot \tau_d\cdot \left(  \lambda^+(x) -  \lambda^-(x) + o(1)   \right)~.
\end{equation*}
A corresponding derivation for $x \leq 0$ yields
\begin{equation*}
	A_n =  \min\left\{1, \frac{|x|}{c} \right\} \cdot \tau_d\cdot \left(  \lambda^+(-x) -  \lambda^-(-x) + o(1)   \right)~.
\end{equation*}

Next consider $B_n$. By derivations that is similar to \eqref{equation--mu_conv_discont_constant}, we have that {\small
\begin{align*}
 \frac{1}{|R_n^+(xh_n)|} \int_{R^+_n(xh_n)} \left(\left(I - G_n\right)^{-1} \circ l_n(z) \right)\, dF(z) =
	 & \gamma(0) \cdot \left(\lambda^+(x) - \mathbf{1}\{x \geq 0\}\right)+ O\left( \frac{2\bar{\delta}^{\lfloor1/r_n\rfloor-|a|}}{1-\bar{\delta}}\right)
\end{align*}}

\noindent which implies that
\begin{equation*}
	B_n = \min\left\{1, \frac{|x|}{c} \right\} \cdot \gamma(0)\cdot \left(  \tilde{\lambda}^+(x) - \tilde{\lambda}^-(x) + o(1)   \right)~.
\end{equation*}
when $z \geq 0$. When $z < 0$,
\begin{equation*}
	B_n = \min\left\{1, \frac{|x|}{c} \right\} \cdot \gamma(0)\cdot \left( \tilde{\lambda}^+(-x) -  \tilde{\lambda}^-(-x) + o(1)   \right)~.
\end{equation*}

\subsubsection*{Case 4: $r_n/h_n \to 0$}

Write
\begin{align*}
	\mu_d(xh_n) & = \frac{1}{|R_n(xh_n)|} \int_{R_n(xh_n)} g^c_n(z) \, dF(z) \\
	& \qquad + \frac{\tau_d}{|R_n(xh_n)|} \int_{R_n(xh_n)} \left(\left(I - G_n\right)^{-1} \circ \mathbf{1}\{y \geq 0\} \right)(z) \, dF(z) \\
	& \qquad + \frac{1}{|R_n(xh_n)|} \int_{R_n(xh_n)} l_n(z) \, dF(z)
\end{align*}
As before, the limit of the first term above is given by Equation \eqref{equation--mu_conv_cont}.

Next, for the second term, we apply Lemma \ref{lemma--delta0approx}, with $h_n$ in the lemma set to bandwidth. Next, let $K = \lfloor w_n/r_n \rfloor + 1$. Then, uniformly for $x \in [-1,1]$,
\begin{align*}
	& \frac{\tau_d}{|R_n(xh_n)|} \int_{R_n(xh_n)} \left(\left(I - G_n\right)^{-1} \circ \mathbf{1}\{y \geq 0\} \right)(z) \, dF(z) \\
	& \qquad = \frac{\tau_d}{|R_n(xh_n)|} \int_{R_n(xh_n)} \left( \sum_{k=0}^K \tilde{G}^k_n \circ \mathbf{1}\{y \geq 0\} \right)(z) \, dF(z) + O\left(\frac{C_\delta h_n}{(1-\bar{\delta})^2} + \frac{\bar{\delta}^K}{1-\bar{\delta}}\right)
\end{align*}
Furthermore, for $|z| > Kr_n$, we have that
\begin{equation*}
	\left(\tilde{G}^k_n \circ \mathbf{1}\{y \geq 0\} \right)(z) = \delta(0)^k \mathbf{1}\{z \geq 0\}~.
\end{equation*}
As such, for $x \in [w_n, 1]$, we have that:
\begin{align}\label{equation--lemma--unifapprox--small_r}
	& \frac{\tau_d}{|R_n(xh_n)|} \int_{R_n(xh_n)} \left( \sum_{k=0}^K \tilde{G}^k_n \circ \mathbf{1}\{y \geq 0\} \right)(z) \, dF(z) \nonumber \\
	& \qquad \qquad  = \tau_d \cdot \sum_{k=0}^K \delta(0)^k = \frac{\tau_d}{1-\delta(0)} + O\left(\frac{\bar{\delta}^k}{1-\bar{\delta}}\right)~.
\end{align}
As such, the second term of $\mu_d(xh_n)$ is
\begin{align*}
	& \frac{\tau_d}{|R_n(xh_n)|} \int_{R_n(xh_n)} \left(\left(I - G_n\right)^{-1} \circ \mathbf{1}\{y \geq 0\} \right)(z) \, dF(z) \\
	& \qquad =  \frac{\tau_d}{1-\delta(0)} + O\left(\bar{\delta}^K\right) + O\left(h_n \right)
\end{align*}

Finally, for the third term in $\mu_d(xh_n)$. Applying Lemma \ref{lemma--delta0approx} again, we have that when $\delta(0) \neq 0$, that for $x \in [-1,1]$, {\small
\begin{align*}
	& \frac{1}{|R_n(xh_n)|} \int_{R_n(xh_n)} l_n(z) \, dF(z) \\
	& \qquad = \frac{1}{\delta(0)} \frac{1}{|R_n(xh_n)|} \int_{R_n(xh_n)} \left(\left(I - \tilde{G}_n\right)^{-1} \circ \left(\gamma(y) \cdot  \left(G_n \circ \mathbf{1}\{y \geq 0\}\right)\right) \right)(z) \, dF(z) \\
	& \qquad = \frac{1}{\delta(0)}  \frac{1}{|R_n(xh_n)|} \int_{R_n(xh_n)} \left( \sum_{k=0}^K \tilde{G}^k_n \circ \left(\gamma(y) \cdot  \left(\tilde{G}_n \circ \mathbf{1}\{y \geq 0\}\right)\right) \right)(z) \, dF(z) + O\left(h_n + \bar{\delta}^K\right)\\
	& \qquad = \frac{1}{\delta(0)}  \frac{1}{|R_n(xh_n)|} \int_{R_n(xh_n)} \left( \sum_{k=0}^K \tilde{G}^k_n \circ \left(\gamma(0) \cdot  \left(\tilde{G}_n \circ \mathbf{1}\{y \geq 0\}\right)\right) \right)(z) \, dF(z) + O\left(h_n + \bar{\delta}^K\right)
\end{align*}}

\noindent where the last equality follows because for $k \leq h_n/r_n$, $\tilde{G}_n^k$ evaluated on $[-h_n, h_n]$ does not depend on values of $\gamma(y)$ outside of $[-2h_n, 2h_n]$. The error from replacing $\gamma(y)$ with $\gamma(0)$ on this latter interval is $O(h_n)$ uniformly in $x\in [-1,1]$. By the same reasoning, so is the error from replacing $\delta(y)$ with $\delta(0)$. Now, suppose $\delta(0) \neq 0$. Then write
\begin{align*}
	& \frac{1}{|R_n(xh_n)|} \int_{R_n(xh_n)} \left( \sum_{k=0}^K \tilde{G}^k_n \circ \left(\gamma(0) \cdot  \left(\tilde{G}_n \circ \mathbf{1}\{y \geq 0\}\right)\right) \right)(z) \, dF(z) \\
	& \qquad = \frac{\gamma(0)}{|R_n(xh_n)|} \int_{R_n(xh_n)} \left( \sum_{k=1}^{K+1} \tilde{G}^k_n \mathbf{1}\{y \geq 0\} \right)(z) \, dF(z) \\
	& \qquad = \gamma(0) \Bigg( \frac{1}{1-\delta(0)}  -   \underbrace{\frac{1}{|R_n(xh_n)|} \int_{R_n(xh_n)} \mathbf{1}\{z \geq 0\} \, dF(z)}_{= 1} \Bigg) + O\left( \bar{\delta}^{K+1}\right) \\
	& \qquad = \frac{\delta(0)\gamma(0)}{1-\delta(0)} + O\left( \bar{\delta}^{K+1}\right)
\end{align*}
where the second to last equality above follows from Equation \eqref{equation--lemma--unifapprox--small_r}. That is, when $\delta(0) \neq 0$,
\begin{align*}
	\frac{1}{|R_n(xh_n)|} \int_{R_n(xh_n)} l_n(z) \, dF(z) & = \frac{\gamma(0)}{1-\delta(0)} + O\left( \bar{\delta}^{K+1}\right)
\end{align*}
uniformly for $x \in [w_n, 1]$. When $\delta(0) = 0$,
\begin{align*}
	\frac{1}{|R_n(xh_n)|} \int_{R_n(xh_n)} l_n(z) \, dF(z) & = \frac{1}{|R_n(xh_n)|} \int_{R_n(xh_n)} \gamma(z) \tilde{d}(z/r_n) \, dF(z) + O\left(h_n + \bar{\delta}^{K}\right) \\
	& = \gamma(0) + O\left(h_n + \bar{\delta}^{K}\right) ~.
\end{align*}
again uniformly for $x \in [w_n, 1]$.

Putting our bounds together, we therefore have that
\begin{align*}
	\mu_d(xh_n) = \frac{m^+(0) + \gamma(0)}{1-\delta(0)} + O \left(h_n + \bar{\delta}^{w_n/r_n}\right)~.
\end{align*}
With Lemma \ref{lemma-mu0_approx}, we then have that for $x \in [w_n, 1]$,
\begin{align*}
	\mu_d(xh_n) - \mu_d(0) = \frac{\tau_d + \gamma(0)}{2\left(1-\delta(0)\right)} + O \left(h_n + \bar{\delta}^{w_n/r_n}\right)
\end{align*}
By a similar argument, when $x \in [-1, -w_n]$
\begin{align*}
	\mu_d(xh_n) - \mu_d(0) = - \frac{\tau_d +\gamma(0)}{2\left(1-\delta(0)\right)} + O \left(h_n + \bar{\delta}^{w_n/r_n}\right)~.
\end{align*}
\hfill\qed

\subsection{Proof of
Lemma \ref{lem--spilloverreg-inference-std-gauss-sufficient-an}}

Define
\begin{equation}
  \begin{split}
    \eta_{X, 1} \left( \xi_{1} \right) =
    & \, \varepsilon_{1} \E \left[ K \left( \frac{Z_{0}}{h} \right) \left(
    \delta (0) + \delta^{\prime} (0) Z_{0} \right) X \left( Z_{0} \right)
    \left\{ \frac{R \left( \frac{Z_{1}}{r} \right)}{\pi_{r} (0)} - \frac{R
    \left( \frac{Z_{1} - Z_{0}}{r} \right)}{\pi_{r} \left( Z_{0} \right)}
    \right\} \middle| \xi_{1} \right] \\
    =
    & \, \varepsilon_{1} \varphi_{\eta, 1} \left( Z_{1} \right), \\
    \varphi_{\eta, 1} \left( Z_{1} \right)
    =
    & \,
    \E \left[ K \left( \frac{Z_{0}}{h} \right) \left( \delta (0) +
    \delta^{\prime} (0) Z_{0} \right) X \left( Z_{0} \right) \left\{ \frac{R
    \left( \frac{Z_{1}}{r} \right)}{\pi_{r} (0)} - \frac{R \left( \frac{Z_{1} -
    Z_{0}}{r} \right)}{\pi_{r} \left( Z_{0} \right)} \right\} \middle| Z_{1}
    \right], \\
    \eta_{X, 2} \left( \xi_{1} \right) =
    & \, \E \left[ K \left( \frac{Z_{0}}{h} \right) \left( \delta (0) +
    \delta^{\prime} (0) Z_{0} \right) X \left( Z_{0} \right) \left( Y \left(
    Z_{1} \right) - \mu_{d} (0) \right) \frac{R \left( \frac{Z_{1}}{r}
    \right)}{\pi_{r} (0)} \middle| \xi_{1} \right] \\
    & - \E \left[ K \left( \frac{Z_{0}}{h} \right) \left( \delta (0) +
    \delta^{\prime} (0) Z_{0} \right) X \left( Z_{0} \right) \left( Y \left(
    Z_{1} \right) - \mu_{d} \left( Z_{0} \right) \right) \frac{R \left(
    \frac{Z_{1} - Z_{0}}{r} \right)}{\pi_{r} \left( Z_{0} \right)} \middle|
    \xi_{1} \right], \\
    \dot{\eta}_{X, 2} \left( \xi_{1} \right) =
    & \, \eta_{X, 2} \left( \xi_{1}
    \right) - \E \left[ \eta_{X, 2} \left( \xi_{1} \right) \right].
  \end{split}
  \label{eqn--etaX12-def}
\end{equation}
By \eqref{eqn--etaX-def},
\begin{equation*}
  \eta_{X} = \eta_{X, 1} + \eta_{X, 2}, \quad \text{and} \quad \dot{\eta}_{X} =
  \eta_{X, 1} + \dot{\eta}_{X, 2}.
\end{equation*}

By mean independence (\(\E \left[ \varepsilon_{1} \middle| Z_{1} \right] = 0\)),
\begin{equation*}
  \Cov \left( K \left( \frac{Z_{i}}{h} \right) X_{i} \varepsilon_{i}, \eta_{X,
  2} \left( \xi_{i} \right) \right) = 0, \quad \text{and} \quad \Cov \left(
  \eta_{X, 1} \left( \xi_{i} \right), \eta_{X, 2} \left( \xi_{i} \right) \right)
  = 0.
\end{equation*}
Thus,
\begin{equation}
  \begin{split}
    \bm{\Omega} =
    & \, \Var \left[ K \left( \frac{Z_{i}}{h} \right) X_{i} \varepsilon_{i} +
    \eta_{X} \left( \xi_{i} \right) \right] = \Var \left[ K \left(
    \frac{Z_{i}}{h} \right) X_{i} \varepsilon_{i} + \eta_{X, 1} \left(
    \xi_{i} \right) \right] + \Var \left[ \eta_{X, 2} \left( \xi_{i} \right)
    \right] \\
    \succeq
    & \, \Var \left[ K \left( \frac{Z_{i}}{h} \right) X_{i} \varepsilon_{i} +
    \eta_{X, 1} \left( \xi_{i} \right) \right] =: \bm{\Omega}_{1},
  \end{split}
  \label{eqn--Omega-dominates-Omega1}
\end{equation}
where \(\succeq\) denotes the Loewner order (\(A \succeq B\) iff \(A - B\) is
positive semi-definite).
This implies that
\(\left\| \bm{\Omega}^{- 1 / 2} t \right\| \leq \left\| \bm{\Omega}_{1}^{- 1 /
2} t \right\|\) for any vector \(t\).
In what follows, define
\begin{equation}
  t_{\ast, i} = K \left( \frac{Z_{i}}{h} \right) X_{i} \varepsilon_{i} +
  \dot{\eta}_{X} \left( \xi_{i} \right) \quad \text{and} \quad \alpha_{\ast, i}
  = \left( \frac{\bm{\Omega}}{n} \right)^{- 1/2} t_{i}.
  \label{eqn--t-alpha-ast-def}
\end{equation}
Clearly \(\E \left[ \alpha_{i} \right] = \mathbf{0}_{8}\) and \(\sum_{i = 1}^{n}
\Var \left[ \alpha_{\ast, i} \right] = \mathbb{I}_{8}\).
Furthermore, \(\bm{\Omega}^{- 1 / 2} \sqrt{n} \widetilde{\mathbf{T}}_{\ast} =
\sum_{i = 1}^{n} \alpha_{\ast, i}\).
It suffices to show that Liapunov's condition holds:
\begin{equation}
  \lim_{n \to \infty} \sum_{i = 1}^{n} \E \left[ \left\| \alpha_{\ast, i}
  \right\|^{3} \right] = 0
  \label{eqn--alpha-ast-3mom-lim-0}
\end{equation}

Following calculations similar to those for \(\widetilde{\mathbf{Q}}\) in the
proofs of Theorems \ref{theorem--spilloverreg_direct} and
\ref{theorem--spilloverreg_indirect} in section \ref{prf--thm--spilloverreg},
\begin{equation}
  \bm{\Omega}_{1} = h \cdot f (0) \sigma_{\varepsilon}^{2} (0) \cdot H
  \left( \bm{\Omega}_{1}^{\ast} + o (1) \right)
  H,
  \label{eqn--Omega1-limit-rate}
\end{equation}
where \(\bm{\Omega}_{1}^{\ast}\) is a limiting positive definite matrix, and
\begin{equation*}
  H = \mathbb{I}_{4} \otimes \mathrm{diag} (1, h) =
  \left[
  \begin{array}{ccccc}
    1      & 0      & \cdots & 0      & 0      \\
    0      & h      & \cdots & 0      & 0      \\
    \vdots & \vdots & \ddots & \vdots & \vdots \\
    0      & 0      & \cdots & 1      & 0      \\
    0 & 0 & \cdots & 0 & h
  \end{array}
  \right].
\end{equation*}
By \eqref{eqn--t-alpha-ast-def}, \eqref{eqn--Omega-dominates-Omega1}, and
\eqref{eqn--Omega1-limit-rate}
it follows that
\begin{align*}
  \sum_{i = 1}^{n} \E \left[ \left\| \alpha_{\ast, i}
  \right\|^{3} \right] =
  & \, n \E \left[ \left\| \alpha_{\ast, 1}
  \right\|^{3} \right] \\
  =
  & \, \frac{1}{\sqrt{n}} \E \left[ \left\| \bm{\Omega}^{- 1 / 2} t_{\ast, 1}
  \right\|^{3} \right] \\
  =
  & \, \frac{1}{\sqrt{n}} \E \left[ \left\| \bm{\Omega}_{1}^{- 1 / 2} t_{\ast,
  1} \right\|^{3} \right] \\
  =
  & \, \frac{f (0) \sigma_{\varepsilon}^{2} (0)}{\sqrt{n h}} \E \left[ \left\|
  \left( H \left( \bm{\Omega}_{1}^{\ast} + o (1) \right) H \right)^{- 1 / 2}
  t_{\ast, 1} \right\|^{3} \right]
  \\
  \leq
  & \,
  \frac{f (0) \sigma_{\varepsilon}^{2} (0)}{\left( \sqrt{\mathrm{eigmin} \left(
  \bm{\Omega}_{1}^{\ast} \right)} + o (1) \right) \cdot \sqrt{n h}} \E \left[
  \left\| H^{- 1} t_{\ast, 1} \right\|^{3} \right].
\end{align*}
Denoting \(\bm{\omega}_{1}^{\ast} = \sqrt{\mathrm{eigmin} \left(
\bm{\Omega}_{1}^{\ast} \right)}\), we get
\begin{equation*}
  \sum_{i = 1}^{n} \E \left[ \left\| \alpha_{\ast, i}
  \right\|^{3} \right] = \left( \frac{f (0) \sigma_{\varepsilon}^{2}
  (0)}{\bm{\omega}_{1}^{\ast}} + o (1) \right) \frac{1}{\sqrt{n h}} \E \left[
  \left\| H^{- 1} t_{\ast, 1} \right\|^{3} \right] \\
  =
  O (1) \cdot \frac{1}{\sqrt{n h}},
\end{equation*}
since \(\E \left[ \left\| H^{- 1} t_{\ast, 1} \right\|^{3} \right] = O (1)\)
when \(\E \left[ \varepsilon^{3} \middle| Z = z \right]\) is uniformly bounded
in a neighborhood of zero.
Thus Liapunov's condition \eqref{eqn--alpha-ast-3mom-lim-0} holds, and so
\(\bm{\Omega}^{- 1 / 2} \sqrt{n} \widetilde{\mathbf{T}}_{\ast} \rightsquigarrow
\N \left( \mathbf{0}_{8}, \mathbb{I}_{8} \right)\).

\subsection{Proof of
Lemma \ref{lem--spilloverreg-inference-std-gauss-sufficient-op1}}

Write
\begin{equation}
  \widetilde{\mathbf{T}}_{\ast, \varepsilon} = \frac{1}{n} \sum_{i = 1}^{n} K
  \left( \frac{Z_{i}}{h} \right) X_{i} \varepsilon_{i}, \quad \text{and} \quad
  \widetilde{\mathbf{T}}_{\ast, X} = \frac{1}{n} \sum_{i = 1}^{n} \left\{
  \eta_{X} \left( \xi_{i} \right) - \E \left[ \eta_{X} \left( \xi_{2} \right)
  \right] \right\}.
  \label{eqn--Tasteps-TastX-def}
\end{equation}
Then,
\begin{equation}
  \begin{split}
    \widetilde{\mathbf{T}}_{\varepsilon} =
    & \, \widetilde{\mathbf{T}}_{\ast, \varepsilon} +
    \widetilde{\mathbf{S}}_{\varepsilon}, \\
    \text{where} \quad
    \widetilde{\mathbf{S}}_{\varepsilon} =
    & \, \frac{1}{n} \sum_{i = 1}^{n} K \left( Z_{i} / h \right) \left(
    \widetilde{X}_{i} - X_{i} \right) \varepsilon_{i}.
  \end{split}
  \label{eqn--wtTeps-breakdown}
\end{equation}
Similarly,
\begin{equation}
  \begin{split}
    \widetilde{\mathbf{T}}_{X} =
    & \,
    \widetilde{\mathbf{T}}_{\ast, X} +
    \widetilde{\mathbf{T}}^{(1)}_{\ast, X} - \widetilde{\mathbf{T}}_{\ast, X} +
    \widetilde{\mathbf{S}}^{(1)}_{X}, \\
    \text{where} \quad
    \widetilde{\mathbf{T}}^{(1)}_{\ast, X} =
    & \, \frac{1}{n} \sum_{i = 1}^{n} K \left(
    Z_{i} / h \right) X_{i} \left( X_{i} - \widetilde{X}_{i} \right)^{\prime}
    \beta \\
    \widetilde{\mathbf{S}}_{X}^{(1)}
    =
    & \, - \frac{1}{n} \sum_{i = 1}^{n} K \left( Z_{i} / h \right) \left(
    \widetilde{X}_{i} - X_{i} \right) \left( \widetilde{X}_{i} - X_{i}
    \right)^{\prime} \beta.
  \end{split}
  \label{eqn--wtTX-breakdown}
\end{equation}
We wish to then show that
\begin{equation}
  \widetilde{\mathbf{S}}_{\varepsilon} = o_{\mathrm{p}} \left(
  \widetilde{\mathbf{T}}_{\ast, \varepsilon} \right)
  \quad \text{and} \quad
  \underbrace{\widetilde{\mathbf{T}}^{(1)}_{\ast, X} -
  \widetilde{\mathbf{T}}_{\ast, X} +
  \widetilde{\mathbf{S}}^{(1)}_{X}}_{\widetilde{\mathbf{S}}_{X}} =
  o_{\mathrm{p}} \left( \widetilde{\mathbf{T}}_{\ast, X} \right).
  \label{eqn-spilloverreg-inference-std-gauss-sufficient-op1-main}
\end{equation}

Define the following statistic
\begin{equation}
  T_{g} = \frac{1}{n} \sum_{i = 1}^{n} K \left( \frac{Z_{i}}{h} \right) g \left(
  \xi_{i} \right) \left( \widehat{\mu}_{i} \left( Z_{i} \right) - \mu \left(
  Z_{i} \right) - \widehat{\mu}_{i} (0) + \mu (0) \right).
  \label{eqn--Tg-def}
\end{equation}
Then,
\(\widetilde{\mathbf{T}}_{\ast, X} + \widetilde{\mathbf{S}}_{X} = T_{g_{X}}\),
where
\begin{equation}
  g_{X} (\xi) = - \left( \delta (0) + \delta^{\prime} (0) Z \right) \cdot X
  \left( Z, \mu_{r} (Z), \mu_{r} (0), \nu_{r} (Z), \nu_{r} (0) \right).
  \label{eqn--Tg-X-def}
\end{equation}
Similarly,
\(\widetilde{\mathbf{S}}_{\varepsilon}\) in \eqref{eqn--wtTeps-breakdown} can be
written as \(\widetilde{\mathbf{S}}_{\varepsilon} = T_{g_{\varepsilon}}\), where
\begin{equation}
  g_{\varepsilon} (\xi) = \left( 0, 0, 0, 0, Y - Y (Z), Z \cdot (Y - Y (Z)), 0,
  0 \right).
\end{equation}
We can show that \(T_{g}\) is the sum of mean-zero i.i.d. random variables with
an additional remainder term (see Lemma \ref{lem--Tg-linear-rep}).
To that end, define
\begin{equation}
  \begin{split}
    \dot{\eta}_{g, 2, h, r} \left( \xi_{1}, \xi_{2} \right) =
    & \, K \left( Z_{1} / h \right) g \left( \xi_{1} \right) \left(
    \dot{\varphi}_{r} \left( \xi_{2}, Z_{1} \right) -
    \dot{\varphi}_{r} \left( \xi_{2},
    0 \right) \right) \\
    & + K \left( Z_{2} / h \right) g \left( \xi_{2} \right)
    \left( \dot{\varphi}_{r} \left( \xi_{1}, Z_{2} \right) -
    \dot{\varphi}_{r} \left( \xi_{1}, 0 \right) \right), \\
    \text{and} \quad \dot{\eta}_{g, 1, h, r} \left( \xi_{1} \right) =
    & \, \E \left[ \dot{\eta}_{g, 2, h, r} \left( \xi_{1}, \xi_{2}
    \right) \middle| \xi_{1} \right] \\
    =
    & \, \E \left[ K \left( Z_{2} / h \right) g \left( \xi_{2} \right)
    \left( \dot{\varphi}_{r} \left( \xi_{1}, Z_{2} \right) -
    \dot{\varphi}_{r} \left( \xi_{1}, 0 \right) \right) \middle|
    \xi_{1} \right], \\
    \text{and for brevity, } \dot{\eta}_{g, j} \equiv
    & \, \dot{\eta}_{g, j, h, r} \quad \text{for } j \in \{1, 2\}.
  \end{split}
  \label{eqn--dot-eta-g-def}
\end{equation}
\(\dot{\eta}_{g, 1}\) is the first-order Hajek projection of \(\dot{\eta}_{g,
2}\),
and \(\dot{\varphi}_{r}\) (defined in
\eqref{eqn--muhat-loo-mu-diff-influence-varphi-def}) is an influence function
for the difference
\(\widehat{\mu}_{i} (z) - \mu (z)\) based on a first order Taylor approximation
of reciprocals.

The negligibility claim in
\eqref{eqn-spilloverreg-inference-std-gauss-sufficient-op1-main} follows
from Lemmas \ref{lem--Tg-linear-rep},
\ref{lem--Tg-linear-rep-Tast-rate-normality} and
\ref{lem--Tg-linear-rep-rate-UR} applied componentwise to each component of
\(g_{\varepsilon}\) and \(g_{X}\).

\subsection{Proof of Lemma \ref{lem--Tg-linear-rep}}
\label{secprf--lem--Tg-linear-rep}

For this proof, write \(\varphi \equiv \varphi_{r}\), \(\dot{\varphi} \equiv
\dot{\varphi}_{r}\), \(\mu \equiv \mu_{r}\), \(\nu \equiv \nu_{r}\) and \(\pi
\equiv \pi_{r}\).

\subsubsection{Proof of Lemma \ref{lem--Tg-linear-rep}: break down of
\texorpdfstring{\(T_{g}\)}{Tg}}

Break \(T_{g}\) in \eqref{eqn--Tg-def} into parts and use the representation in
\eqref{eqn--muhat-loo-mu-taylor-remainder-varphi-dot},
\begin{equation*}
  \begin{split}
    T_{g} =
    & \, \frac{1}{n} \sum_{i = 1}^{n} K \left( Z_{i} / h \right) g \left(
    \xi_{i} \right) \left( \widehat{\mu}_{i} \left( Z_{i} \right) - \mu \left(
    Z_{i} \right) \right) - \frac{1}{n} \sum_{i = 1}^{n} K \left( Z_{i} / h
    \right) g \left( \xi_{i} \right) \left( \widehat{\mu}_{i} (0) - \mu (0)
    \right) \\
    =
    & \, \frac{1}{n} \sum_{i = 1}^{n} K \left( Z_{i} / h \right) g \left(
    \xi_{i} \right) \left( \widehat{P}_{- i} \left[
    \dot{\varphi} \left( \cdot, Z_{i} \right) \right] +
    \widehat{\mathrm{Rem}}_{i, r} \left( Z_{i} \right) \right) \\
    & - \frac{1}{n} \sum_{i = 1}^{n} K \left( Z_{i} / h \right) g \left( \xi_{i}
    \right) \left( \widehat{P}_{- i} \left[ \dot{\varphi}
    (\cdot, 0) \right] + \widehat{\mathrm{Rem}}_{i, r} (0) \right).
  \end{split}
\end{equation*}
Thus,
\begin{equation}
  \begin{split}
    T_{g} =
    & \, T_{g}^{(1)} + R_{g}, \\
    \text{where} \quad T_{g}^{(1)} =
    & \, \frac{1}{n (n - 1)} \sum_{i = 1}^{n} \sum_{j = 1, j \neq i}^{n} K
    \left( Z_{i} / h \right) g
    \left( \xi_{i} \right) \left( \dot{\varphi} \left( \xi_{j},
    Z_{i} \right) - \dot{\varphi} \left( \xi_{j},
    0 \right) \right), \\
    R_{g} =
    & \, \frac{1}{n} \sum_{i = 1}^{n} K \left( Z_{i} / h \right) g \left(
    \xi_{i} \right) \left( \widehat{\mathrm{Rem}}_{i, r} \left( Z_{i} \right) -
    \widehat{\mathrm{Rem}}_{i, r} (0) \right)
  \end{split}
  \label{eqn--Tg-non-symmetric-U-stat-plus-remainder}
\end{equation}
Turn the expression for \(T_{g}^{(1)}\) into that of a second order
U-statistic with a symmetric kernel: combine
\eqref{eqn--Tg-non-symmetric-U-stat-plus-remainder} with the identity
\(\sum_{i, j: i \neq j} A_{i j} = \sum_{i, j: i < j} \left( A_{i j} + A_{j i}
\right) = 2 \sum_{i, j: i < j} \frac{1}{2} \left( A_{i j} + A_{j i} \right)\)
and \(\dot{\eta}_{g, 2}\) in \eqref{eqn--dot-eta-g-def} to get
\begin{equation}
  \begin{split}
    T_{g} =
    & \, T_{g}^{(1)} + R_{g}, \\
    \text{where} \quad T_{g}^{(1)} =
    & \, \frac{1}{n (n - 1)} \sum_{i = 1}^{n - 1} \sum_{j = i + 1}^{n}
    \dot{\eta}_{g, 2} \left( \xi_{i}, \xi_{j} \right), \\
    R_{g} =
    & \, \frac{1}{n} \sum_{i = 1}^{n} K \left( Z_{i} / h \right) g \left(
    \xi_{i} \right) \left( \widehat{\mathrm{Rem}}_{i, r} \left( Z_{i} \right) -
    \widehat{\mathrm{Rem}}_{i, r} (0) \right).
  \end{split}
  \label{eqn--Tg-symmetric-U-stat-plus-remainder}
\end{equation}
Now \eqref{eqn--Tg-linear-rep} follows from
\eqref{eqn--Tg-symmetric-U-stat-plus-remainder} and noting that
\(U_{g} = T_{g}^{(1)} - T_{\ast, g}\) (by
definition in \eqref{eqn--Tg-linear-rep}) so that \(T_{g}^{(1)} =
T_{\ast, g} + U_{g}\).

\subsection{Proof of Lemma \ref{lem--Tg-linear-rep-Tast-rate-normality}}
\label{secprf--lem--Tg-linear-rep-Tast-rate-normality}

Recall from \eqref{eqn--Tg-linear-rep} and
\eqref{eqn--dot-eta-g-def} that
\begin{equation*}
  \begin{split}
    T_{\ast, g} =
    & \, \frac{1}{n} \sum_{i = 1}^{n} \dot{\eta}_{g, 1} \left( \xi_{i} \right),
    \text{ where } \dot{\eta}_{g, 1} \left( \xi_{1} \right) =
    \E \left[ K \left( Z_{2} / h \right) g \left( \xi_{2}
    \right) \left( \dot{\varphi} \left( \xi_{1}, Z_{2} \right) -
    \dot{\varphi} \left( \xi_{1}, 0 \right) \right) \middle|
    \xi_{1} \right]
  \end{split}
\end{equation*}
Since \(\E \left[ \dot{\eta}_{g, 1} \left( \xi_{1} \right) \right] = 0\) and
\(\xi_{i}\) are independent, we get
\begin{equation*}
  \E \left[ T_{\ast, g}^{2} \right] = \frac{\E \left[ \dot{\eta}_{g 1} \left(
  \xi_{1} \right)^{2} \right]}{n}.
\end{equation*}

Define the ``un-centered'' version of \(\dot{\eta}_{g, 1}\) by
\begin{equation}
  \eta_{g, 1} \left( \xi_{1} \right) =
  \E \left[ K \left( Z_{2} / h \right) g \left( \xi_{2} \right) \left(
  \varphi \left( \xi_{1}, Z_{2} \right) - \varphi \left( \xi_{1}, 0
  \right) \right) \middle| \xi_{1} \right]
  \label{eqn--eta-g1-def}
\end{equation}
Recall from \eqref{eqn--muhat-loo-mu-diff-influence-varphi-def-alt-1} that
\begin{equation*}
  \varphi \left( \xi_{1}, z \right) = \left( Y_{1} - \mu (z)
  \right) \frac{R \left( \frac{Z_{1} - z}{r} \right)}{\pi (z)}.
\end{equation*}
Hence combining \eqref{eqn--muhat-loo-mu-diff-influence-varphi-def-alt-1} with
\eqref{eqn--eta-g1-def} yields
\begin{equation}
  \eta_{g, 1} \left( \xi_{1} \right) =
  \E \left[ K \left( \frac{Z_{2}}{h} \right) g \left( \xi_{2} \right)
  \left( \left( Y_{1} - \mu \left( Z_{2} \right) \right) \frac{R \left(
  \frac{Z_{1} - Z_{2}}{r} \right)}{\pi \left( Z_{2} \right)}
  - \left( Y_{1} - \mu (0) \right) \frac{R \left( \frac{Z_{1}}{r}
  \right)}{\pi (0)} \right) \middle| \xi_{1} \right].
\end{equation}

Break this down as follows
\begin{equation}
  \begin{split}
    \eta_{g, 1, h, r} \left( \xi_{1} \right) =
    & \, \alpha_{1, h, r} \left( \xi_{1} \right) + \alpha_{2, h, r} \left(
    \xi_{1} \right) - \alpha_{3, h, r} \left( \xi_{1} \right), \\
    \alpha_{1, h, r} \left( \xi_{1} \right) =
    & \, \varepsilon_{1} \E \left[ K \left( \frac{Z_{2}}{h} \right) g
    \left( \xi_{2} \right) \left( \frac{R \left( \frac{Z_{1} - Z_{2}}{r}
    \right)}{\pi \left( Z_{2} \right)} - \frac{R \left( \frac{Z_{1}}{r}
    \right)}{\pi (0)} \right) \middle| Z_{1} \right], \\
    \alpha_{2, h, r} \left( \xi_{1} \right) =
    & \, \E \left[ K \left( \frac{Z_{2}}{h} \right) g \left( \xi_{2}
    \right) \left( Y \left( Z_{1} \right) - \mu \left( Z_{2} \right) \right)
    \left( \frac{R \left( \frac{Z_{1} - Z_{2}}{r} \right)}{\pi \left( Z_{2}
    \right)} - \frac{R \left( \frac{Z_{1}}{r} \right)}{\pi (0)} \right)
    \middle| Z_{1} \right], \\
    \alpha_{3, h, r} \left( \xi_{1} \right) =
    & \, \E \left[ K \left( \frac{Z_{2}}{h} \right) g \left( \xi_{2} \right)
    \left( \mu \left( Z_{2} \right) - \mu (0) \right) \right] \frac{R
    \left( \frac{Z_{1}}{r} \right)}{\pi (0)}.
  \end{split}
  \label{eqn--alpha-breakdown}
\end{equation}
As before, we suppress reference to \(h, r\) and furthermore write \(\alpha_{j}
= \alpha_{j, h, r} \left( \xi_{1} \right)\).
We now derive moments of the \(\alpha_{j}\)'s and their approximate
representations.
By \(\E \left[ \varepsilon_{1} \middle| Z_{1} \right] = 0\),
\(\E \left[ \alpha_{1} \right] = 0\) and \(\E \left[ \alpha_{1} \cdot
\alpha_{j} \right] = 0\) for each \(j \in \{2, 3\}\).

Throughout, the following approximation result will be useful:
\begin{equation}
  f \text{ bounded and uniformly continuous } \implies
  \lim_{r \to 0} \sup_{z} \left| \frac{\pi (z)}{2 r} - f (z) \right| = 0.
  \label{eqn--pi-approximates-f}
\end{equation}
Our use of \eqref{eqn--pi-approximates-f} will be that normalization by division
by a factor \(2 r\) results in an ``asymptotic stabilization'' of \(\pi\).

\subsubsection{Moments for \texorpdfstring{\(\alpha_{1}\)}{alpha1}}

Use \eqref{eqn--alpha-breakdown} to write
\begin{equation}
  \begin{split}
    \alpha_{1} \left( \xi_{1} \right) =
    & \, \varepsilon \left( \xi_{1} \right) \cdot \dot{\varphi}_{\alpha, 1}
    \left( Z_{1} \right), \\
    \text{where } \dot{\varphi}_{\alpha, 1} \left( Z_{1} \right) =
    & \, \E \left[ K \left( \frac{Z_{2}}{h} \right) g \left( \xi_{2} \right)
    \left( \frac{R \left( \frac{Z_{1} - Z_{2}}{r} \right)}{\pi \left( Z_{2}
    \right)} - \frac{R \left( \frac{Z_{1}}{r} \right)}{\pi (0)} \right)
    \middle| Z_{1} \right].
  \end{split}
\end{equation}
In what follows, define
\begin{equation}
  m_{g} (z) = \E [g (\xi) | Z = z].
  \label{eqn--mg-def}
\end{equation}
For fixed \(z_{1}\), by a standard change of variables,
\begin{equation}
  \dot{\varphi}_{\alpha, 1} \left( z_{1} \right) = \frac{h}{2 r} \int f (u h) K
  (u) m_{g} (u h) \left( \frac{R \left( \frac{z_{1}}{r} - u \frac{h}{r}
  \right)}{\pi (u h) / 2 r} - \frac{R \left( \frac{z_{1}}{r}
  \right)}{\pi (0) / 2 r} \right) \mathrm{d} u.
\end{equation}
Then
\begin{equation*}
  \dot{\varphi}_{\alpha, 1} (v r) =
  \frac{h}{2 r} \int f (u h) K (u)
  m_{g} (u h) \left( \frac{R \left( v - u \frac{h}{r}
  \right)}{\pi (u h) / 2 r} - \frac{R (v)}{\pi (0) / 2 r} \right)
  \mathrm{d} u.
\end{equation*}
Define
\begin{equation}
  \begin{split}
    \varphi_{\alpha, 1} (v; h, r) =
    & \, \int f (u h) K (u) m_{g} (u h) \left( \frac{R \left( v - u \frac{h}{r}
    \right)}{\pi (u h) / 2 r} - \frac{R (v)}{\pi (0) / 2 r} \right)
    \mathrm{d} u, \\ \text{so that} \quad
    \dot{\varphi}_{\alpha, 1} (v r) =
    & \, \frac{h}{2 r} \varphi_{\alpha, 1} (v; h, r).
  \end{split}
  \label{eqn--alpha-1-rate-stable-breakdown}
\end{equation}
Along sequences \((h, r) = \left( h_{n}, r_{n} \right)\) such that \(\liminf (h
/ r) \in (0, \infty)\),
the quantity \(\varphi_{\alpha, 1}\) is asymptotically stable.

Denote \(\sigma^{2} \left( Z_{1} \right) = \E \left[ \varepsilon_{1}^{2}
\middle| Z_{1} \right]\).
Then
\begin{align*}
  \Var \left[ \alpha_{1} \right] =
  & \, \E \left[ \alpha_{1}^{2} \right] =
  \E \left[ \varepsilon_{1}^{2} \cdot \dot{\varphi}_{\alpha, 1} \left( Z_{1}
  \right)^{2} \right] = \E \left[ \E \left[ \varepsilon_{1}^{2} \middle| Z_{1}
  \right] \cdot \dot{\varphi}_{\alpha, 1} \left( Z_{1} \right)^{2} \right] \\
  =
  & \, \E \left[ \sigma^{2} \left( Z_{1} \right) \dot{\varphi}_{\alpha, 1}
  \left( Z_{1} \right)^{2} \right] = \int f \left( z_{1} \right) \sigma^{2}
  \left( z_{1} \right) \dot{\varphi}_{\alpha, 1} \left( z_{1} \right)^{2} \;
  \mathrm{d} z_{1}
  \\
  =
  & \, r \int f (v r) \sigma^{2} (v r) \dot{\varphi}_{\alpha, 1} (v r)^{2} \;
  \mathrm{d} v.
\end{align*}
Combine this with \eqref{eqn--alpha-1-rate-stable-breakdown} to get
\begin{equation}
  \Var \left[ \alpha_{1} \right] = h \frac{h}{4 r} \int f
  (v r) \sigma^{2} (v r) \varphi_{\alpha, 1} (v; h, r)^{2} \mathrm{d} v.
  \label{eqn--Var-alpha1}
\end{equation}
Next denote \(\rho_{3} \left( Z_{1} \right) = \E \left[ \left| \varepsilon_{1}
\right|^{3} \middle| Z_{1} \right]\).
Then
\begin{align*}
  \E \left[ \left| \alpha_{1} \right|^{3} \right] =
  & \, \E \left[ \left| \varepsilon_{1} \right|^{3} \left|
  \dot{\varphi}_{\alpha, 1} \left( Z_{1} \right) \right|^{3} \right] = \E \left[
  \rho_{3} \left( Z_{1} \right) \left| \dot{\varphi}_{\alpha, 1} \left( Z_{1}
  \right) \right|^{3} \right] \\
  =
  & \, \int f \left( z_{1} \right) \rho_{3} \left( z_{1} \right) \left|
  \dot{\varphi}_{\alpha, 1} \left( z_{1} \right) \right|^{3} \mathrm{d} z_{1}
  = r \int f (v r) \rho_{3} (v r) \left| \dot{\varphi}_{\alpha, 1} (v r)
  \right|^{3} \mathrm{d} v \\
  =
  & \, r \int f (v r) \rho_{3} (v r) \left| \dot{\varphi}_{\alpha, 1} (v r)
  \right|^{3} \mathrm{d} v.
\end{align*}
Combine this with \eqref{eqn--alpha-1-rate-stable-breakdown} to get
\begin{equation}
  \E \left[ \left| \alpha_{1} \right|^{3} \right] = h \frac{h^{2}}{8 r^{2}} \int
  f (v r) \rho_{3} (v r) \left| \varphi_{\alpha, 1} (v; h, r) \right|^{3}
  \mathrm{d} v.
  \label{eqn--Eabs3-alpha1}
\end{equation}
Thus combining \(\E \left[ \alpha_{1} \right] = 0\) with
\eqref{eqn--Var-alpha1} and \eqref{eqn--Eabs3-alpha1}, we get
\begin{equation}
  \begin{gathered}
    \E \left[ \alpha_{1} \right] = 0, \qquad
    \E \left[ \alpha_{1}^{2} \right] = h \frac{h}{4 r} \int
    f (v r) \sigma^{2} (v r) \varphi_{\alpha, 1} (v; h, r)^{2} \mathrm{d} v, \\
    \E \left[ \left| \alpha_{1} \right|^{3} \right] = h \frac{h^{2}}{8 r^{2}}
    \int f (v r) \rho_{3} (v r) \left| \varphi_{\alpha, 1} (v; h, r) \right|^{3}
    \mathrm{d} v.
    \label{eqn--alpha1-moments}
  \end{gathered}
\end{equation}

\subsubsection{Moments for \texorpdfstring{\(\alpha_{2}\)}{alpha2}}

Use \eqref{eqn--alpha-breakdown} and \eqref{eqn--mg-def} to write
\begin{equation}
  \alpha_{2} =
  \alpha_{2} \left( Z_{1} \right) :=
  \E \left[ K \left( \frac{Z_{2}}{h} \right) m_{g} \left( Z_{2} \right) \left( Y
  \left( Z_{1} \right) - \mu \left( Z_{2} \right) \right) \left( \frac{R
  \left( \frac{Z_{1} - Z_{2}}{r} \right)}{\pi \left( Z_{2} \right)} -
  \frac{R \left( \frac{Z_{1}}{r} \right)}{\pi (0)} \right) \middle| Z_{1}
  \right]
\end{equation}
Then
\begin{equation*}
  \alpha_{2} \left( z_{1} \right) =
  \int K \left( \frac{z_{2}}{h} \right) m_{g} \left( z_{2} \right)
  \left( Y \left( z_{1} \right) - \mu \left( z_{2} \right) \right) \left(
  \frac{R \left( \frac{z_{1} - z_{2}}{r} \right)}{\pi \left( z_{2} \right)}
  - \frac{R \left( \frac{z_{1}}{r} \right)}{\pi (0)} \right) f \left( z_{2}
  \right) \; \mathrm{d} z_{2}
\end{equation*}
By the change of variables \(z_{2} = u h\),
\begin{equation*}
  \alpha_{2} \left( z_{1} \right) = \frac{h}{2 r} \int K (u) f (u h) m_{g} (u h)
  \left( Y \left( z_{1} \right) - \mu (u h) \right) \left( \frac{R \left(
  \frac{z_{1}}{r} - u \frac{h}{r} \right)}{\pi (u h) / 2 r} - \frac{R \left(
  \frac{z_{1}}{r} \right)}{\pi (0) / 2 r} \right) \; \mathrm{d} u.
\end{equation*}
Therefore,
\begin{equation*}
  \alpha_{2} (v r) = \frac{h}{2 r} \int K (u) f (u h) m_{g} (u h) \left( Y
  (v r) - \mu (u h) \right) \left( \frac{R \left( v - u \frac{h}{r}
  \right)}{\pi (u h) / 2 r} - \frac{R (v)}{\pi (0) / 2 r} \right) \;
  \mathrm{d} u.
\end{equation*}
Let
\begin{equation*}
  \varphi_{\alpha, 2} (v; h, r) = \int K (u) f (u h) m_{g} (u h) \left( Y
  (v r) - \mu (u h) \right) \left( \frac{R \left( v - u \frac{h}{r}
  \right)}{\pi (u h) / 2 r} - \frac{R (v)}{\pi (0) / 2 r} \right) \;
  \mathrm{d} u.
\end{equation*}
Then
\begin{equation}
  \alpha_{2} (v r) = \frac{h}{2 r} \varphi_{\alpha, 2} (v; h, r).
  \label{eqn--alpha-2-rate-stable-breakdown}
\end{equation}
Along sequences \((h, r) = \left( h_{n}, r_{n} \right)\) such that \(\liminf (h
/ r) \in (0, \infty)\),
the quantity \(\varphi_{\alpha, 2}\) is asymptotically stable.

By the change of variables \(z_{2} = v r\),
\begin{equation*}
  \E \left[ \alpha_{2} \right] = \int \alpha_{2} \left( z_{1} \right) f \left(
  z_{1} \right) \; \mathrm{d} z_{1} = r \int \alpha_{2} (v r) f (v r) \;
  \mathrm{d} v.
\end{equation*}
Therefore using \eqref{eqn--alpha-2-rate-stable-breakdown},
\begin{equation*}
  \E \left[ \alpha_{2} \right] = \frac{h}{2} \int \varphi_{\alpha, 2} (v; h, r)
  f (v r) \; \mathrm{d} v.
\end{equation*}
For the second moment
\begin{equation*}
  \E \left[ \alpha_{2}^{2} \right] = \int \alpha_{2} \left( z_{1} \right)^{2} f
  \left( z_{1} \right) \; \mathrm{d} z_{1} = r \int \alpha_{2} (v r)^{2} f (v r)
  \; \mathrm{d} v,
\end{equation*}
and so, \eqref{eqn--alpha-2-rate-stable-breakdown} yields
\begin{equation*}
  \E \left[ \alpha_{2}^{2} \right] = h \frac{h}{4 r}  \int \varphi_{\alpha, 2}
  (v; h, r)^{2} f (v r) \; \mathrm{d} v.
\end{equation*}
A similar calculation for the third absolute moment using
\eqref{eqn--alpha-2-rate-stable-breakdown} yields
\begin{equation*}
  \E \left[ \left| \alpha_{2} \right|^{3} \right] = h \frac{h^{2}}{8 r^{2}}
  \int \left| \varphi_{\alpha, 2} (v; h, r) \right|^{3} f (v r) \; \mathrm{d}
  v.
\end{equation*}

Combine all of the above to get
\begin{equation}
  \begin{gathered}
    \E \left[ \alpha_{2} \right] = \frac{h}{2} \int \varphi_{\alpha, 2} (v; h,
    r) f (v r) \; \mathrm{d} v, \quad
    \E \left[ \alpha_{2}^{2} \right] = h \frac{h}{4 r}  \int \varphi_{\alpha, 2}
    (v; h, r)^{2} f (v r) \; \mathrm{d} v \\
    \E \left[ \left| \alpha_{2} \right|^{3} \right] = h \frac{h^{2}}{8 r^{2}}
    \int \left| \varphi_{\alpha, 2} (v; h, r) \right|^{3} f (v r) \; \mathrm{d}
    v.
  \end{gathered}
  \label{eqn--alpha2-moments}
\end{equation}

\subsubsection{Moments for \texorpdfstring{\(\alpha_{3}\)}{alpha3}}

From \eqref{eqn--alpha-breakdown},
\begin{equation*}
  \alpha_{3} = \dot{\varphi}_{\alpha, 3} (h, r) \cdot \frac{R (Z / r)}{\pi
  (0)}, \quad \text{where} \quad \dot{\varphi}_{\alpha, 3} = \E \left[ K \left(
  \frac{Z}{h} \right) m_{g} (Z) \left( \mu (Z) - \mu (0) \right)
  \right].
\end{equation*}
Therefore, since \(\E \left[ \frac{R (Z / r)}{\pi (0)} \right] = 1\),
\begin{equation*}
  \E \left[ \alpha_{3} \right] = \dot{\varphi}_{\alpha, 3}, \quad
  \E \left[ \alpha_{3}^{2} \right] = \dot{\varphi}_{\alpha, 3}^{2} \frac{\E
  \left[ R (Z / r)^{2} \right]}{\pi (0)^{2}}, \quad \text{and} \quad
  \E \left[ \left| \alpha_{3} \right|^{3} \right] = \left|
  \dot{\varphi}_{\alpha, 3} \right|^{3} \frac{\E \left[ R (Z / r)^{3}
  \right]}{\pi (0)^{3}}.
  \label{eqn--alpha3-moments-prelim}
\end{equation*}
Furthermore
\begin{equation*}
  \dot{\varphi}_{\alpha, 3} = \E \left[ K \left( \frac{Z}{h} \right) m_{g}
  (Z) \left( \mu (Z) - \mu (0) \right) \right] = \int K \left(
  \frac{z}{h} \right) m_{g} (z) \left( \mu (z) - \mu (0) \right) f (z)
  \mathrm{d} z,
\end{equation*}
and so by change of variables \(z = u h\),
\begin{equation*}
  \dot{\varphi}_{\alpha, 3} = h \int K (u) m_{g} (u h) \left( \mu (u h) -
  \mu (0) \right) f (u h) \mathrm{d} u.
\end{equation*}
Therefore
\begin{equation}
  \begin{split}
    \dot{\varphi}_{\alpha, 3} =
    & \, h \cdot \frac{\pi (0)}{2 r} \cdot \varphi_{\alpha, 3} (h, r), \quad
    \text{where} \\
    \varphi_{\alpha, 3} (h, r) =
    & \, \frac{2 r}{\pi (0)} \int K (u) m_{g} (u h) \left( \mu (u h) -
    \mu (0) \right) f (u h) \mathrm{d} u.
  \end{split}
  \label{eqn--E-alpha3}
\end{equation}

For the second moment, from \eqref{eqn--alpha3-moments-prelim},
\begin{equation*}
  \E \left[ \alpha_{3}^{2} \right] = \frac{h^{2}}{4 r^{2}} \varphi_{\alpha,
  3}^{2} (h, r) \int R \left( \frac{z}{r} \right)^{2} f (z) \; \mathrm{d} z,
\end{equation*}
and so
\begin{equation}
  \E \left[ \alpha_{3}^{2} \right] = h \frac{h}{4 r} \varphi_{\alpha, 3}^{2} (h,
  r) \int R (v)^{2} f (v r) \; \mathrm{d} v.
  \label{eqn--E-alpha3-2}
\end{equation}
For the third moment, essentially the same calculation as that for the
second moment yields
\begin{equation}
  \E \left[ \left| \alpha_{3} \right|^{3} \right] = h \frac{h^{2}}{8 r^{2}}
  \left| \varphi_{\alpha, 3} (h, r) \right|^{3} \int R (v)^{3} f (v r) \;
  \mathrm{d} v.
  \label{eqn--E-alpha3-3}
\end{equation}

Combine \eqref{eqn--E-alpha3}, \eqref{eqn--E-alpha3-2} and
\eqref{eqn--E-alpha3-3}:
\begin{equation}
  \begin{gathered}
    \E \left[ \alpha_{3} \right] = h \frac{\pi (0)}{2 r} \varphi_{\alpha, 3}
    (h, r), \quad
    \E \left[ \alpha_{2}^{2} \right] = h \frac{h}{4 r} \varphi_{\alpha, 3} (h,
    r)^{2} \int R (v)^{2} f (v r) \; \mathrm{d} v, \\
    \E \left[ \left| \alpha_{3} \right|^{3} \right] = h \frac{h^{2}}{8 r^{2}}
    \left| \varphi_{\alpha, 3} (h, r) \right|^{3} \int R (v)^{3} f (v r) \;
    \mathrm{d} v.
  \end{gathered}
  \label{eqn--alpha3-moments}
\end{equation}

\subsection{Proof of Lemma \ref{lem--Tg-linear-rep-rate-UR}}
\label{secprf--lem--Tg-linear-rep-rate-UR}

\subsubsection*{Proof of Lemma \ref{lem--Tg-linear-rep-rate-UR}: Variance of
\texorpdfstring{\(U_{g}\)}{Ug}}

For \eqref{eqn--Tg-linear-rep-Ug-rate}, first note that \(\dot{\eta}_{g, 1}\) is
the first order projection of \(\dot{\eta}_{g, 2}\), since
\begin{align*}
  \E \left[ \dot{\eta}_{g, 2} \left( \xi_{1}, \xi_{2} \right) \middle|
  \xi_{1} \right] =
  & \, \E \left[ K \left( Z_{1} / h \right) g \left( \xi_{1}
  \right) \left( \dot{\varphi} \left( \xi_{2}, Z_{1} \right) -
  \dot{\varphi} \left( \xi_{2}, 0 \right) \right) \middle|
  \xi_{1} \right] \\
  & + \E \left[ K \left( Z_{2} / h \right) g \left( \xi_{2}
  \right) \left( \dot{\varphi} \left( \xi_{1}, Z_{2} \right) -
  \dot{\varphi} \left( \xi_{1}, 0 \right) \right) \middle|
  \xi_{1} \right] \\
  =
  & \, K \left( Z_{1} / h \right) g \left( \xi_{1} \right) \left( \E
  \left[ \dot{\varphi} \left( \xi_{2}, Z_{1} \right) \middle|
  Z_{1} \right] - \E \left[ \dot{\varphi} \left(
  \xi_{2}, 0 \right) \middle| \xi_{1} \right] \right)
  \\
  & + \E \left[ K \left( Z_{2} / h \right) g \left( \xi_{2}
  \right) \left( \dot{\varphi} \left( \xi_{1}, Z_{2} \right) -
  \dot{\varphi} \left( \xi_{1}, 0 \right) \right) \middle|
  \xi_{1} \right].
\end{align*}
Since \(\E \left[ \dot{\varphi} \left( \xi_{2}, z
\right) \right] = 0\) for every \(z\),
\begin{equation*}
  \E \left[ \dot{\eta}_{g, 2} \left( \xi_{1}, \xi_{2} \right) \middle|
  \xi_{1} \right] = \E \left[ K \left( Z_{2} / h \right) g
  \left( \xi_{2} \right) \left( \dot{\varphi} \left( \xi_{1},
  Z_{2} \right) - \dot{\varphi} \left( \xi_{1}, 0 \right)
  \right) \middle| \xi_{1} \right] = \dot{\eta}_{g, 1} \left( \xi_{1} \right).
\end{equation*}
Thus \(\dot{\eta}_{g, 1}\) is the first order projection of \(\dot{\eta}_{g,
2}\).
Hence \eqref{eqn--Tg-linear-rep-Ug-rate} follows from
\eqref{eqn--L2-dist-proj-2} in Lemma \ref{lem--L2-dist-proj}.

\begin{lemma}
\label{lem--L2-dist-proj}
Let \(P\) be a probability measure and for \(m \in \mathbb{N} \setminus \{1\}\),
let \(\eta \in \mathscr{L}_{2} \left( P^{m} \right)\) be a
\(m\)\textsuperscript{th} order kernel that is permutation symmetric.
In the case of \(m = 2\), we have the equality
\begin{equation}
  \mathrm{E} \left[ \left\| U_{n} (\eta) - \widehat{U}_{n} (\eta) \right\|^{2}
  \right] = \frac{2}{n (n - 1)} \mathrm{tr} \left( \mathrm{E} \left[
  \mathrm{Var} \left[ \eta \left( \xi_{1}, \xi_{2} \right) \middle| \xi_{1}
  \right] \right] - \mathrm{Var} \left[ \mathrm{E} \left[ \eta \left( \xi_{1},
  \xi_{2} \right) \middle| \xi_{1} \right] \right] \right).
  \label{eqn--L2-dist-proj-2}
\end{equation}
More generally, for any \(m \geq 2\), we have the bound
\begin{equation}
  \begin{split}
  \mathrm{E} \left[ \left\| U_{n} (\eta) - \widehat{U}_{n} (\eta)
  \right\|^{2} \right] \leq
  & \, \frac{m^{2} (m - 1)^{2} \left( 1 + \frac{m - 1}{n - m + 1}
  \right)}{n^{2}} \zeta_{1} (\eta) \\
  & + \sum_{c = 2}^{m} \frac{1}{\left| \mathrm{Inj}_{n, c} \right|} \cdot
  \frac{(m !)^{2}}{((m - c)!)^{2} c!}
  \left( 1 + \delta_{n, m, c} \right) \zeta_{c} (\eta),
  \end{split}
  \label{eqn--L2-dist-proj-m}
\end{equation}
where for each \(c \in \{2, \dots, m\}\), the quantities \(\delta_{n, m,
c}\) do not depend on \(\eta\), \(P\) or the dimension of \(\xi_{i}\)
(except possibly through \(n\) and \(m\)) and satisfy \(\lim_{n \to \infty}
\delta_{n, m, c} = 0\).
\end{lemma}

The result \eqref{eqn--L2-dist-proj-2} and the definitions of the covariances
\(\zeta_{c}\) can be found in the chapter on U-statistics in
\citet{1998vandervaartAsymptoticStatistics}.
The result \eqref{eqn--L2-dist-proj-m} is not difficult to derive.

\subsubsection*{Proof of Lemma \ref{lem--Tg-linear-rep-rate-UR}: rate of
convergence of \texorpdfstring{\(U_{g}\)}{Ug}}

Now for the rate of convergence, first write
\begin{equation*}
  \begin{split}
    \dot{\eta}_{g, 2} \left( \xi_{1}, \xi_{2} \right) =
    & \, \dot{\eta}_{\ast} \left( \xi_{1}, \xi_{2} \right)
    + \dot{\eta}_{\ast} \left( \xi_{2}, \xi_{1} \right), \\
    \dot{\eta}_{\ast} \left( \xi_{1}, \xi_{2} \right) = \dot{\eta}_{\ast, h, r}
    \left( \xi_{1}, \xi_{2} \right) =
    & \, K \left( Z_{1} / h \right) g \left( \xi_{1} \right) \left(
    \dot{\varphi} \left( \xi_{2}, Z_{1} \right) -
    \dot{\varphi} \left( \xi_{2}, 0 \right) \right). \\
  \end{split}
\end{equation*}
Since \(\E \left[ \dot{\varphi} \left( \xi_{2}, z_{1} \right)
\right] = 0\) for every \(z_{1}\) (so that \(\E \left[ \dot{\varphi} \left(
\xi_{2}, Z_{1} \right) \middle| Z_{1} \right] = 0\)),
\begin{equation*}
  \Var \left[ \dot{\eta}_{g, 2} \right] = \E \left[ \dot{\eta}_{g, 2}^{2}
  \right] \quad \text{and} \quad
  \Var \left[ \dot{\eta}_{\ast} \right] = \E \left[ \dot{\eta}_{\ast}^{2}
  \right].
\end{equation*}
By the \(C_{r}\) inequality,
\begin{equation*}
  \Var \left[ \dot{\eta}_{g, 2} \left( \xi_{1}, \xi_{2} \right) \right] = \E
  \left[ \dot{\eta}_{g, 2} \left( \xi_{1}, \xi_{2} \right)^{2} \right]
  \leq 2 \left( \E \left[ \dot{\eta}_{\ast}
  \left( \xi_{1}, \xi_{2} \right)^{2} \right] + \E \left[ \dot{\eta}_{\ast}
  \left( \xi_{2}, \xi_{1} \right)^{2} \right] \right).
\end{equation*}
and so
\begin{equation}
  \Var \left[ \dot{\eta}_{g, 2} \left( \xi_{1}, \xi_{2} \right) \right] \leq 4
  \E \left[ \dot{\eta}_{\ast} \left( \xi_{1}, \xi_{2} \right)^{2} \right].
  \label{eqn--Var-eta-g2}
\end{equation}
Denote
\begin{equation*}
  m_{g^{2}} \left( Z_{1} \right) = \E \left[ g \left( \xi_{1} \right)^{2}
  \middle| Z_{1} \right].
\end{equation*}
Expanding the moment in the upper bound in \eqref{eqn--Var-eta-g2},
\begin{align*}
  \E \left[ \dot{\eta}_{\ast} \left( \xi_{1}, \xi_{2} \right)^{2} \right] =
  & \, \E \left[ K \left( \frac{Z_{1}}{h} \right)^{2} g \left( \xi_{1}
  \right)^{2} \left\{ \dot{\varphi} \left( \xi_{2}, Z_{1} \right) -
  \dot{\varphi} \left( \xi_{2}, 0 \right) \right\}^{2} \right] \\
  =
  & \, \E \left[ K \left( \frac{Z_{1}}{h} \right)^{2} m_{g^{2}} \left( Z_{1}
  \right) \left\{ \dot{\varphi} \left( \xi_{2}, Z_{1} \right) -
  \dot{\varphi} \left( \xi_{2}, 0 \right) \right\}^{2} \right] \\
  =
  & \, \E \left[ K \left( \frac{Z_{1}}{h} \right)^{2} m_{g^{2}} \left( Z_{1}
  \right) \Var \left[ \varphi \left( \xi_{2}, Z_{1} \right) -
  \varphi \left( \xi_{2}, 0 \right) \middle| Z_{1} \right]
  \right].
\end{align*}
Denote
\begin{equation}
  v^{2} \left( Z_{1} \right) = \Var \left[ \varphi \left(
  \xi_{2}, Z_{1} \right) - \varphi \left( \xi_{2}, 0 \right)
  \middle| Z_{1} \right].
  \label{eqn--cond-Var-eta-g2-Z1}
\end{equation}
Then,
\begin{equation*}
  \E \left[ \dot{\eta}_{\ast} \left( \xi_{1}, \xi_{2} \right)^{2} \right] =
  \E \left[ K \left( \frac{Z_{1}}{h} \right)^{2} m_{g^{2}} \left( Z_{1} \right)
  v^{2} \left( Z_{1} \right) \right]
  = \int f \left( z_{1} \right) K \left( \frac{z_{1}}{h} \right) m_{g^{2}}
  \left( z_{1} \right) v^{2} \left( z_{1} \right) \; \mathrm{d} z_{1},
\end{equation*}
and so by the change of variables \(z_{1} = u h\),
\begin{equation}
  \E \left[ \dot{\eta}_{\ast} \left( \xi_{1}, \xi_{2} \right)^{2} \right] =
  h \int K (u)^{2} f (u h) m_{g^{2}} (u h) v^{2} (u h) \; \mathrm{d} u.
  \label{eqn--Var-eta-g2-intermediate}
\end{equation}

Recall from \eqref{eqn--muhat-loo-mu-diff-influence-varphi-def-alt-1} that
\begin{equation*}
  \varphi \left( \xi_{2}, z_{1} \right) = \left( Y_{2} - \mu
  \left( z_{1} \right) \right) \frac{R \left( \frac{Z_{2} - z_{1}}{r}
  \right)}{\pi \left( z_{1} \right)}.
\end{equation*}
With some algebra, we can show
\begin{equation}
  \begin{split}
    \varphi \left( \xi_{2}, z_{1} \right) -
    \varphi \left( \xi_{2}, 0 \right) =
    & \, \varepsilon_{2} \left\{ \frac{R \left( \frac{Z_{2} - z_{1}}{r}
    \right)}{\pi \left( z_{1} \right)} - \frac{R \left( \frac{Z_{2}}{r}
    \right)}{\pi (0)} \right\} \\
    & + \left\{ Y \left( Z_{2} \right) - \mu \left( z_{1} \right) \right\}
    \left\{ \frac{R \left( \frac{Z_{2} - z_{1}}{r} \right)}{\pi \left( z_{1}
    \right)} - \frac{R \left( \frac{Z_{2}}{r} \right)}{\pi (0)} \right\} \\
    & + \left\{ \mu (0) - \mu \left( z_{1} \right) \right\} \frac{R
    \left( \frac{Z_{2}}{r} \right)}{\pi (0)}
  \end{split}
  \label{eqn--varphi-z1-diff-0}
\end{equation}

Revisiting \eqref{eqn--cond-Var-eta-g2-Z1},
\begin{equation*}
  v^{2} \left( z_{1} \right) = \Var \left[ \varphi \left(
  \xi_{2}, Z_{1} \right) - \varphi \left( \xi_{2}, 0 \right)
  \middle| Z_{1} = z_{1} \right] = \Var \left[ \varphi \left(
  \xi_{2}, z_{1} \right) - \varphi \left( \xi_{2}, 0 \right)
  \right],
\end{equation*}
and so,
\begin{equation}
  v^{2} \left( z_{1} \right) = \E \left[ \left\{ \varphi \left(
  \xi_{2}, z_{1} \right) - \varphi \left( \xi_{2}, 0 \right)
  \right\}^{2} \right] - \left( \E \left[ \varphi \left(
  \xi_{2}, z_{1} \right) - \varphi \left( \xi_{2}, 0 \right)
  \right] \right)^{2}.
  \label{eqn--cond-Var-eta-g2-Z1-to-uncond}
\end{equation}
Therefore using \eqref{eqn--varphi-z1-diff-0}, the two moments in
\eqref{eqn--cond-Var-eta-g2-Z1-to-uncond}
can be written as
\begin{equation*}
  \begin{split}
    \E \left[ \varphi \left( \xi_{2}, z_{1} \right) -
    \varphi \left( \xi_{2}, 0 \right) \right] =
    & \, \E \left[ \left( Y \left( Z_{2} \right) - \mu \left( z_{1} \right)
    \right) \left( \frac{R \left( \frac{Z_{2} - z_{1}}{r} \right)}{\pi
    \left( z_{1} \right)} - \frac{R \left( \frac{Z_{2}}{r} \right)}{\pi (0)}
    \right) \right] \\
    & + \left( \mu (0) - \mu \left( z_{1} \right) \right) \E \left[
    \frac{R \left( \frac{Z_{2}}{r} \right)}{\pi (0)} \right] \\
    \E \left[ \left\{ \varphi \left( \xi_{2}, z_{1} \right) -
    \varphi \left( \xi_{2}, 0 \right) \right\}^{2} \right] =
    & \, \E \left[ \sigma^{2} \left( Z_{2} \right) \left( \frac{R \left(
    \frac{Z_{2} - z_{1}}{r} \right)}{\pi \left( z_{1} \right)} - \frac{R
    \left( \frac{Z_{2}}{r} \right)}{\pi (0)} \right)^{2} \right] \\
    & + \E \left[ \left\{
    \begin{array}{c}
      \left( Y \left( Z_{2} \right) - \mu \left( z_{1} \right)
      \right) \left( \frac{R \left( \frac{Z_{2} - z_{1}}{r} \right)}{\pi
      \left( z_{1} \right)} - \frac{R \left( \frac{Z_{2}}{r} \right)}{\pi (0)}
      \right) \\
      + \left( \mu (0) - \mu \left( z_{1} \right) \right)
      \frac{R \left( \frac{Z_{2}}{r} \right)}{\pi (0)}
    \end{array}
    \right\}^{2} \right]
  \end{split}
\end{equation*}
We can then bound the second moment so that
\begin{equation}
  \begin{split}
    \E \left[ \varphi \left( \xi_{2}, z_{1} \right) -
    \varphi \left( \xi_{2}, 0 \right) \right] =
    & \, \E \left[ \left( Y \left( Z_{2} \right) - \mu \left( z_{1} \right)
    \right) \left( \frac{R \left( \frac{Z_{2} - z_{1}}{r} \right)}{\pi
    \left( z_{1} \right)} - \frac{R \left( \frac{Z_{2}}{r} \right)}{\pi (0)}
    \right) \right] \\
    & + \E \left[ \left( \mu (0) - \mu \left( z_{1} \right) \right)
    \frac{R \left( \frac{Z_{2}}{r} \right)}{\pi (0)} \right] \\
    \E \left[ \left\{ \varphi \left( \xi_{2}, z_{1} \right) -
    \varphi \left( \xi_{2}, 0 \right) \right\}^{2} \right] \leq
    & \, \E \left[ \sigma^{2} \left( Z_{2} \right) \left( \frac{R \left(
    \frac{Z_{2}
    - z_{1}}{r} \right)}{\pi \left( z_{1} \right)} - \frac{R \left(
    \frac{Z_{2}}{r} \right)}{\pi (0)} \right)^{2} \right] \\
    & + 2 \E \left[ \left( Y \left( Z_{2} \right) - \mu \left( z_{1}
    \right) \right)^{2} \left( \frac{R \left( \frac{Z_{2} - z_{1}}{r}
    \right)}{\pi \left( z_{1} \right)} - \frac{R \left( \frac{Z_{2}}{r}
    \right)}{\pi (0)} \right)^{2} \right] \\
    & + 2  \left( \mu (0) - \mu \left( z_{1} \right) \right)^{2}
    \frac{\E \left[ R \left( \frac{Z_{2}}{r} \right)^{2} \right]}{\pi (0)^{2}}.
  \end{split}
  \label{eqn--cond-Var-eta-g2-Z1-to-uncond-intermediate}
\end{equation}

For the first moment,
\begin{equation*}
  \begin{split}
    \E \left[ \varphi \left( \xi_{2}, z_{1} \right) - \varphi \left( \xi_{2}, 0
    \right) \right] =
    & \, \int f \left( z_{2} \right) \left( Y \left( z_{2} \right) - \mu \left(
    z_{1} \right) \right) \left( \frac{R \left( \frac{z_{2} - z_{1}}{r}
    \right)}{\pi \left( z_{1} \right)} - \frac{R \left( \frac{z_{2}}{r}
    \right)}{\pi (0)} \right) \; \mathrm{d} z_{2} \\
    & + \left( \mu (0) - \mu \left( z_{1} \right) \right) \\
    =
    & \, r \int f (v) \left( Y (v r) - \mu \left( z_{1} \right) \right)
    \left( \frac{R \left( v - \frac{z_{1}}{r} \right)}{\pi \left( z_{1} \right)}
    - \frac{R (v)}{\pi (0)} \right) \; \mathrm{d} v \\
    & + \left( \mu (0) - \mu \left( z_{1} \right) \right),
  \end{split}
\end{equation*}
where the last line follows by the change of variables \(z_{2} = v r\).
Normalizing \(\pi\) to \(\pi / 2 r\) for stability, we get
\begin{equation}
  \begin{split}
    \E \left[ \varphi \left( \xi_{2}, z_{1} \right) - \varphi \left( \xi_{2}, 0
    \right) \right] =
    & \, \frac{1}{2} \int f (v) \left( Y (v r) - \mu \left( z_{1} \right)
    \right) \left( \frac{R \left( v - \frac{z_{1}}{r} \right)}{\pi \left( z_{1}
    \right) / 2 r} - \frac{R (v)}{\pi (0) / 2 r} \right) \; \mathrm{d} v \\
    & + \left( \mu (0) - \mu \left( z_{1} \right) \right)
  \end{split}
  \label{eqn--cond-Var-eta-g2-Z1-to-uncond-1}
\end{equation}

For the variance, from \eqref{eqn--cond-Var-eta-g2-Z1-to-uncond-intermediate}
\begin{align*}
  \E \left[ \left\{ \varphi \left( \xi_{2}, z_{1} \right) - \varphi \left(
  \xi_{2}, 0 \right) \right\}^{2} \right] \leq
  & \, \int f \left( z_{2} \right) \sigma^{2} \left( z_{2} \right)
  \left( \frac{R \left( \frac{z_{2} - z_{1}}{r} \right)}{\pi \left( z_{1}
  \right)} - \frac{R \left( \frac{z_{2}}{r} \right)}{\pi (0)} \right)^{2}
  \; \mathrm{d} z_{2} \\
  & +
  2 \int f \left( z_{2} \right) \left( Y \left( z_{2} \right) - \mu \left( z_{1}
  \right) \right)^{2} \left( \frac{R \left( \frac{z_{2} - z_{1}}{r} \right)}{\pi
  \left( z_{1} \right)} - \frac{R \left( \frac{z_{2}}{r} \right)}{\pi (0)}
  \right)^{2} \; \mathrm{d} z_{2} \\
  & + 2 \left( \mu (0) - \mu \left( z_{1} \right) \right)^{2}
  \frac{\int f \left( z_{2} \right) R \left( \frac{z_{2}}{r} \right)^{2} \;
  \mathrm{d} z_{2}}{\pi (0)} \\
  =
  & \, r \int f (v r) \sigma^{2} (v r) \left( \frac{R \left( v - \frac{z_{1}}{r}
  \right)}{\pi \left( z_{1} \right)} - \frac{R (v)}{\pi (0)} \right)^{2}
  \; \mathrm{d} v \\
  & + 2 r \int f (v r) \left( Y (v r) - \mu \left( z_{1} \right) \right)^{2}
  \left( \frac{R \left( v - \frac{z_{1}}{r} \right)}{\pi \left( z_{1} \right)} -
  \frac{R (v r)}{\pi (0)} \right)^{2} \; \mathrm{d} v \\
  & + 2 \left( \mu (0) - \mu \left( z_{1} \right) \right)^{2}
  \frac{r \int f (v r) R (v)^{2} \; \mathrm{d} v}{\pi (0)^{2}}
\end{align*}
Thus, normalizing \(\pi\) to \(\pi / 2 r\),
\begin{equation}
  \begin{split}
    \E \left[ \left\{ \varphi \left( \xi_{2}, z_{1} \right) - \varphi \left(
    \xi_{2}, 0 \right) \right\}^{2} \right] =
    & \,
    \frac{1}{4 r} \int f (v r) \sigma^{2} (v r) \left( \frac{R \left( v -
    \frac{z_{1}}{r} \right)}{\pi \left( z_{1} \right) / 2 r} - \frac{R (v)}{\pi
    (0) / 2 r} \right)^{2} \; \mathrm{d} v \\
    & + \frac{1}{2 r} \int f (v r) \left( Y (v r) - \mu \left( z_{1} \right)
    \right)^{2} \left( \frac{R \left( v - \frac{z_{1}}{r} \right)}{\pi \left(
    z_{1} \right) / 2 r} - \frac{R (v r)}{\pi (0) / 2 r} \right)^{2} \;
    \mathrm{d} v \\
    & + \frac{1}{2 r} \left( \mu (0) - \mu \left( z_{1} \right) \right)^{2}
    \frac{\int f (v r) R (v)^{2} \; \mathrm{d} v}{(\pi (0) / 2 r)^{2}}
  \end{split}
  \label{eqn--cond-Var-eta-g2-Z1-to-uncond-2}
\end{equation}

Thus, plugging \eqref{eqn--cond-Var-eta-g2-Z1-to-uncond-1} and
\eqref{eqn--cond-Var-eta-g2-Z1-to-uncond-2}, into
\eqref{eqn--cond-Var-eta-g2-Z1-to-uncond}
\begin{equation}
  \begin{split}
    v^{2} \left( z_{1} \right) \leq
    & \, \frac{1}{4 r} \int f (v r) \sigma^{2} (v r) \left( \frac{R \left( v -
    \frac{z_{1}}{r} \right)}{\pi \left( z_{1} \right) / 2 r} - \frac{R (v)}{\pi
    (0) / 2 r} \right)^{2} \; \mathrm{d} v \\
    & + \frac{1}{2 r} \int f (v r) \left( Y (v r) - \mu \left( z_{1} \right)
    \right)^{2} \left( \frac{R \left( v - \frac{z_{1}}{r} \right)}{\pi \left(
    z_{1} \right) / 2 r} - \frac{R (v r)}{\pi (0) / 2 r} \right)^{2} \;
    \mathrm{d} v \\
    & + \frac{1}{2 r} \left( \mu (0) - \mu \left( z_{1} \right) \right)^{2}
    \frac{\int f (v r) R (v)^{2} \; \mathrm{d} v}{(\pi (0) / 2 r)^{2}} \\
    & - \left(
    \begin{array}{l}
      \frac{1}{2} \int f (v) \left( Y (v r) - \mu \left( z_{1} \right) \right)
      \left( \frac{R \left( v - \frac{z_{1}}{r} \right)}{\pi \left( z_{1}
      \right) / 2 r} - \frac{R (v)}{\pi (0) / 2 r} \right) \; \mathrm{d} v \\
      + \left( \mu (0) - \mu \left( z_{1} \right) \right)
    \end{array}
    \right)^{2}.
  \end{split}
  \label{eqn--cond-Var-eta-g2-Z1-to-uncond-bound}
\end{equation}

Thus evaluate the upper bound in \eqref{eqn--cond-Var-eta-g2-Z1-to-uncond-bound}
along sequences \(z_{1} = u h\) to conclude that
\begin{equation}
  \begin{split}
    v^{2} \left( u h \right) \leq
    & \, \frac{1}{r} C_{v} (u; h, r) \\
    C_{v} (u; h, r) =
    & \, \frac{1}{4} \int f (v r) \sigma^{2} (v r) \left( \frac{R \left( v -
    u \frac{h}{r} \right)}{\pi (u h) / 2 r} - \frac{R (v)}{\pi
    (0) / 2 r} \right)^{2} \; \mathrm{d} v \\
    & + \frac{1}{2} \int f (v r) \left( Y (v r) - \mu (u h)
    \right)^{2} \left( \frac{R \left( v - u \frac{h}{r} \right)}{\pi (u h) / 2
    r} - \frac{R (v r)}{\pi (0) / 2 r} \right)^{2} \;
    \mathrm{d} v \\
    & + \frac{1}{2} \left( \mu (0) - \mu (u h) \right)^{2}
    \frac{\int f (v r) R (v)^{2} \; \mathrm{d} v}{(\pi (0) / 2 r)^{2}} \\
    & - r \left(
    \begin{array}{l}
      \frac{1}{2} \int f (v) \left( Y (v r) - \mu (u h) \right)
      \left( \frac{R \left( v - u \frac{h}{r} \right)}{\pi \left(u h\right) / 2
      r} - \frac{R (v)}{\pi (0) / 2 r} \right) \; \mathrm{d} v \\
      + \left( \mu (0) - \mu \left( u h \right) \right)
    \end{array}
    \right)^{2}.
  \end{split}
  \label{eqn--cond-Var-eta-g2-Z1-to-uncond-rate-and-constant}
\end{equation}
Combine \eqref{eqn--cond-Var-eta-g2-Z1-to-uncond-rate-and-constant} with
\eqref{eqn--Var-eta-g2-intermediate} to get
\begin{equation}
  \E \left[ \dot{\eta}_{\ast} \left( \xi_{1}, \xi_{2} \right)^{2} \right] \leq
  \frac{h}{r} \int K (u)^{2} f (u h) m_{g^{2}} (u h) C_{v} (u; h r) \;
  \mathrm{d} u.
\end{equation}
Furthermore, by \eqref{eqn--Var-eta-g2}
\begin{equation*}
  \Var \left[ \dot{\eta}_{g, 2} \left( \xi_{1}, \xi_{2} \right) \right] \leq 4
  \frac{h}{r} \int K (u)^{2} f (u h) m_{g^{2}} (u h) C_{v} (u; h r) \;
  \mathrm{d} u.
\end{equation*}
The inequality in
\eqref{eqn--Tg-linear-rep-Ug-rate} now follows from the above and the fact that
\begin{equation*}
  \begin{split}
    & \mathrm{E} \left[ \mathrm{Var} \left[ \dot{\eta}_{g, 2}
    \left( \xi_{1}, \xi_{2} \right) \middle| \xi_{1} \right] \right] -
    \mathrm{Var} \left[ \mathrm{E} \left[ \dot{\eta}_{g, 2} \left( \xi_{1},
    \xi_{2} \right) \middle| \xi_{1} \right] \right] \\
    \leq
    & \,
    \mathrm{E} \left[ \mathrm{Var} \left[ \dot{\eta}_{g, 2} \left( \xi_{1},
    \xi_{2} \right) \middle| \xi_{1} \right] \right] + \mathrm{Var} \left[
    \mathrm{E} \left[ \dot{\eta}_{g, 2} \left( \xi_{1},
    \xi_{2} \right) \middle| \xi_{1} \right] \right] \\
    =
    & \, \Var \left[ \dot{\eta}_{g, 2} \left( \xi_{1}, \xi_{2} \right) \right].
  \end{split}
\end{equation*}

\subsubsection*{Proof of Lemma \ref{lem--Tg-linear-rep-rate-UR}: rate of
convergence of \texorpdfstring{\(U_{g}\)}{Ug} relative to
\texorpdfstring{\(T_{g, \ast}\)}{Tgstar}}

Take the ratio of second moments:
By the expression for \(\E \left[ U_{g}^{2} \right]\) in
\eqref{eqn--Tg-linear-rep-Ug-rate} and for
\(\E \left[ T_{g, \ast}^{2} \right]\) in \eqref{eqn--Tg-linear-rep-Tg-ast-rate},
\begin{align*}
  \frac{\E \left[ U_{g}^{2} \right]}{\E \left[ T_{\ast, g}^{2} \right]} \leq
  & \, \frac{8}{(n - 1) h} \cdot \frac{\int K (u)^{2} f (u h) m_{g^{2}} (u h)
  C_{v} (u; h r) \; \mathrm{d} u}{(1 + o (1)) \left( \int f (v r) \sigma^{2} (v
  r) \varphi_{\alpha, 1} (v; h, r)^{2} \mathrm{d} v + \int \varphi_{\alpha, 2}
  (v; h, r)^{2} f (v r) \; \mathrm{d} v \right)}.
\end{align*}
Therefore, \(U_{g} = o_{\mathrm{p}} \left( T_{\ast, g} \right)\) if \(n h \to
\infty\) since the variance ratio tends to zero.

\subsubsection*{Proof of Lemma \ref{lem--Tg-linear-rep-rate-UR}: rate of
convergence of \texorpdfstring{\(R_{g}\)}{Rg} relative to
\texorpdfstring{\(U_{g, \ast}\)}{Ugstar}}

Write
\begin{equation*}
  \mathcal{E} = \left\{ \max_{i = 1, \dots, n} \sup_{z \in [- 1, 1]} \left|
  \widehat{\pi}_{i} (z) - \pi (z) \right| \leq \frac{r}{2} \inf f \right\}.
\end{equation*}
We want to apply \eqref{eqn--pi-hat-concentration-i-max} in
Lemma \ref{lem--pi-hat-concentration-main} with \(y = \frac{r}{16} \inf f\)
(so that \(8 y = \frac{r}{2} \inf f\)).
First, the side condition \(y \geq \sqrt{\frac{C r}{n}}\) is satisfied for \(n\)
large enough since \(\frac{r}{16} \inf f \geq \sqrt{\frac{C r}{n}}\) if and only
if \(\frac{1}{\sqrt{n r}} \leq \frac{\inf f}{16 \sqrt{C}}\) and the latter
occurs since \(1 / (n r) \to 0\).
Then by \eqref{eqn--pi-hat-concentration-i-max} with \(y = \frac{r}{16} \inf
f\),
\begin{equation*}
  \begin{split}
    \Pr \left\{ \mathcal{E}^{c} \right\} \leq
    & \, A \exp \left\{- \frac{(n - 1) (\inf f)^{2} r}{265 C_{\ast}} + 2 \log (1
    16 / (r \inf f)) + \log n \right\} \\
    & + A \exp \left\{ - C (n - 1) r + 2 \log (1 / (C r)) + \log n \right\} \\
    \to
    & \, 0.
  \end{split}
\end{equation*}
since \((\log n) / (n r) \to 0\).

Now split
\begin{equation}
  \left| R_{g} \right| = \left| R_{g} \right| \mathbf{1}_{\mathcal{E}} +
  \left| R_{g} \right| \left( 1 - \mathbf{1}_{\mathcal{E}} \right) = \left( 1 +
  o_{\mathrm{p}} (1) \right) \left| R_{g} \right| \mathbf{1}_{\mathcal{E}},
  \label{eqn--Rg-breakdown}
\end{equation}
where the latter equality follows since \(1 - \mathbf{1}_{\mathcal{E}}
\overset{\mathrm{p}}{\to} 0\) and \(1 - \mathbf{1}_{\mathcal{E}}\) is a 0-1
sequence, the second term in the first equality above is negligible at any rate
(i.e. for arbitrary norming sequences).

Recall from \eqref{eqn--Tg-linear-rep} that
\begin{equation*}
  R_{g} = \frac{1}{n} \sum_{i = 1}^{n} K \left( Z_{i} / h \right) g
  \left( \xi_{i} \right) \left( \widehat{\mathrm{Rem}}_{i, r} \left( Z_{i}
  \right) - \widehat{\mathrm{Rem}}_{i, r} (0) \right),
\end{equation*}
and from \eqref{eqn--muhat-loo-mu-taylor-expand-remainder-magnitude}:
\begin{equation*}
  \begin{gathered}
    \text{If} \quad \left| \widehat{\pi}_{i, r} (z) - \pi_{r} (z) \right| =
    \left| \left( \widehat{P}_{- i} - P \right) \left[ \psi_{2, r, z} \right]
    \right| \leq \frac{r}{2} \inf f, \\
    \begin{aligned}
      \text{then} \quad \left| \widehat{\mathrm{Rem}}_{i, r, z} \right| \leq
      & \,
      2 \left| \overline{\psi}_{1, r} (z) \right| \cdot \frac{\left[ \left(
      \widehat{P}_{- i} - P \right) \left[ \psi_{2, r, z} \right]
      \right]^{2}}{r^{3} (\inf f)^{3}} \\
      & + \frac{\left[ \left( \widehat{P}_{- i} - P \right) \left[ \psi_{1, r,
      z} \right] \right]^{2} +  \left[ \left( \widehat{P}_{- i} - P \right)
      \left[ \psi_{2, r, z} \right] \right]^{2}}{r^{2} (\inf f)^{2}}.
    \end{aligned}
  \end{gathered}
\end{equation*}

Combine \eqref{eqn--Rg-breakdown} with the above to get
\begin{align*}
  \left| R_{g} \right| =
  & \, \left( 1 + o_{\mathrm{p}} (1) \right) \mathbf{1}_{\mathcal{E}} \left|
  R_{g} \right| \leq
  \frac{\left( 1 + o_{\mathrm{p}} (1) \right) \mathbf{1}_{\mathcal{E}}}{n}
  \sum_{i = 1}^{n} K \left( \frac{Z_{i}}{h} \right) \left| g \left(
  \xi_{i} \right) \right| \left( \left| \widehat{\mathrm{Rem}}_{i, r} \left(
  Z_{i} \right) \right| + \left| \widehat{\mathrm{Rem}}_{i, r} (0) \right|
  \right) \\
  \leq
  & \, \frac{\left( 2 + o_{\mathrm{p}} (1) \right) \mathbf{1}_{\mathcal{E}}}{n}
  \sum_{i = 1}^{n} K \left( \frac{Z_{i}}{h} \right) \left| g \left(
  \xi_{i} \right) \right| \left| \overline{\psi}_{1, r} \left( Z_{i} \right)
  \right| \cdot \frac{\left[ \left( \widehat{P}_{- i} - P \right) \left[
  \psi_{2, r} \left( \cdot; Z_{i} \right) \right] \right]^{2}}{r^{3} (\inf
  f)^{3}} \\
  & + \frac{\left( 2 + o_{\mathrm{p}} (1) \right) \mathbf{1}_{\mathcal{E}}}{n}
  \sum_{i = 1}^{n} K \left( \frac{Z_{i}}{h} \right) \left| g \left(
  \xi_{i} \right) \right| \left| \overline{\psi}_{1, r} \left( 0 \right)
  \right| \cdot \frac{\left[ \left( \widehat{P}_{- i} - P \right) \left[
  \psi_{2, r} \left( \cdot; 0 \right) \right] \right]^{2}}{r^{3} (\inf f)^{3}}
  \\
  & + \frac{\left( 1 + o_{\mathrm{p}} (1) \right) \mathbf{1}_{\mathcal{E}}}{n}
  \sum_{i = 1}^{n} K \left( \frac{Z_{i}}{h} \right) \left| g \left(
  \xi_{i} \right) \right| \frac{\left[ \left( \widehat{P}_{- i} - P \right)
  \left[ \psi_{1, r} \left( \cdot; Z_{i} \right) \right] \right]^{2}}{r^{2}
  (\inf f)^{2}} \\
  & + \frac{\left( 1 + o_{\mathrm{p}} (1) \right) \mathbf{1}_{\mathcal{E}}}{n}
  \sum_{i = 1}^{n} K \left( \frac{Z_{i}}{h} \right) \left| g \left(
  \xi_{i} \right) \right| \frac{\left[ \left( \widehat{P}_{- i} - P \right)
  \left[ \psi_{2, r} \left( \cdot; Z_{i} \right) \right] \right]^{2}}{r^{2}
  (\inf f)^{2}} \\
  & + \frac{\left( 1 + o_{\mathrm{p}} (1) \right) \mathbf{1}_{\mathcal{E}}}{n}
  \sum_{i = 1}^{n} K \left( \frac{Z_{i}}{h} \right) \left| g \left(
  \xi_{i} \right) \right| \frac{\left[ \left( \widehat{P}_{- i} - P \right)
  \left[ \psi_{1, r} \left( \cdot; 0 \right) \right] \right]^{2}}{r^{2}
  (\inf f)^{2}} \\
  & + \frac{\left( 1 + o_{\mathrm{p}} (1) \right) \mathbf{1}_{\mathcal{E}}}{n}
  \sum_{i = 1}^{n} K \left( \frac{Z_{i}}{h} \right) \left| g \left(
  \xi_{i} \right) \right| \frac{\left[ \left( \widehat{P}_{- i} - P \right)
  \left[ \psi_{2, r} \left( \cdot; 0 \right) \right] \right]^{2}}{r^{2}
  (\inf f)^{2}}
\end{align*}

Each term is a \(V\) statistic which can be separated into a \(U\) statistic and
a term \(V - U\) both of which have variances that evolve at maximal rate of
\(n^{- 2} O (h / r)\).
This proves \eqref{eqn--Tg-linear-rep-Rg-rate}.

\subsection{Proofs of auxiliary lemmas used for the proof of Theorem
\ref{thm--spilloverreg-inference}}

\subsubsection{A useful leave-one-out result}

Let \(\widehat{P}_{- i}\) denote the empirical leave-\(i\)-out distribution,
i.e.
\begin{equation*}
  \widehat{P}_{- i} [g (\xi)] = \int g (x) \widehat{P}_{- i} (\mathrm{d} x) =
  \frac{1}{n - 1} \sum_{j = 1, j \neq i}^{n} g \left( \xi_{j} \right).
\end{equation*}
Following usual empirical process notation, we denote \(P g = P [g (\xi)] = \int
g (x) P (\mathrm{d} x)\).

To estimate \(\mu\), we can use its empirical analogue.
To that end, define
\begin{equation}
  \begin{split}
    \psi_{1, r, z} \left( \xi_{1} \right) :=
    & \, \psi_{1, r} \left( \xi_{1}, z \right) = Y_{1} R \left( \frac{Z_{1} -
    z}{r} \right), \\
    \psi_{2, r, z} \left( \xi_{1} \right) :=
    & \, \psi_{2, r} \left( \xi_{1}, z \right) = R \left( \frac{Z_{1} - z}{r}
    \right).
  \end{split}
\end{equation}
Furthermore,
\begin{equation}
  \begin{split}
    \text{define} \quad
    & \overline{\psi}_{1, r} (z) = P \left[ \psi_{1, r, z} \right] = \int
    \psi_{1, r} \left( x_{1}, z \right) P \left( \mathrm{d} x_{1} \right) \\
    \text{and note that} \quad
    & \pi_{r} (z) = P \left[ \psi_{2, r, z} \right] = \int
    \psi_{2, r} \left( x_{1}, z \right) P \left( \mathrm{d} x_{1} \right) \\
    \text{and} \quad & \mu_{r} (z) =
    \frac{\overline{\psi}_{1, r} (z)}{\pi_{r} (z)} \quad \text{by } \E_{P} [Y -
    Y (Z; D) | Z] = 0.
  \end{split}
\end{equation}
An analogue estimator for \(\mu (z)\) is then
\begin{equation}
  \widehat{\mu}_{i} (z) = \frac{\widehat{P}_{- i} \left[ \psi_{1, r, z}
  \right]}{\widehat{P}_{- i} \left[ \psi_{2, r, z} \right]}.
  \label{eqn--endog-spill-muhat-def}
\end{equation}
The following lemma helps us to characterize the behavior of the difference
\(\widehat{\mu}_{i} (z) - \mu (z)\).

\begin{lemma}
\label{lem--muhat-loo-mu-taylor-expand}
Let
\begin{equation}
  \varphi_{r, z} \left( \xi_{1} \right) := \varphi_{r}
  \left( \xi_{1}, z \right) := \frac{1}{\pi_{r} (z)} \cdot
  \psi_{1, r, z} \left( \xi_{1} \right) - \frac{\overline{\psi}_{1, r}
  (z)}{\pi_{r} (z)^{2}} \cdot \psi_{2, r, z} \left( \xi_{1}
  \right),
  \label{eqn--muhat-loo-mu-diff-influence-varphi-def}
\end{equation}
and let \(\widehat{\mathrm{Rem}}_{i, r, z}\) be defined by
\begin{equation}
  \widehat{\mathrm{Rem}}_{i, r, z} := \widehat{\mathrm{Rem}}_{i, r} (z) :=
  \widehat{\mu}_{i} (z) - \mu (z) - \left( \widehat{P}_{- i} - P \right) \left[
  \varphi_{r} (\cdot, z) \right].
  \label{eqn--muhat-loo-mu-taylor-remainder}
\end{equation}
If \(\inf f := \inf_{z \in [-1 , 1]} f (z) > 0\), then
\(\widehat{\mathrm{Rem}}_{i, r, z}\) satisfies the following:
\begin{equation}
  \begin{gathered}
    \text{If} \quad \left| \left( \widehat{P}_{- i} - P \right) \left[ \psi_{2,
    r, z} \right] \right| \leq \frac{r}{2} \inf f, \\
    \begin{aligned}
      \text{then} \quad \left| \widehat{\mathrm{Rem}}_{i, r, z} \right| \leq
      & \,
      2 \left| P \psi_{1, r, z} \right| \cdot \frac{\left| \left( \widehat{P}_{-
      i} - P \right) \left[ \psi_{2, r, z} \right] \right|^{2}}{r^{3} (\inf
      f)^{3}} \\
      & + 2 \frac{\left| \left( \widehat{P}_{- i} - P \right) \left[ \psi_{1, r,
      z} \right] \right| \cdot \left| \left( \widehat{P}_{- i} - P \right)
      \left[ \psi_{2, r, z} \right] \right|}{r^{2} (\inf f)^{2}}.
    \end{aligned}
  \end{gathered}
  \label{eqn--muhat-loo-mu-taylor-expand-remainder-magnitude}
\end{equation}
\end{lemma}

\begin{lemma}
\label{lem--Ppsi2-denominator-control}
If \(\inf f := \inf_{z \in [-1 , 1]} f (x) > 0\), then for every \(r \in (0,
1]\),
\begin{equation*}
  \inf_{z \in [-1, 1]} P \left[ \psi_{2, r, z} \right] \geq r \inf f.
\end{equation*}
\end{lemma}

Lemma \ref{lem--reciprocal-approx-k-order} below is an error bound on the Taylor
approximation for reciprocals of positive numbers.
We omit its straightforward proof.

\begin{lemma}
\label{lem--reciprocal-approx-k-order}
Let \(k \in \mathbb{N}\) and \(a_{0} \geq a_{\min} > 0\).
Then for any \(a > 0\),
\begin{equation}
  \frac{1}{a} - \frac{1}{a_{0}} - \frac{1}{a_{0}} \sum_{j = 1}^{k} \left(
  \frac{a_{0} - a}{a_{0}} \right)^{j} = \left( \frac{a_{0} - a}{a_{0}}
  \right)^{k} \left( \frac{1}{a} - \frac{1}{a_{0}} \right).
  \label{eqn--reciprocal-taylor-and-rem-use-bounds}
\end{equation}
Thus for \(a > 0\) such that \(\left| a - a_{0} \right| < a_{\min}\),
\begin{equation}
  \left| \frac{1}{a} - \frac{1}{a_{0}} - \frac{1}{a_{0}} \sum_{j = 1}^{k} \left(
  \frac{a_{0} - a}{a_{0}} \right)^{j} \right| \leq \frac{1}{1 - \left( \left| a
  - a_{0} \right| / a_{\min} \right)} \cdot \frac{\left| a - a_{0}
  \right|^{k + 1}}{a_{\min}^{k + 2}}.
  \label{eqn--reciprocal-approx-k-order}
\end{equation}
Furthermore, if \(\left| a - a_{0} \right| \leq \frac{1}{2} a_{\min}\),
\begin{equation}
  \left| \frac{1}{a} - \frac{1}{a_{0}} - \frac{1}{a_{0}} \sum_{j = 1}^{k} \left(
  \frac{a_{0} - a}{a_{0}} \right)^{j} \right| \leq \frac{2 \left| a - a_{0}
  \right|^{k + 1}}{a_{\min}^{k + 2}}.
  \label{eqn--reciprocal-approx-k-order-2}
\end{equation}
\end{lemma}

\(\varphi_{r}\) in
\eqref{eqn--muhat-loo-mu-diff-influence-varphi-def} has the following
alternative representation:
\begin{equation}
  \varphi_{r} \left( \xi_{1}, z \right) = \left( Y_{1} -
  \frac{\overline{\psi}_{1, r} (z)}{\pi_{r} (z)} \right)
  \frac{\psi_{2, r} \left( \xi_{1}, z \right)}{\pi_{r} (z)} =
  \left( Y_{1} - \mu_{r} (z) \right) \frac{R \left( \frac{Z_{1} - z}{r}
  \right)}{\pi_{r} (z)}.
  \label{eqn--muhat-loo-mu-diff-influence-varphi-def-alt-1}
\end{equation}
Define the centered version of \(\varphi_{r}\) by
\begin{equation}
  \dot{\varphi}_{r, z} (\xi) := \dot{\varphi}_{r} (\xi,
  z) := \varphi_{r} (\xi, z) - P \left[ \varphi_{r}
  (\cdot, z) \right].
\end{equation}
Then
\begin{equation}
  P \left[ \dot{\varphi}_{r, z} \right] = \E_{\xi \sim P} \left[
  \dot{\varphi}_{r, z} (\xi) \right] = 0 \quad \text{for all } z
  \text{ and } r,
  \label{eqn--muhat-loo-mu-diff-influence-varphi-dot-mean-zero}
\end{equation}
and \(\widehat{\mathrm{Rem}}_{i, r} (z)\) in
\eqref{eqn--muhat-loo-mu-taylor-remainder} also satisfies
\begin{equation}
  \widehat{\mu}_{i} (z) - \mu (z) = \widehat{P}_{- i} \left[
  \dot{\varphi}_{r} (\cdot, z) \right] + \widehat{\mathrm{Rem}}_{i,
  r} (z).
  \label{eqn--muhat-loo-mu-taylor-remainder-varphi-dot}
\end{equation}

\subsubsection{Proof of Lemma \ref{lem--muhat-loo-mu-taylor-expand}}
For the purpose of this proof, let \(r, z\) be fixed and denote \(\psi_{j} =
\psi_{j, r, z}\) for \(j \in \{1, 2\}\).
From \eqref{eqn--muhat-loo-mu-diff-influence-varphi-def} and
\eqref{eqn--muhat-loo-mu-taylor-remainder},
\begin{equation*}
  \widehat{\mathrm{Rem}}_{i, r, z} =
  \frac{\widehat{P}_{- i} \left[ \psi_{1} \right]}{\widehat{P}_{- i} \left[
  \psi_{2} \right]} - \frac{P \left[ \psi_{1} \right]}{P \left[ \psi_{2}
  \right]} - \frac{\left( \widehat{P}_{- i} - P \right) \left[ \psi_{1}
  \right]}{P \left[ \psi_{2} \right]} + \frac{P \left[ \psi_{1} \right]}{\left(
  P \left[ \psi_{2} \right] \right)^{2}} \cdot \left( \widehat{P}_{- i} - P
  \right) \left[ \psi_{2} \right]
\end{equation*}
This is a first order Taylor approximation remainder.
For the bound in \eqref{eqn--muhat-loo-mu-taylor-expand-remainder-magnitude},
combine Lemmas \ref{lem--Ppsi2-denominator-control} and
\ref{lem--reciprocal-approx-k-order} below.

\subsubsection{Proof of Lemma \ref{lem--Ppsi2-denominator-control}}
\label{sec--prf--lem--Ppsi2-denominator-control}

Assume \(0 < r \leq 1\).
For every \(z \in [- 1, 1]\),
\begin{align*}
  P \left[ \psi_{2, r, z} \right] =
  & \, \int R \left( \frac{u - z}{r} \right) f (u) \, \mathrm{d} u \\
  \geq
  & \, (\inf f) \int_{- 1}^{1} R \left( \frac{u - z}{r} \right) \, \mathrm{d} u
  \\
  =
  & \, r \inf f \int_{(- 1 - z) / r}^{(1 - z) / r} R (v) \, \mathrm{d} v.
\end{align*}
Switching variables back to \(u\), we have
\begin{equation*}
  P \left[ \psi_{2, r} (\xi, z) \right] \geq r \inf f \int_{(- 1 - z) / r}^{(1 -
  z) / r} R (u) \, \mathrm{d} u.
\end{equation*}
By \(R (u) = \mathbf{1}_{[- 1, 1]} (u)\),
\begin{align*}
  P \left[ \psi_{2, r, z} \right]
  \geq
  & \, r \inf f \int_{\max \{- 1, (- 1 - z) / r\}}^{\min \{1, (1 - z) / r\}} \,
  \mathrm{d} u \\
  =
  & \, r \inf f \cdot \left( \min \{1, (1 - z) / r\} - \max \{- 1, (- 1 - z) /
  r\} \right),
\end{align*}

Start with \(z \geq 0\).
Then \(\max \{- 1, (- 1 - z) / r\} = - 1\).
Hence
\begin{align*}
  P \left[ \psi_{2, r, z} \right]
  \geq
  & \, r \inf f \cdot \left( 1 + \min \{1, (1 - z) / r\} \right) \\
  \geq
  & \, r \inf f,
\end{align*}
since \(\min \{1, (1 - z) / r\} \geq 0\) for every \(z \in [0, 1]\).

Next consider \(- 1 \leq z < 0\).
Then \(\min \{1, (1 - z) / r\} = 1\).
Hence
\begin{align*}
  P \left[ \psi_{2, r, z} \right]
  \geq
  & \, r \inf f \cdot \left( 1 - \max \{- 1, (- 1 - z) / r\} \right) \\
  =
  & \, r \inf f \cdot \left( 1 + \min \{1, (1 + z) / r\} \right) \\
  \geq
  & \, r \inf f
\end{align*}
since \(\min \{1, (1 + z) / r\} \geq 0\) for every \(z \in [-1, 0)\).

\section{Concentration Inequalities}

\begin{theorem}
\label{lem--vc-prob-inequality}
Let \(\widehat{F}_{n}\) denote empirical measure under an iid
sample of size \(n\) from a probability measure \(F\).
Let \(\mathcal{G}\) be a Vapnik-{\u C}ervonenkis (VC) class of measurable
functions with VC-dimension \(\mathcal{V} (\mathcal{G})\) such that \(\sup_{g
\in \mathcal{G}} \|g\|_{\infty} \leq 1\).
Then there exists a universal positive constant \(K \in (0, \infty)\) not
depending on \(\mathcal{G}\), \(P\) and \(n\) such that
\small
\begin{equation}
  \begin{split}
    & \Pr \left\{ \left\| \widehat{F}_{n} - F \right\|_{\mathcal{G}} > 8 y
    \right\} \\
    \leq
    & \, 16 K \cdot \mathcal{V} (\mathcal{G}) (16 e)^{\mathcal{V} (\mathcal{G})}
    \left[ \exp \left\{- \frac{n y^{2}}{128 t^{2}} + \mathcal{V} (\mathcal{G})
    \log (1 / y) \right\} + \exp \left\{ - n t^{2} + 2 \mathcal{V} (\mathcal{G})
    \log (1 / t) \right\} \right],
  \end{split}
  \label{eqn--vc-prob-inequality}
\end{equation}
\normalsize
for every \(y, t > 0\) such that
\begin{equation}
  y \geq \frac{1}{\sqrt{8 n}} \sup_{g \in \mathcal{G}} \sqrt{F \left[ g^{2}
  \right] - (F g)^{2}} \quad \text{and} \quad t \geq \sup_{g \in \mathcal{G}}
  \sqrt{F \left[ g^{2} \right]}.
  \label{eqn--vc-prob-inequality-premise}
\end{equation}
\end{theorem}

\begin{remark}
Recall that \(\|\widehat{F}_{n} - F\|_{\mathcal{G}} = \sup_{g \in \mathcal{G}}
|\widehat{F}_{n} [g] - F [g]|\).
In cases where measurability of this quantity is of concern,
\eqref{eqn--vc-prob-inequality} still holds with outer probability \(\Pr^{\ast}
\left\{ \cdot \right\}\) in place of \(\Pr \left\{ \cdot \right\}\).
\end{remark}

\begin{proof}[Proof of Lemma \ref{lem--vc-prob-inequality}]
Combine the covering number results for VC classes in Theorem 2.6.7 in
\citet[p. 206]{2023vandervaartWeakConvergenceEmpirical} with the
proof of Theorem II.37 in
\citet[pp. 34 and 35]{1984pollardConvergenceStochasticProcesses}.
\end{proof}

For our use, the following parameterized classes will be of interest:
\begin{equation}
  \mathcal{G}_{r} = \left\{ g_{z, r} : z \in [- 1, 1] \right\} \quad
  \text{where} \quad g_{z, r} \left( z_{1} \right) = R \left( \frac{z_{1} -
  z}{r} \right).
  \label{eqn--Gr-gzr-def}
\end{equation}
Recall that \(R (u) = \mathbf{1} \{|u| \leq 1\}\), but all subsequent arguments
apply to functions of the form \(R (u) = \psi (|u|)\) for \(\psi
(\cdot)\) non-increasing on \([0, \infty)\).
We parameterize the classes by \(r\) so that subsequent concentration
inequalities depend on \(r\).
For any fixed \(r > 0\), \(\mathcal{G}_{r}\) satisfies \(\mathcal{V} \left(
\mathcal{G}_{r} \right) \leq 2\) --- see \citet[Chapter II Problem
28]{1984pollardConvergenceStochasticProcesses}.
Now, denote
\begin{equation}
  \widehat{\pi}_{i, r} (z) = \frac{1}{n - 1} \sum_{j = 1, j \neq i}^{n}
  \mathbf{1} \left\{ \left| Z_{j} - z \right| \leq r \right\} \quad \text{and}
  \quad \pi_{r} (z) = \int R \left( \frac{z_{1} - z}{r} \right) f \left( z_{1}
  \right) \; \mathrm{d} z_{1}.
  \label{eqn-pi-hat-and-pi-redef}
\end{equation}

\begin{remark}
In \eqref{eqn-pi-hat-and-pi-redef},
\begin{enumerate}
  \item \(\widehat{\pi}_{i, r} (z)\) is the empirical probability
    of being \(i\)'s neighbor if \(i\) had \(z\);
  \item \(\pi_{r} (z)\) is the population probability of being neighbors with
    someone with running variable \(z\).
\end{enumerate}
\end{remark}

\begin{remark}
Denote the probability measure with Lebesgue density \(f\) by \(F\).
Denote the leave-\(i\)-out empirical measure integrating \(Z_{j}\)'s with \(j
\in \{1, \dots, n\} \setminus \{i\}\) by \(\widehat{F}_{- i}\), i.e.
\begin{equation*}
  \widehat{F}_{- i} [g] := \frac{1}{n - 1} \sum_{j = 1, i \neq j}^{n} g \left(
  Z_{i} \right).
\end{equation*}
Then \(\widehat{\pi}_{i, r} (z)\) and \(\pi_{r} (z)\) in
\eqref{eqn-pi-hat-and-pi-redef} can be rewritten in empirical process notation
via \eqref{eqn--Gr-gzr-def} as
\begin{equation}
  \pi_{i, r} (z) = \widehat{F}_{- i} \left[ g_{z, r} \right] \quad \text{and}
  \quad \pi_{r} (z) = F \left[ g_{z, r} \right].
  \label{eqn--pi-hat-and-pi-emp-proc}
\end{equation}
\end{remark}

\begin{lemma}
\label{lem--pi-hat-concentration-main}
Assume \(\|f\|_{\infty} < \infty\).
Take \(C \geq \|f\|_{\infty} \int R (u)^{2} \; \mathrm{d} u\), \(C_{\ast} = 128
C\) and \(y > 0\) such that \(y \geq \sqrt{\frac{C r}{8 n}}\).
Then there is a universal constant \(A\) not depending on \(r\) or \(n\) such
that
\begin{equation}
  \begin{split}
    & \max_{i = 1, \dots, n} \Pr \left\{ \sup_{z \in [- 1, 1]} \left|
    \widehat{\pi}_{i, r} (z) - \pi_{r} (z)
    \right| > 8 y \right\} \\
    \leq
    & \, A \left[ \exp \left\{- \frac{(n - 1) y^{2}}{C_{\ast} r} + 2 \log (1 /
    y) \right\} + \exp \left\{ - C (n - 1) r + 2 \log (1 / (C r)) \right\}
    \right].
  \end{split}
  \label{eqn--pi-hat-concentration-i}
\end{equation}
and therefore,
\begin{equation}
  \begin{split}
    & \Pr \left\{ \max_{i = 1, \dots, n} \sup_{z \in [- 1, 1]} \left|
    \widehat{\pi}_{i, r} (z) - \pi_{r} (z)
    \right| > 8 y \right\} \\
    \leq
    & \, A \exp \left\{- \frac{(n - 1) y^{2}}{C_{\ast} r} + 2 \log (1 /
    y) + \log n \right\} \\
    & + A \exp \left\{ - C (n - 1) r + 2 \log (1 / (C r)) + \log n \right\}.
  \end{split}
  \label{eqn--pi-hat-concentration-i-max}
\end{equation}
For sequences \(r = r_{n}\) such that \(r \downarrow 0\) and \(\log n / (n r)
\downarrow 0\), the first exponential term (dependent on \(y\)) dominates in
both \eqref{eqn--pi-hat-concentration-i} and
\eqref{eqn--pi-hat-concentration-i-max}.
\end{lemma}

\begin{proof}[Proof of Lemma \ref{lem--pi-hat-concentration-main}]
It is straightforward to show that
\begin{equation}
  \sup_{g \in \mathcal{G}} \left( F \left[ g^{2} \right] - (F g)^{2} \right)
  \leq \sup_{g \in \mathcal{G}} F \left[ g^{2} \right] \leq C r.
  \label{eqn--kappa2-crude-bound-Lebesgue}
\end{equation}

Following the notation conventions in \eqref{eqn--pi-hat-and-pi-emp-proc}, and
using \eqref{eqn--vc-prob-inequality},
\begin{equation*}
  \begin{split}
    & \Pr \left\{ \left\| \widehat{F}_{- i} - F \right\|_{\mathcal{G}_{r}} > 8 y
    \right\} \\
    \leq
    & \, 32 K (16 e)^{2}
    \left[ \exp \left\{- \frac{(n - 1) y^{2}}{128 t^{2}} + 2 \log (1 / y)
    \right\} +
    \exp \left\{ - (n - 1) t^{2} + 4 \log (1 / t) \right\} \right],
  \end{split}
\end{equation*}
for every \(y, t > 0\) such that
\begin{equation*}
  y \geq \sqrt{\frac{C r}{8 n}} \quad
  \text{and} \quad t \geq \sqrt{C r},
\end{equation*}
by \eqref{eqn--vc-prob-inequality-premise} and
\eqref{eqn--kappa2-crude-bound-Lebesgue}.
Since this holds for every \(i \in \{1, \dots, n\}\), set \(A = 32 K (16
e)^{2}\), \(t = \sqrt{C r}\), \(C_{\ast} = 128 C\) to get
\begin{equation*}
  \begin{split}
    & \max_{i = 1, \dots, n} \Pr \left\{ \left\| \widehat{F}_{- i} - F
    \right\|_{\mathcal{G}_{r}} > 8 y \right\} \\
    \leq
    & \, A \left[ \exp \left\{- \frac{(n - 1) y^{2}}{C_{\ast} r} + 2 \log (1 /
    y) \right\} + \exp \left\{ - C (n - 1) r + 2 \log (1 / (C r)) \right\}
    \right],
  \end{split}
\end{equation*}
which is exactly \eqref{eqn--pi-hat-concentration-i}.

Then, \eqref{eqn--pi-hat-concentration-i-max} follows from \(\Pr \left\{ \max_{j
= 1, \dots, N} V_{j} > \varepsilon \right\} \leq N \max_{j = 1, \dots, N} \Pr
\left\{ V_{j} > \varepsilon \right\}\) for arbitrary random variables \(V_{1},
\dots, V_{N}\).
The claim about the dominant first term stems from the fact that
for \(r_{n} \downarrow 0\),
\begin{equation*}
  \frac{\log n}{n r_{n}} \to 0 \quad \text{if and only if} \quad \frac{\log
  \left( 1 / r_{n} \right)}{n r_{n}} \to 0.
\end{equation*}
Therefore for the second term independent of \(y\) for example in
\eqref{eqn--pi-hat-concentration-i-max}
\begin{equation*}
  \begin{split}
    & \exp \left\{ - C (n - 1) r + 2 \log (1 / (C r)) + \log n \right\} \\
    =
    & \, \exp \left\{ - C (n - 1) r \left( 1 + \frac{2 \log (1 / (C r))}{n} +
    \frac{\log n}{n r} \right) \right\} = \exp \left\{ - C (n - 1) r \left( 1 +
    o (1) \right) \right\},
  \end{split}
\end{equation*}
and \(\log n / (n r) \downarrow 0\) implies \(n r \uparrow \infty\).
This term tends to zero regardless of the chosen \(y > 0\) with \(y \geq
\sqrt{(C r) / (8 n)}\).
\end{proof}

\begin{lemma}
\label{lem--maxpihat-diff-unif-opr}
Assume \(\|f\|_{\infty} < \infty\).
Let \(r_{n}\) be a sequence such that \(r_{n} \downarrow 0\) and \(\log n /
\left( n r_{n} \right) \downarrow 0\).
Then for any \(\varepsilon > 0\),
\begin{equation*}
  \Pr \left\{ \max_{i = 1, \dots, n} \sup_{z \in [- 1, 1]} \left|
  \widehat{\pi}_{i, r_{n}} (z) - \pi_{r_{n}} (z) \right| > r \varepsilon
  \right\} \to 0, \quad \text{as } n \to \infty.
\end{equation*}
Therefore, \(\max_{i = 1, \dots, n} \sup_{z \in [- 1, 1]} \left|
\widehat{\pi}_{i, r_{n}} (z) - \pi_{r_{n}} (z) \right| = o_{\mathrm{p}} \left(
r_{n} \right)\).
\end{lemma}

\begin{proof}[Proof of Lemma \ref{lem--maxpihat-diff-unif-opr}]
Write \(r = r_{n}\) and let \(A\), \(C\) and \(C_{\ast}\) be as defined in Lemma
\ref{lem--pi-hat-concentration-main}.
Set \(y = (\varepsilon r) / 8\).
Then \(y \geq \sqrt{\frac{C r}{8 n}}\) for \(n\) sufficiently large.
To see this note that \(y \geq \sqrt{\frac{C r}{8 n}}\) iff
\(\frac{\varepsilon}{\sqrt{8 C}} \geq \sqrt{\frac{1}{n r}}\).
By \(\log n / (n r) \to 0\), \(1 / (n r) \to 0\) and so, \(y \geq \sqrt{\frac{C
r}{8 n}}\) for \(n\) sufficiently large.
By Lemma \ref{lem--pi-hat-concentration-main},
\begin{align*}
  & \Pr \left\{ \max_{i = 1, \dots, n} \sup_{z \in [- 1, 1]} \left|
  \widehat{\pi}_{i, r_{n}} (z) - \pi_{r_{n}} (z) \right| > r \varepsilon
  \right\} \\
  =
  & \,  \Pr \left\{ \max_{i = 1, \dots, n} \sup_{z \in [- 1, 1]} \left|
  \widehat{\pi}_{i, r_{n}} (z) - \pi_{r_{n}} (z) \right| > 8 y
  \right\} \\
  \leq
  & \, A \exp \left\{- \frac{(n - 1) y^{2}}{C_{\ast} r} + 2 \log (1 /
  y) + \log n \right\} \\
  & + A \exp \left\{ - C (n - 1) r + 2 \log (1 / (C r)) + \log n \right\} \\
  =
  & \, A \exp \left\{- \frac{(n - 1) \varepsilon r}{64 C_{\ast}} + 2 \log (8 /
  (\varepsilon r)) + \log n \right\} \\
  & + A \exp \left\{ - C (n - 1) r + 2 \log (1 / (C r)) + \log n \right\}.
\end{align*}
Both terms in the final two lines above tend to zero by \(\log n / \left( n
r_{n} \right) \to 0\) (which is equivalent to \(\log \left( 1 / r_{n} \right) /
\left( n r_{n} \right) \to 0\) when \(r_{n} \to 0\)).
\end{proof}

\begin{lemma}\label{lemma--conc_mu}
	Suppose $h_n \gg n^{-1/3}$. Then
	\begin{equation*}
		 \max_{i: Z_i \in [-h_n, h_n]} \left\lvert \tilde{\mu}(Z_i) - \mu(Z_i)\right\rvert = o(h_n) \quad \mbox{ w.p.a. 1}~.
	\end{equation*}
\end{lemma}

\begin{proof}
	Let $\varepsilon = \frac{f(0)h_n}{2r_n}$. Let $\Gamma$ be the event that all $Z_i$ have between $f(0) \cdot nh_n/2$ and $f(0) \cdot 3nh_n/2$ neighbors. Then by Lemma \ref{lem--maxpihat-diff-unif-opr}, $\mathbb{P}(\Gamma) \to 1$.  Let $K$ be the upper bound of $Y_d(z)$. That is, $Y_d(z) \leq K$ for all $z \in [-1,1]$. Applying Hoeffding's Inequality to the conditional distribution of $Y_j \mid Z_j \in R(Z_i)$ and on the event $\Gamma$, we have:
	\begin{align*}
		& \mathbb{P}\left( \left\lvert \frac{1}{(n-1)\hat{\pi}_{i,r}}\sum_{j =1, j \neq i}^n Y_j \mathbf{1}\{ \lvert Z_j - z \rvert \leq r_n \}  - \mu(Z_i) \right\rvert \geq (nh_n)^{-1/2+\eta} \right) \\
		& \leq \exp\left( - \frac{(nh_n)^{1+2\eta}}{(n-1)\hat{\pi}_{i,r} K^2}\right)\\
		& \leq  \exp\left( - \frac{2(nh_n)^{1+2\eta}}{(n-1)h_n K^2}\right)  \quad \mbox{ on } \Gamma~.
	\end{align*}
	Now, w.p.a. 1, there are no more than $3nh_n/2$ observations on $[-h_n, h_n]$. On this event and $\Gamma$, the union bound gives us that:
	\begin{equation*}
		\mathbb{P}\left( \max_{i: Z_i \in [-h_n, h_n]} \left\lvert\tilde{\mu}(Z_i) - \mu(Z_i) \right\rvert \geq (nh_n)^{-1/2+\eta} \right) \leq \frac{3f(0)}{2}\cdot nh_n \exp\left( - \frac{2}{K^2} (nh_n)^{2\eta}\right)  \to 0
	\end{equation*}
	Since $h_n \gg n^{-1/3}$, choose $\eta$ such that $(nh_n)^{-1/2+\eta} \ll h_n$ and we are done.
\end{proof}

\end{document}